\documentclass[aps,prl,twocolumn,superscriptaddress,showpacs,floatfix]{revtex4-2}

\usepackage{graphicx}

\usepackage{bm}
\usepackage{xcolor}
\usepackage{url}
\usepackage{bookmark}
\usepackage{amsmath}
\usepackage{tikz}
\usetikzlibrary{arrows.meta,decorations.markings,calc,3d,positioning}

\newcommand{\panelfont}{\fontfamily{ptm}\selectfont} 

\newcommand{\be}{\begin{equation}}
\newcommand{\ee}{\end{equation}}
\newcommand{\eq}[1]{Eq.~(\ref{#1})}
\newcommand{\fig}[1]{Fig.~\ref{#1}}
\def\bea{\begin{eqnarray}}
\def\eea{\end{eqnarray}}

\newcommand{\dop}{\delta}
\newcommand{\Jpar}{J}
\newcommand{\Jperp}{J_{\perp}}
\newcommand{\qvec}{\mathbf{q}}
\newcommand{\kvec}{\mathbf{k}}
\newcommand{\qc}{\mathbf{q}_{c}}
\newcommand{\zBOP}{z\text{-BOP}}
\newcommand{\rperp}{r_{\perp}}
\newcommand{\Aperp}{A_{\perp}}
\newcommand{\Dperp}{\Delta_{\perp}}
\def\ii{\mathrm{i}}

\begin{document}

\title{Competing Interlayer Loop Currents and Superconductivity \\ in the Bilayer $t$-$J_\perp$-$V$ Model}

\author{Luciano Zinni}
\affiliation{Facultad de Ciencias Exactas, Ingenier\'{\i}a y Agrimensura. Avenida Pellegrini 250, 2000 Rosario, Argentina}

\author{Fabricio G\'omez}
\affiliation{Facultad de Ciencias Exactas, Ingenier\'{\i}a y Agrimensura. Avenida Pellegrini 250, 2000 Rosario, Argentina}

\author{Jun Zhan}
\affiliation{Department of Physics, Nagoya University, Nagoya 464-8602, Japan}

\author{Mat\'{\i}as Bejas}
\affiliation{Facultad de Ciencias Exactas, Ingenier\'{\i}a y Agrimensura and Instituto de F\'{\i}sica Rosario (UNR-CONICET), Avenida Pellegrini 250, 2000 Rosario, Argentina}

\author{Xianxin Wu}
\affiliation{Institute of Theoretical Physics, Chinese Academy of Sciences, Beijing, China}

\author{Andreas P. Schnyder}
\affiliation{Max Planck Institute for Solid State Research, Heisenbergstrasse 1, 70569 Stuttgart, Germany}

\author{Andr\'es Greco}
\affiliation{Facultad de Ciencias Exactas, Ingenier\'{\i}a y Agrimensura and Instituto de F\'{\i}sica Rosario (UNR-CONICET), Avenida Pellegrini 250, 2000 Rosario, Argentina}

\date{\today}

\begin{abstract}
The recent discovery of high-$T_c$ superconductivity in pressurized and thin-film bilayer nickelates, featuring a strong interlayer exchange coupling, and their potential similarities with cuprate superconductors, has made this a very active topic in condensed matter physics. In the present paper we study the strongly correlated one-orbital ($d_{x^2-y^2}$) bilayer $t$-$J_\perp$-$V$ model for nickelates, where $V$ denotes the Coulomb interactions, using a controlled large-$N$ expansion at and beyond the mean-field level. Focusing on the out-of-plane spin exchange interaction ($J_\perp$), we find that it triggers both out-of-plane $s$-wave superconductivity and an out-of-plane bond-order phase ($z$-BOP) instability. The $z$-BOP gives rise to a complex $z$-axis hopping dominated by its imaginary component, which drives out-of-plane currents and induces in-plane ones, spontaneously forming on the vertical plaquettes a loop-current state that breaks time-reversal symmetry. Competition between this loop-current phase and superconductivity yields a dome-shaped superconducting region, with optimal superconductivity occurring near the $z$-BOP quantum critical point. The resulting phase diagram features a pure loop-current region, a low-doping coexistence phase, a pure superconducting state at higher doping, and a correlated metallic state.
\end{abstract}

\maketitle

\textit{Introduction}.-
The recent discovery of superconductivity (SC) at high critical temperature $T_c$ in pressurized~\cite{sun23,hou23} and ambient-pressure thin-film~\cite{ko25,zhou25} bilayer nickelates has motivated huge experimental 
and theoretical interest; models ranging from multiorbital ($d_{z^2}$ and $d_{x^2-y^2}$) to only one active orbital ($d_{x^2-y^2}$) have been proposed
(see Refs.~\cite{Qiu_2026,zhang2026,wang2024normal,zhang2026exp,wang_2025Re} for recent and comprehensive reviews).
Nowadays there is a general consensus that two key features seem to be relevant for analyzing the physics behind bilayer nickelates, i.e., the out-of-plane spin exchange $J_\perp$ between $d_{z^2}$ orbitals and the role of correlations~\cite{Qiu_2026,zhang2026,wang2024normal,zhang2026exp,wang_2025Re} which, similar to cuprates~\cite{keimer15}, can lead to high $T_c$ values. Part of the community has adopted the viewpoint that the 
combination of correlations and the strong Hund's coupling $J_H$ can
transfer the interlayer coupling from the $d_{z^2}$ to the $d_{x^2-y^2}$ orbitals, and thus a minimal one-orbital ($d_{x^2-y^2}$) bilayer $t$-$J$-$J_\perp$ model was proposed~\cite{lu24,qu24} (see also Refs.~\cite{oh2025dop,pan2026review} for recent reviews). 
This model exhibits mixed-dimensional character and can host highly mobile but tightly bound pairs~\cite{bohrdt2022strong}. Owing to its potential realization in experimental platforms such as ultracold atoms in optical lattices and superconducting quantum circuits~\cite{bourgund2025formation}, exploring its exotic states is of significant interest and may provide key insights into the physics of nickelates.

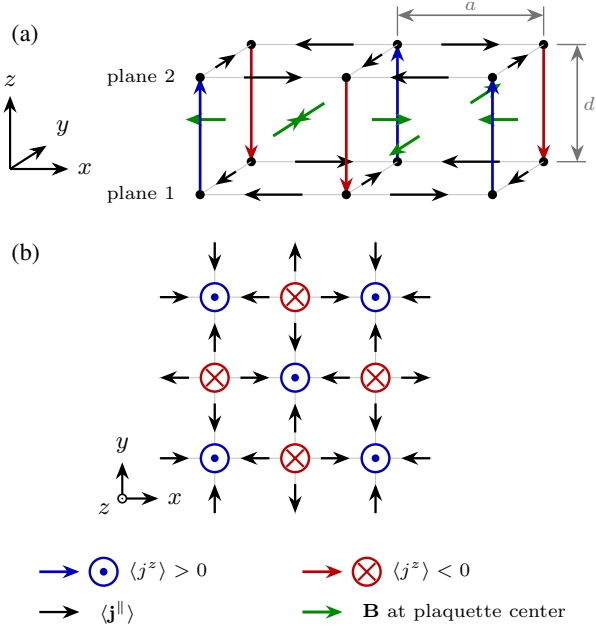
\begin{figure}[!t]
\centering
\def\hz{1.6}  
\resizebox{8cm}{!}{%
\begin{tikzpicture}[
  scale=0.95,
  >={Stealth[length=2.2mm,width=1.7mm]},
  site/.style={circle,fill=black,inner sep=1.2pt},
  jzup/.style={->,thick,blue!75!black,line width=1.0pt},
  jzdn/.style={->,thick,red!75!black,line width=1.0pt},
  jpar/.style={->,thick,black,line width=0.85pt},
  flux/.style={->,very thick,green!55!black,line width=1.1pt},
  ghostbond/.style={black!22,thin},
  dim/.style={<->,black!55,line width=0.6pt},          
  dimtick/.style={black!55,line width=0.5pt},          
  zout/.style={circle,draw=blue!75!black,fill=white,line width=0.9pt,
    inner sep=0pt,minimum size=3.6mm,
    path picture={\fill[blue!75!black] (path picture bounding box.center) circle (0.5mm);}},
  zin/.style={circle,draw=red!75!black,fill=white,line width=0.9pt,
    inner sep=0pt,minimum size=3.6mm,
    path picture={\draw[red!75!black,line width=0.8pt]
      ($(path picture bounding box.center)!0.62!(path picture bounding box.south west)$) --
      ($(path picture bounding box.center)!0.62!(path picture bounding box.north east)$)
      ($(path picture bounding box.center)!0.62!(path picture bounding box.north west)$) --
      ($(path picture bounding box.center)!0.62!(path picture bounding box.south east)$);}}
]
  \begin{scope}[shift={(0,0)}]
  \coordinate (B00) at (0,   0);          \coordinate (B10) at (2,   0);          \coordinate (B20) at (4,   0);
  \coordinate (B01) at (0.7, 0.45);       \coordinate (B11) at (2.7, 0.45);       \coordinate (B21) at (4.7, 0.45);
  \coordinate (T00) at (0,   {\hz});      \coordinate (T10) at (2,   {\hz});      \coordinate (T20) at (4,   {\hz});
  \coordinate (T01) at (0.7, {0.45+\hz}); \coordinate (T11) at (2.7, {0.45+\hz}); \coordinate (T21) at (4.7, {0.45+\hz});
  \draw[ghostbond] (B01)--(B11) (B11)--(B21) (T01)--(T11) (T11)--(T21);
  \draw[jzdn] (T01) -- (B01);
  \draw[jzup] (B11) -- (T11);
  \draw[jzdn] (T21) -- (B21);
  \draw[jpar] ($(B01)!0.30!(B11)$) -- ($(B01)!0.70!(B11)$);
  \draw[jpar] ($(B21)!0.30!(B11)$) -- ($(B21)!0.70!(B11)$);
  \draw[jpar] ($(T11)!0.30!(T01)$) -- ($(T11)!0.70!(T01)$);
  \draw[jpar] ($(T11)!0.30!(T21)$) -- ($(T11)!0.70!(T21)$);
  \coordinate (Pc) at (1.7, {0.45+\hz/2}); \draw[flux] (Pc) -- ($(Pc)-(0.42, 0.27)$);
  \coordinate (Pd) at (3.7, {0.45+\hz/2}); \draw[flux] (Pd) -- ($(Pd)+(0.42, 0.27)$);
  \foreach \p in {B01,B11,B21,T01,T11,T21}{ \node[site] at (\p) {}; }
  \draw[ghostbond] (B00)--(B01) (B10)--(B11) (B20)--(B21) (T00)--(T01) (T10)--(T11) (T20)--(T21);
  \draw[jpar] ($(B01)!0.30!(B00)$) -- ($(B01)!0.70!(B00)$);
  \draw[jpar] ($(B10)!0.30!(B11)$) -- ($(B10)!0.70!(B11)$);
  \draw[jpar] ($(B21)!0.30!(B20)$) -- ($(B21)!0.70!(B20)$);
  \draw[jpar] ($(T00)!0.30!(T01)$) -- ($(T00)!0.70!(T01)$);
  \draw[jpar] ($(T11)!0.30!(T10)$) -- ($(T11)!0.70!(T10)$);
  \draw[jpar] ($(T20)!0.30!(T21)$) -- ($(T20)!0.70!(T21)$);
  \coordinate (Pe) at (0.35, {0.225+\hz/2}); \draw[flux] (Pe) -- ($(Pe)-(0.55, 0)$);
  \coordinate (Pf) at (2.35, {0.225+\hz/2}); \draw[flux] (Pf) -- ($(Pf)+(0.55, 0)$);
  \coordinate (Pg) at (4.35, {0.225+\hz/2}); \draw[flux] (Pg) -- ($(Pg)-(0.55, 0)$);
  \draw[ghostbond] (B00)--(B10) (B10)--(B20) (T00)--(T10) (T10)--(T20);
  \draw[jzup] (B00) -- (T00);
  \draw[jzdn] (T10) -- (B10);
  \draw[jzup] (B20) -- (T20);
  \draw[jpar] ($(B10)!0.30!(B00)$) -- ($(B10)!0.70!(B00)$);
  \draw[jpar] ($(B10)!0.30!(B20)$) -- ($(B10)!0.70!(B20)$);
  \draw[jpar] ($(T00)!0.30!(T10)$) -- ($(T00)!0.70!(T10)$);
  \draw[jpar] ($(T20)!0.30!(T10)$) -- ($(T20)!0.70!(T10)$);
  \coordinate (Pa) at (1.0, {\hz/2});  \draw[flux] (Pa) -- ($(Pa)+(0.42, 0.27)$);
  \coordinate (Pb) at (3.0, {\hz/2});  \draw[flux] (Pb) -- ($(Pb)-(0.42, 0.27)$);
  \foreach \p in {B00,B10,B20,T00,T10,T20}{ \node[site] at (\p) {}; }
  \draw[dimtick] ($(T11)+(0,0.12)$) -- ($(T11)+(0,0.52)$);
  \draw[dimtick] ($(T21)+(0,0.12)$) -- ($(T21)+(0,0.52)$);
  \draw[dim] ($(T11)+(0,0.40)$) -- ($(T21)+(0,0.40)$)
    node[midway,above,font=\scriptsize,inner sep=1.5pt] {$a$};
  \draw[dimtick] ($(B21)+(0.18,0)$) -- ($(B21)+(0.58,0)$);
  \draw[dimtick] ($(T21)+(0.18,0)$) -- ($(T21)+(0.58,0)$);
  \draw[dim] ($(B21)+(0.46,0)$) -- ($(T21)+(0.46,0)$)
    node[midway,right,font=\scriptsize,inner sep=2pt] {$d$};
  \node[anchor=east,font=\scriptsize] at ($(B00)+(-0.22,0)$) {plane $1$};
  \node[anchor=east,font=\scriptsize] at ($(T00)+(-0.22,0)$) {plane $2$};
  \begin{scope}[shift={(-2.6,0.35)}]
    \draw[->,thick] (0,0) -- (0.8,0)        node[right]{$x$};
    \draw[->,thick] (0,0) -- (0.5,0.32)     node[above right]{$y$};
    \draw[->,thick] (0,0) -- (0,1.0)        node[above]{$z$};
  \end{scope}
  \end{scope}
  \begin{scope}[shift={(0.2,-3.6)},scale=0.55]
    \foreach \i in {0,1,2}{\foreach \j in {0,1,2}{\coordinate (s\i\j) at (2*\i,2*\j);}}
    \foreach \y in {0,2,4}{\draw[ghostbond] (0,\y) -- (4,\y);}
    \foreach \x in {0,2,4}{\draw[ghostbond] (\x,0) -- (\x,4);}
    \foreach \x/\y/\ang in {2/0/270, 0/2/180, 4/2/0, 2/4/90}{
      \draw[ghostbond] (\x,\y) -- ++(\ang:1.40);
      \draw[jpar] ($(\x,\y)+(\ang:0.64)$) -- ($(\x,\y)+(\ang:1.36)$);}
    \foreach \x/\y/\ang in {0/0/180,0/0/270,4/0/0,4/0/270,0/4/180,0/4/90,4/4/0,4/4/90}{
      \draw[ghostbond] (\x,\y) -- ++(\ang:1.40);
      \draw[jpar] ($(\x,\y)+(\ang:1.36)$) -- ($(\x,\y)+(\ang:0.64)$);}
    \foreach \a/\b in {s10/s00,s10/s20,s10/s11, s01/s00,s01/s02,s01/s11,
                       s21/s20,s21/s22,s21/s11, s12/s02,s12/s22,s12/s11}{
      \draw[jpar] ($(\a)!0.32!(\b)$) -- ($(\a)!0.68!(\b)$);}
    \foreach \x/\y in {0/0,4/0,2/2,0/4,4/4}{\node[zout] at (\x,\y) {};}
    \foreach \x/\y in {2/0,0/2,4/2,2/4}{\node[zin] at (\x,\y) {};}
    \begin{scope}[shift={(-2.3,-1.0)}]
      \draw[->,thick] (0,0) -- (0.9,0) node[right]{$x$};
      \draw[->,thick] (0,0) -- (0,0.9) node[above]{$y$};
      \node[circle,draw,fill=white,line width=0.6pt,inner sep=1.1pt] at (0,0) {};
      \fill (0,0) circle (0.6pt);
      \node[font=\small] at (-0.45,-0.25) {$z$};
    \end{scope}
  \end{scope}
  \begin{scope}[shift={(-2.2,-5.15)}]
    \draw[jzup] (0,0) -- (0.55,0);            \node[zout] at (0.88,0) {};
    \node[anchor=west,font=\scriptsize] at (1.10,0) {$\langle j^{z}\rangle>0$};
    \draw[jzdn] (3.60,0) -- (4.15,0);         \node[zin] at (4.48,0) {};
    \node[anchor=west,font=\scriptsize] at (4.70,0) {$\langle j^{z}\rangle<0$};
    \draw[jpar] (0,-0.55) -- (0.55,-0.55);
    \node[anchor=west,font=\scriptsize] at (0.70,-0.55) {$\langle\mathbf{j}^{\parallel}\rangle$};
    \draw[flux] (3.60,-0.55) -- (4.15,-0.55);
    \node[anchor=west,font=\scriptsize] at (4.30,-0.55) {$\mathbf{B}$ at plaquette center};
  \end{scope}
  \node[anchor=north west,font=\panelfont] at (-2.70, 2.45) {(a)};
  \node[anchor=north west,font=\panelfont] at (-2.70,-0.55) {(b)};
\end{tikzpicture}%
}
\caption{Schematic current and flux pattern of the $\Aperp$ $\zBOP$ for the
commensurate ordering vector $(\pi,\pi)$.
(a) Three-dimensional view. Blue (red) vertical arrows are the interlayer currents
$\langle j^{z}_{i}\rangle>0$ ($<0$) on opposite sublattices, generated by the staggered
Peierls phase of the effective interlayer hopping. Black
arrows are the induced in-plane currents, opposite in the two layers, which close each
vertical plaquette into a loop. Green arrows give
the local magnetic field $\mathbf{B}$ at the plaquette centers, along $\pm\hat{y}$
($\pm\hat{x}$) for $xz$ ($yz$) plaquettes and alternating with the sublattice, so that its
spatial average vanishes. (b) Top view of plane $1$: $\odot$ ($\otimes$) marks sites where
the interlayer current leaves (enters) the plane, in a checkerboard pattern,
and the in-plane currents flow from each $\langle j^{z}\rangle<0$ site to its four
neighbors, ensuring local charge conservation. All arrows and symbols are reversed in
plane $2$. $a$ and $d$ are the in-plane and intrabilayer lattice constants, respectively.}
\label{fig:pattern}
\end{figure}

In the present paper we study the bilayer $t$-$\Jperp$-$V$ model, i.e., we focus on the relevant out-of-plane spin exchange coupling $\Jperp$ and we do not consider the in-plane exchange interaction $J$. $V$ denotes the in-plane and the out-of-plane Coulomb interactions. Therefore, we adopt a strong-coupling perspective, in contrast to the weak-coupling theories (see Ref.~\cite{jianjian26} for a recent review). We implement a large-$N$ approximation based on the path-integral representation~\cite{foussats04,bejas12} for Hubbard operators~\cite{hubbard63}, and perform calculations at the mean-field and beyond-mean-field levels. We find out-of-plane $s$-wave superconductivity triggered by $\Jperp$. We also obtain an out-of-plane bond-order phase ($\zBOP$) instability~\cite{bejas25} below a critical temperature $T_c^{\zBOP}$, which decreases with increasing doping $\dop$ and reaches $T = 0$ at a quantum critical point (QCP). 
The out-of-plane bond field has a real ($\rperp$) and an imaginary ($\Aperp$) component, whose instabilities are nearly degenerate. However, the imaginary component $\Aperp$ is the leading one.
We solve the $\zBOP$ gap equation and show that the resulting condensate is precisely the imaginary channel $\Aperp$. The imaginary gap imprints a sublattice-staggered Peierls phase on the interlayer hopping, producing staggered interlayer currents and loop currents (LCs) that thread the vertical plaquettes of the bilayer, 
closed by in-plane currents that are automatically induced within the theory~\cite{WuCJ2004}, thereby ensuring local charge conservation (see the schematic in Fig.~\ref{fig:pattern}). This spontaneous interlayer LC state breaks time-reversal symmetry (TRS).

Finally, solving the coupled LC and superconducting gap equations, we obtain a dome-shaped behavior for SC, with the optimal value located at the doping where the LCs develop. With increasing doping from half-filling, the phase diagram shows a broad region in doping and temperature where the LC state is stable, a region where LCs and SC coexist at low temperature, a region with only SC, and a correlated metal at high doping.

\textit{The model}.-
We study the following bilayer model,
{\setlength{\jot}{2pt}%
\begin{align}
H =\;& \sum\nolimits_{i,j,\sigma,\alpha} t_{ij}\,
\tilde{c}^{\dagger}_{i\sigma,\alpha}\tilde{c}_{j\sigma,\alpha}
+ t_{\perp}\sum\nolimits_{i,\sigma,\alpha}
\tilde{c}^{\dagger}_{i\sigma,\alpha}\tilde{c}_{i\sigma,\bar{\alpha}}
\nonumber\\
&+\frac{J_\perp}{2}\sum\nolimits_{i,\alpha}
\Big(\vec{S}_{i,\alpha}\!\cdot\!\vec{S}_{i,\bar{\alpha}}
-\tfrac{1}{4}n_{i,\alpha}n_{i,\bar{\alpha}}\Big)
-\mu\sum\nolimits_{i,\alpha} n_{i,\alpha}
\nonumber\\
&+V_{\parallel}\sum\nolimits_{\langle i,j\rangle,\alpha} n_{i,\alpha}n_{j,\alpha}
+ V_{\perp}\sum\nolimits_{i} n_{i,1}n_{i,2}\,.
\label{eq:H}
\end{align}}
%

In \eq{eq:H} $\alpha=1,2$ labels the planes, $\bar{\alpha}$ is the plane opposite to $\alpha$, and $i$, $j$ run over the sites of the square lattice in each plane.
$\langle i,j \rangle$ indicates a nearest-neighbor pair of sites.
The hopping $t_{ij}$ takes the value $t$ between first-nearest and $t'$ between second-nearest neighbors. $t_\perp$ and $\Jperp$ are the out-of-plane hopping and spin exchange, respectively. $V_\parallel$ and $V_\perp$ are the in-plane and out-of-plane Coulomb interactions, respectively. $\mu$ is the chemical potential. 
$\tilde{c}^{\dagger}_{i\sigma,\alpha}$ ($\tilde{c}_{i\sigma,\alpha}$) creates (annihilates) electrons in the Fock space without double occupancy, and $n_{i,\alpha}$ and $\vec{S}_{i,\alpha}$ are the density and spin operators, respectively. The local constraint is handled by a large-$N$ expansion based on the path-integral representation for Hubbard operators~\cite{foussats04,bejas12} [see Sec.~S1 of the Supplemental Material (SM)~\cite{SM} for the complete formalism].
Energies are measured in units of $t$.

\textit{Results and discussions}.-
Without loss of generality and motivated by bilayer nickelates, we present results for $\Jperp=0.4t$, $t_\perp=0.05t$, and $t'=-0.15t$~\cite{luo23} (see Sec.~S6 of the SM~\cite{SM} for other values of $\Jperp$).
The temperature $T$ and the remaining parameters are indicated in each figure.

The Hamiltonian in \eq{eq:H} defines the six-component bosonic field 
\begin{align}
\delta X^{a} =& (\delta R_1,\;\delta{\lambda_1},\;
                \delta R_2,\;\delta{\lambda_2},\;
                \rperp,\; \Aperp)\, ,
\label{eq:boson-fieldp}
\end{align}
where $\delta R_\alpha$ describes the fluctuations of the number of holes at a given site in plane $\alpha$ and is therefore related to the on-site charge fluctuations; $\delta \lambda_\alpha$ is the fluctuation of the Lagrange multiplier introduced to enforce the constraint that prohibits double occupancy at any site of each plane; and $\rperp$ and $\Aperp$ are the real and imaginary parts, respectively, of the fluctuations of the out-of-plane bond field coming from the $\Jperp$ term. This field defines a $6 \times 6$ bosonic propagator $D_{ab}(\qvec, \ii\omega_n)$, which contains information on usual charge and $\zBOP$ fluctuations. See Sec.~S1 of the SM~\cite{SM} for technical details. There are two main sectors, a $4 \times 4$ sector with $a,b = 1$-$4$, which involves mainly usual charge excitations, and a $2 \times 2$ sector with $a,b=5$-$6$, which corresponds to $\zBOP$ excitations. Although both sectors are coupled, we will see later that this coupling is weak. 

Besides the parameters above, we choose $V_\parallel=1.0t$ and $V_\perp=0.3t$. The former avoids phase separation (PS), and the latter any vestige of charge-density-wave (CDW) order (see Fig.~S1 and Sec.~S2 of the SM~\cite{SM}). Thus, we can discuss the presence of $\zBOP$, SC, and their competition starting from a correlated homogeneous metallic state. It is possible to see that the instabilities as a function of doping and temperature computed from the full $6 \times 6$ propagator are the same as those obtained by projecting the full propagator onto the $2 \times 2$ $\zBOP$ sector (see Sec.~S2 of the SM~\cite{SM}). This fact has two important implications. (i) As in the usual $t$-$J$ model in two dimensions (2D)~\cite{bejas17}, there is a dual structure that decouples the local charge excitations from the $\Jperp$ sector that leads to $\zBOP$ excitations. (ii) Since the Coulomb terms $V_\parallel$ and $V_\perp$ belong to the $4 \times 4$ charge sector, the results show that the Coulomb interaction is irrelevant for the formation and stability of the $\zBOP$ and the corresponding LCs. 

The $\zBOP$ instability is signaled by the divergence of the corresponding
susceptibility. Figure~S2 of the SM~\cite{SM} shows the resulting critical temperature versus
doping for both the real ($\rperp$) and the imaginary ($\Aperp$) channels;
the two are nearly degenerate, with $\Aperp$ marginally the leading
one. The instability takes place at a momentum $\qc$ that depends on
$\dop$ and $T$, as indicated by the color scale, but
remains close to $(\pi,\pi)$. This fact shows that translational symmetry is also broken. In particular, at $T=0.02t$ it is already commensurate,
$\qc=(\pi,\pi)$, and the critical doping is $\dop \sim 0.15$.

\begin{figure}[ht]
\centering
\includegraphics[]{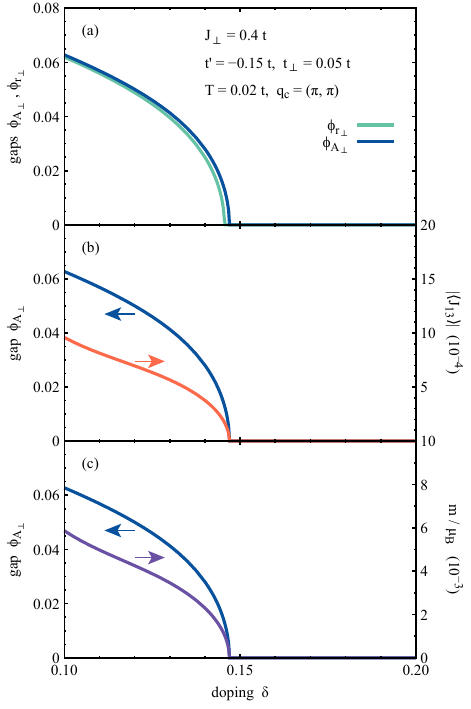}
\caption{(a) Real ($\phi_{\rperp}$) and imaginary ($\phi_{\Aperp}$) $\zBOP$ gaps vs doping
$\dop$, obtained separately. The two channels are nearly degenerate, $\Aperp$
being marginally the leading one. (b) Full complex gap
$\phi=\phi_{\rperp}+\ii\phi_{\Aperp}$, allowing competition: the solution is purely
imaginary ($\phi_{\rperp}$ negligible). The induced in-plane current
$|\langle J_{13}\rangle|$ (right axis, in units of $10^{-4}$) follows
$\phi_{\Aperp}$ (left axis), signaling the LCs of \fig{fig:pattern}. (c) Orbital
magnetic moment $m$ of a single loop (right axis, in units of $10^{-3}\mu_{B}$),
which follows $\phi_{\Aperp}$ (left axis) and vanishes at the QCP; see Sec.~S4 of the SM~\cite{SM} for details on its estimation.}
\label{fig:gaps_J13}
\end{figure}

To compute the complex $\zBOP$ gap $\phi=\phi_{\rperp}+\ii\phi_{\Aperp}$, we introduce the spinor
\begin{equation}
{\psi}^{\dagger}_{\kvec}= (f_{\kvec \sigma,1}^{\dagger},\;f_{\kvec \sigma,2}^{\dagger},\; f_{\kvec+\qc \sigma,1}^{\dagger},\;f_{\kvec+\qc \sigma,2}^{\dagger})\, .
\label{spinorzBOP}
\end{equation}
In \fig{fig:gaps_J13}(a) we show the real and imaginary gaps versus $\dop$ at $T=0.02t$, computed separately (see Sec.~S3 of the SM~\cite{SM}). 
Although the difference between the two gaps is very small, the imaginary one $\phi_{\Aperp}$ is larger. This small difference is due to the small value $t_\perp=0.05t$. For $t_\perp=0.12t$, the imaginary channel is more clearly favored (see Fig.~S3 of the SM~\cite{SM}).

However, when the full gap $\phi=\phi_{\rperp}+\ii\phi_{\Aperp}$ is considered, allowing the competition between the real and the imaginary gaps, the self-consistent solution is purely imaginary, with $\phi_{\rperp}$ negligible. This result is shown in \fig{fig:gaps_J13}(b), where we also show that the in-plane current is induced by the imaginary gap. Thus, the interlayer and in-plane currents lead to the LCs (see Sec.~S4 of the SM~\cite{SM}), as shown schematically in \fig{fig:pattern}. The magnetic moment of a single loop is a few times $10^{-3}\mu_{B}$, $\mu_{B}$ being the Bohr magneton; it decreases with doping and vanishes at the QCP [\fig{fig:gaps_J13}(c), see Sec.~S4 of the SM~\cite{SM}].

Loop currents were discussed long ago in the context of cuprates~\cite{affleck88a,chakravarty01,cappelluti99,varma97,kaminski02,fauque06}, but those studies were not conclusive. Recently, the topic has attracted renewed and active interest in cuprates~\cite{paul2026,bounoua_2026}, kagome materials~\cite{jiang2021unconventional,mielke2022time,guo22,wangka,fu2025exotic,Tazai_2026,jun26}, iridates~\cite{jeong17}, 
and on the surface of Sr$_2$RuO$_4$~\cite{fittipaldi21}. In addition, a recent report suggests a TRS-breaking phase in nickelates~\cite{ji2026}, which might signal the presence of LCs. Interestingly, specific STM signatures of an LC state were proposed very recently~\cite{morisseau26}.

\begin{figure}[ht]
\centering
\includegraphics[]{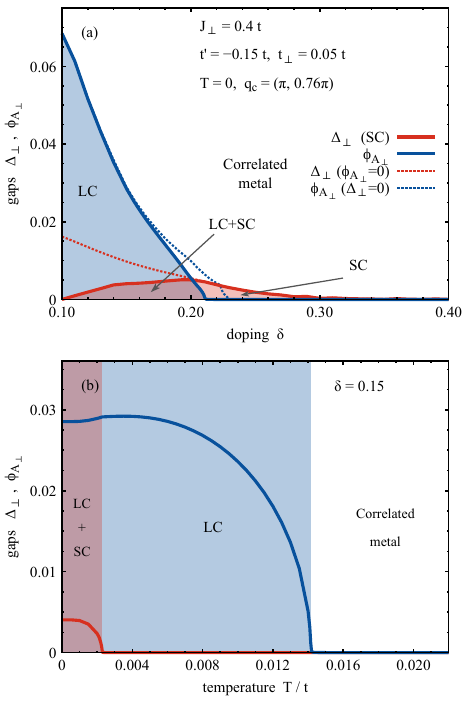}
\caption{(a) Superconducting gap $\Dperp$ and $\zBOP$ gap $\phi_{\Aperp}$ vs doping
$\delta$ at $T=0$, from the coupled gap equations (solid). Dashed lines are the gaps
without competition, $\Dperp$ at $\phi_{\Aperp}=0$ and $\phi_{\Aperp}$ at $\Dperp=0$. $\Dperp$ shows a dome
with optimal doping close to the $\zBOP$ QCP, which the competition shifts to lower
$\delta$. Labels indicate the LC, LC+SC, SC, and correlated-metal regions.
(b) Both gaps vs temperature at $\delta=0.15$: $\Dperp$ is BCS-like
($\Dperp=1.76\,T_{c}$), while $\phi_{\Aperp}=2.02\,T_{c}^{\zBOP}$.}
\label{fig:competion_temp}
\end{figure}

Now we discuss the competition between LCs and SC. In the present model SC shows out-of-plane $s$-wave symmetry (see Sec.~S5 of the SM~\cite{SM}), which corresponds to $s^{\pm}$ superconductivity in the band basis. 
Figure \ref{fig:competion_temp}(a) shows results for the competition between $\phi_{\Aperp}$ and the superconducting gap $\Dperp$ at $T=0$ (see Sec.~S5 of the SM~\cite{SM}). 
$\Dperp$ shows a dome behavior with an optimal doping $\dop \sim 0.20$, close to the $\zBOP$ QCP. In the absence of competition, $\Dperp$ and $\phi_{\Aperp}$ follow the dashed lines. Note the shift of the $\zBOP$ QCP between the two cases, also observed in the two-dimensional $t$-$J$ model when SC competes with a $d$-charge-density wave~\cite{zeyher03}. 
The corresponding critical temperatures for the LC and SC phases (not shown) follow the same trend as the gaps. 
Figure \ref{fig:competion_temp}(b) shows $\phi_{\Aperp}$ and $\Dperp$ versus temperature at $\dop=0.15$. $\Dperp$ follows a BCS-like behavior, with the BCS ratio $\Dperp=1.76\,T_c$, whereas $\phi_{\Aperp}=2.02\,T_c^{\zBOP}$. The temperature dependence of $\phi_{\Aperp}$ deviates slightly downwards when $\Dperp$ opens, a feature also found in the 2D $t$-$J$ model~\cite{greco04}. Figures~\ref{fig:competion_temp}(a) and~\ref{fig:competion_temp}(b) show a broad region of the $\dop$-$T$ phase diagram where only LCs exist (light-blue region), a coexistence of LCs and SC at low temperature below the QCP (grayish-red region), a region of pure SC above the QCP (light-red region), and a correlated metallic state at large doping and high temperature (white region).

A superconducting dome has been recently reported in nickelates~\cite{hao2025,Wang_2026}. A density matrix renormalization group study of the type-II $t$-$J$ model for bilayer nickelates~\cite{oh25} also finds such a dome, as we do here. As in cuprates~\cite{keimer15}, a pseudogap (PG)~\cite{Shen_2026} and a strange-metal state~\cite{sun23,craco24,zhang23} have also been reported. Whether the LC gap $\phi_{\Aperp}$ can be associated with a PG still requires further experimental study. It would be desirable for future experiments to improve sample quality and to allow these systems to be studied as a function of doping; attempts in this direction exist in the literature~\cite{hao2025,fengNd,zhongNd,qiu2025,zhong2025e,gao2026,shi26}. Furthermore, the existence of LCs and the associated TRS breaking should be studied experimentally, for instance along the lines proposed in Ref.~\cite{morisseau26}.

We have also examined whether the LC$+$SC state supports topologically protected
surface states, which would show up as a zero-bias peak in tunneling spectroscopy,
and find that it does not: no states appear inside the gap and the Chern number
vanishes over the whole coexistence region (see Sec.~S7 of the SM~\cite{SM}). The absence is
structural rather than numerical, since the interlayer pairing carries no momentum
dependence and the state remains invariant under the product of time reversal and a
one-site translation. Should such a peak nevertheless be observed in these materials,
it would therefore have to originate from a different pairing symmetry, and not from
the loop currents themselves.

Finally, it is worth clarifying the distinction between our work and the model studied in Ref.~\cite{fan2026}. That work considers an unconstrained $t$-$V_\perp$-$\Jperp$ Hamiltonian of standard fermions, where $\Jperp$ effectively acts as a four-fermion interaction, whereas our framework explicitly accounts for strong local correlations through the no-double-occupancy constraint. This fundamental difference leads to distinct physical mechanisms. In particular, the Coulomb interaction is required there to induce loop currents near half-filling, while in our approach the screening driven by the local constraint makes any additional Coulomb repulsion irrelevant for loop-current formation.

\textit{Conclusion}.- We have studied the effective strongly correlated one-orbital ($d_{x^2-y^2}$) bilayer $t$-$\Jperp$-$V$ model for nickelates. We performed a controlled large-$N$ expansion, at the mean-field and beyond-mean-field levels, focusing on the out-of-plane spin exchange $\Jperp$, the coupling considered key to the physics of nickelates. We found tendencies towards out-of-plane $s$-wave superconductivity and an out-of-plane $\zBOP$ instability, both triggered by $\Jperp$. In principle, the $\zBOP$ instability opens a complex gap. However, the competition between the real and imaginary components selects the imaginary one, which renders the hopping along the $z$ direction inside the bilayer complex. This complex hopping carries a current along the $z$ direction and induces in-plane currents. As a consequence, a loop-current state that breaks time-reversal symmetry is spontaneously formed on the vertical plaquettes of the bilayer. The competition between superconductivity and the loop-current phase leads to a dome-shaped superconducting region as a function of doping, with the optimal doping close to the quantum critical point where the $\zBOP$ vanishes. The resulting phase diagram shows a broad region in doping and temperature where pure loop currents exist, a coexistence phase of superconductivity and loop currents at low doping below the quantum critical point, a pure superconducting state at larger doping, and a correlated metallic state.

As mentioned in the introduction, the debate persists on whether a single-orbital or multiorbital model is most suitable for nickelates, but there is broad consensus regarding the essential role of strong correlations. Consequently, our strong-coupling calculations within a single-orbital framework provide valuable insights into the fundamental physics that are likely to remain relevant even in a multiorbital picture.

Beyond its possible implications for the physics of nickelates, our work provides a concrete model in which a loop-current state is spontaneously generated.
This model is well-suited for quantum simulation on platforms such as ultracold atoms in optical lattices and superconducting circuits. These platforms offer a unique advantage: they allow for the precise tuning of Hamiltonian parameters and provide site-resolved probes that are often unavailable in bulk materials. Consequently, such simulators could provide a controlled environment to directly observe the competition between interlayer pairing and loop-current orders, offering a new window into the many-body physics of mixed-dimensional systems.

The authors thank M. Hepting for illuminating discussions and D. Manske for reading the manuscript.
A part of the results
presented in this work were obtained by using the facilities of
the CCT-Rosario Computational Center, member of the High
Performance Computing National System (SNCAD, MincyT-Argentina). A.~Greco and L.~Zinni acknowledge the Max-Planck-Institute for Solid State Research in Stuttgart for hospitality and financial support. This work was supported by the CONICET–DFG Bilateral Cooperation Program under the 2023 CONICET–DFG call, National Key R\&D Program of China (Grant No. 2023YFA1407300) and the National Natural
Science Foundation of China (Grant No. 12447103, No. 12574151).

\bibliography{main}

\begin{thebibliography}{70}%
\makeatletter
\providecommand \@ifxundefined [1]{%
 \@ifx{#1\undefined}
}%
\providecommand \@ifnum [1]{%
 \ifnum #1\expandafter \@firstoftwo
 \else \expandafter \@secondoftwo
 \fi
}%
\providecommand \@ifx [1]{%
 \ifx #1\expandafter \@firstoftwo
 \else \expandafter \@secondoftwo
 \fi
}%
\providecommand \natexlab [1]{#1}%
\providecommand \enquote  [1]{``#1''}%
\providecommand \bibnamefont  [1]{#1}%
\providecommand \bibfnamefont [1]{#1}%
\providecommand \citenamefont [1]{#1}%
\providecommand \href@noop [0]{\@secondoftwo}%
\providecommand \href [0]{\begingroup \@sanitize@url \@href}%
\providecommand \@href[1]{\@@startlink{#1}\@@href}%
\providecommand \@@href[1]{\endgroup#1\@@endlink}%
\providecommand \@sanitize@url [0]{\catcode `\\12\catcode `\$12\catcode
  `\&12\catcode `\#12\catcode `\^12\catcode `\_12\catcode `\%12\relax}%
\providecommand \@@startlink[1]{}%
\providecommand \@@endlink[0]{}%
\providecommand \url  [0]{\begingroup\@sanitize@url \@url }%
\providecommand \@url [1]{\endgroup\@href {#1}{\urlprefix }}%
\providecommand \urlprefix  [0]{URL }%
\providecommand \Eprint [0]{\href }%
\providecommand \doibase [0]{https://doi.org/}%
\providecommand \selectlanguage [0]{\@gobble}%
\providecommand \bibinfo  [0]{\@secondoftwo}%
\providecommand \bibfield  [0]{\@secondoftwo}%
\providecommand \translation [1]{[#1]}%
\providecommand \BibitemOpen [0]{}%
\providecommand \bibitemStop [0]{}%
\providecommand \bibitemNoStop [0]{.\EOS\space}%
\providecommand \EOS [0]{\spacefactor3000\relax}%
\providecommand \BibitemShut  [1]{\csname bibitem#1\endcsname}%
\let\auto@bib@innerbib\@empty
\bibitem [{\citenamefont {Sun}\ \emph {et~al.}(2023)\citenamefont {Sun},
  \citenamefont {Huo}, \citenamefont {Hu}, \citenamefont {Li}, \citenamefont
  {Liu}, \citenamefont {Han}, \citenamefont {Tang}, \citenamefont {Mao},
  \citenamefont {Yang}, \citenamefont {Wang}, \citenamefont {Cheng},
  \citenamefont {Yao}, \citenamefont {Zhang},\ and\ \citenamefont
  {Wang}}]{sun23}%
  \BibitemOpen
  \bibfield  {author} {\bibinfo {author} {\bibfnamefont {H.}~\bibnamefont
  {Sun}}, \bibinfo {author} {\bibfnamefont {M.}~\bibnamefont {Huo}}, \bibinfo
  {author} {\bibfnamefont {X.}~\bibnamefont {Hu}}, \bibinfo {author}
  {\bibfnamefont {J.}~\bibnamefont {Li}}, \bibinfo {author} {\bibfnamefont
  {Z.}~\bibnamefont {Liu}}, \bibinfo {author} {\bibfnamefont {Y.}~\bibnamefont
  {Han}}, \bibinfo {author} {\bibfnamefont {L.}~\bibnamefont {Tang}}, \bibinfo
  {author} {\bibfnamefont {Z.}~\bibnamefont {Mao}}, \bibinfo {author}
  {\bibfnamefont {P.}~\bibnamefont {Yang}}, \bibinfo {author} {\bibfnamefont
  {B.}~\bibnamefont {Wang}}, \bibinfo {author} {\bibfnamefont {J.}~\bibnamefont
  {Cheng}}, \bibinfo {author} {\bibfnamefont {D.-X.}\ \bibnamefont {Yao}},
  \bibinfo {author} {\bibfnamefont {G.-M.}\ \bibnamefont {Zhang}},\ and\
  \bibinfo {author} {\bibfnamefont {M.}~\bibnamefont {Wang}},\ }\bibfield
  {title} {\bibinfo {title} {{Signatures of superconductivity near
  80{\thinspace}K in a nickelate under high pressure}},\ }\href
  {https://doi.org/10.1038/s41586-023-06408-7} {\bibfield  {journal} {\bibinfo
  {journal} {Nature}\ }\textbf {\bibinfo {volume} {621}},\ \bibinfo {pages}
  {493} (\bibinfo {year} {2023})}\BibitemShut {NoStop}%
\bibitem [{\citenamefont {Hou}\ \emph {et~al.}(2023)\citenamefont {Hou},
  \citenamefont {Yang}, \citenamefont {Liu}, \citenamefont {Li}, \citenamefont
  {Shan}, \citenamefont {Ma}, \citenamefont {Wang}, \citenamefont {Wang},
  \citenamefont {Guo}, \citenamefont {Sun}, \citenamefont {Uwatoko},
  \citenamefont {Wang}, \citenamefont {Zhang}, \citenamefont {Wang},\ and\
  \citenamefont {Cheng}}]{hou23}%
  \BibitemOpen
  \bibfield  {author} {\bibinfo {author} {\bibfnamefont {J.}~\bibnamefont
  {Hou}}, \bibinfo {author} {\bibfnamefont {P.-T.}\ \bibnamefont {Yang}},
  \bibinfo {author} {\bibfnamefont {Z.-Y.}\ \bibnamefont {Liu}}, \bibinfo
  {author} {\bibfnamefont {J.-Y.}\ \bibnamefont {Li}}, \bibinfo {author}
  {\bibfnamefont {P.-F.}\ \bibnamefont {Shan}}, \bibinfo {author}
  {\bibfnamefont {L.}~\bibnamefont {Ma}}, \bibinfo {author} {\bibfnamefont
  {G.}~\bibnamefont {Wang}}, \bibinfo {author} {\bibfnamefont {N.-N.}\
  \bibnamefont {Wang}}, \bibinfo {author} {\bibfnamefont {H.-Z.}\ \bibnamefont
  {Guo}}, \bibinfo {author} {\bibfnamefont {J.-P.}\ \bibnamefont {Sun}},
  \bibinfo {author} {\bibfnamefont {Y.}~\bibnamefont {Uwatoko}}, \bibinfo
  {author} {\bibfnamefont {M.}~\bibnamefont {Wang}}, \bibinfo {author}
  {\bibfnamefont {G.-M.}\ \bibnamefont {Zhang}}, \bibinfo {author}
  {\bibfnamefont {B.-S.}\ \bibnamefont {Wang}},\ and\ \bibinfo {author}
  {\bibfnamefont {J.-G.}\ \bibnamefont {Cheng}},\ }\bibfield  {title} {\bibinfo
  {title} {{Emergence of High-Temperature Superconducting Phase in Pressurized
  La3Ni2O7 Crystals}},\ }\href {https://doi.org/10.1088/0256-307X/40/11/117302}
  {\bibfield  {journal} {\bibinfo  {journal} {Chinese Physics Letters}\
  }\textbf {\bibinfo {volume} {40}},\ \bibinfo {pages} {117302} (\bibinfo
  {year} {2023})}\BibitemShut {NoStop}%
\bibitem [{\citenamefont {Ko}\ \emph {et~al.}(2025)\citenamefont {Ko},
  \citenamefont {Yu}, \citenamefont {Liu}, \citenamefont {Bhatt}, \citenamefont
  {Li}, \citenamefont {Thampy}, \citenamefont {Kuo}, \citenamefont {Wang},
  \citenamefont {Lee}, \citenamefont {Lee}, \citenamefont {Lee}, \citenamefont
  {Goodge}, \citenamefont {Muller},\ and\ \citenamefont {Hwang}}]{ko25}%
  \BibitemOpen
  \bibfield  {author} {\bibinfo {author} {\bibfnamefont {E.~K.}\ \bibnamefont
  {Ko}}, \bibinfo {author} {\bibfnamefont {Y.}~\bibnamefont {Yu}}, \bibinfo
  {author} {\bibfnamefont {Y.}~\bibnamefont {Liu}}, \bibinfo {author}
  {\bibfnamefont {L.}~\bibnamefont {Bhatt}}, \bibinfo {author} {\bibfnamefont
  {J.}~\bibnamefont {Li}}, \bibinfo {author} {\bibfnamefont {V.}~\bibnamefont
  {Thampy}}, \bibinfo {author} {\bibfnamefont {C.-T.}\ \bibnamefont {Kuo}},
  \bibinfo {author} {\bibfnamefont {B.~Y.}\ \bibnamefont {Wang}}, \bibinfo
  {author} {\bibfnamefont {Y.}~\bibnamefont {Lee}}, \bibinfo {author}
  {\bibfnamefont {K.}~\bibnamefont {Lee}}, \bibinfo {author} {\bibfnamefont
  {J.-S.}\ \bibnamefont {Lee}}, \bibinfo {author} {\bibfnamefont {B.~H.}\
  \bibnamefont {Goodge}}, \bibinfo {author} {\bibfnamefont {D.~A.}\
  \bibnamefont {Muller}},\ and\ \bibinfo {author} {\bibfnamefont {H.~Y.}\
  \bibnamefont {Hwang}},\ }\bibfield  {title} {\bibinfo {title} {{Signatures of
  ambient pressure superconductivity in thin film La3Ni2O7}},\ }\href
  {https://doi.org/10.1038/s41586-024-08525-3} {\bibfield  {journal} {\bibinfo
  {journal} {Nature}\ }\textbf {\bibinfo {volume} {638}},\ \bibinfo {pages}
  {935} (\bibinfo {year} {2025})}\BibitemShut {NoStop}%
\bibitem [{\citenamefont {Zhou}\ \emph {et~al.}(2025)\citenamefont {Zhou},
  \citenamefont {Lv}, \citenamefont {Wang}, \citenamefont {Nie}, \citenamefont
  {Chen}, \citenamefont {Li}, \citenamefont {Huang}, \citenamefont {Chen},
  \citenamefont {Sun}, \citenamefont {Xue},\ and\ \citenamefont
  {Chen}}]{zhou25}%
  \BibitemOpen
  \bibfield  {author} {\bibinfo {author} {\bibfnamefont {G.}~\bibnamefont
  {Zhou}}, \bibinfo {author} {\bibfnamefont {W.}~\bibnamefont {Lv}}, \bibinfo
  {author} {\bibfnamefont {H.}~\bibnamefont {Wang}}, \bibinfo {author}
  {\bibfnamefont {Z.}~\bibnamefont {Nie}}, \bibinfo {author} {\bibfnamefont
  {Y.}~\bibnamefont {Chen}}, \bibinfo {author} {\bibfnamefont {Y.}~\bibnamefont
  {Li}}, \bibinfo {author} {\bibfnamefont {H.}~\bibnamefont {Huang}}, \bibinfo
  {author} {\bibfnamefont {W.-Q.}\ \bibnamefont {Chen}}, \bibinfo {author}
  {\bibfnamefont {Y.-J.}\ \bibnamefont {Sun}}, \bibinfo {author} {\bibfnamefont
  {Q.-K.}\ \bibnamefont {Xue}},\ and\ \bibinfo {author} {\bibfnamefont
  {Z.}~\bibnamefont {Chen}},\ }\bibfield  {title} {\bibinfo {title}
  {{Ambient-pressure superconductivity onset above 40{\thinspace}K in
  (La,Pr)3Ni2O7 films}},\ }\href {https://doi.org/10.1038/s41586-025-08755-z}
  {\bibfield  {journal} {\bibinfo  {journal} {Nature}\ }\textbf {\bibinfo
  {volume} {640}},\ \bibinfo {pages} {641} (\bibinfo {year}
  {2025})}\BibitemShut {NoStop}%
\bibitem [{\citenamefont {Qiu}\ and\ \citenamefont {Yao}(2026)}]{Qiu_2026}%
  \BibitemOpen
  \bibfield  {author} {\bibinfo {author} {\bibfnamefont {W.}~\bibnamefont
  {Qiu}}\ and\ \bibinfo {author} {\bibfnamefont {D.-X.}\ \bibnamefont {Yao}},\
  }\bibfield  {title} {\bibinfo {title} {Progress of ambient-pressure
  superconductivity in bilayer nickelate thin films},\ }\href
  {https://doi.org/10.15302/frontphys.2026.115301} {\bibfield  {journal}
  {\bibinfo  {journal} {Frontiers of Physics}\ }\textbf {\bibinfo {volume}
  {21}},\ \bibinfo {pages} {115301} (\bibinfo {year} {2026})}\BibitemShut
  {NoStop}%
\bibitem [{\citenamefont {Zhang}\ \emph {et~al.}(2026)\citenamefont {Zhang},
  \citenamefont {Lin}, \citenamefont {Maier},\ and\ \citenamefont
  {Dagotto}}]{zhang2026}%
  \BibitemOpen
  \bibfield  {author} {\bibinfo {author} {\bibfnamefont {Y.}~\bibnamefont
  {Zhang}}, \bibinfo {author} {\bibfnamefont {L.-F.}\ \bibnamefont {Lin}},
  \bibinfo {author} {\bibfnamefont {T.~A.}\ \bibnamefont {Maier}},\ and\
  \bibinfo {author} {\bibfnamefont {E.}~\bibnamefont {Dagotto}},\ }\href
  {https://arxiv.org/abs/2604.18385} {\bibinfo {title} {Superconductivity in
  ruddlesden-popper nickelates: a review of recent progress, focusing on thin
  films}} (\bibinfo {year} {2026}),\ \Eprint {https://arxiv.org/abs/2604.18385}
  {arXiv:2604.18385 [cond-mat.supr-con]} \BibitemShut {NoStop}%
\bibitem [{\citenamefont {Wang}\ \emph {et~al.}(2024)\citenamefont {Wang},
  \citenamefont {Wen}, \citenamefont {Wu}, \citenamefont {Yao},\ and\
  \citenamefont {Xiang}}]{wang2024normal}%
  \BibitemOpen
  \bibfield  {author} {\bibinfo {author} {\bibfnamefont {M.}~\bibnamefont
  {Wang}}, \bibinfo {author} {\bibfnamefont {H.-H.}\ \bibnamefont {Wen}},
  \bibinfo {author} {\bibfnamefont {T.}~\bibnamefont {Wu}}, \bibinfo {author}
  {\bibfnamefont {D.-X.}\ \bibnamefont {Yao}},\ and\ \bibinfo {author}
  {\bibfnamefont {T.}~\bibnamefont {Xiang}},\ }\bibfield  {title} {\bibinfo
  {title} {{Normal and Superconducting Properties of La3Ni2O7}},\ }\href
  {https://doi.org/10.1088/0256-307X/41/7/077402} {\bibfield  {journal}
  {\bibinfo  {journal} {Chinese Physics Letters}\ }\textbf {\bibinfo {volume}
  {41}},\ \bibinfo {pages} {077402} (\bibinfo {year} {2024})}\BibitemShut
  {NoStop}%
\bibitem [{\citenamefont {Zhang}\ and\ \citenamefont
  {Yan}(2026)}]{zhang2026exp}%
  \BibitemOpen
  \bibfield  {author} {\bibinfo {author} {\bibfnamefont {M.}~\bibnamefont
  {Zhang}}\ and\ \bibinfo {author} {\bibfnamefont {X.}~\bibnamefont {Yan}},\
  }\href {https://arxiv.org/abs/2605.11584} {\bibinfo {title} {Experimental
  progress in ambient-pressure superconducting bilayer nickelate films}}
  (\bibinfo {year} {2026}),\ \Eprint {https://arxiv.org/abs/2605.11584}
  {arXiv:2605.11584 [cond-mat.supr-con]} \BibitemShut {NoStop}%
\bibitem [{\citenamefont {Wang}\ \emph {et~al.}(2025)\citenamefont {Wang},
  \citenamefont {Jiang}, \citenamefont {Ying}, \citenamefont {Wu},
  \citenamefont {Cheng}, \citenamefont {Hu},\ and\ \citenamefont
  {Chen}}]{wang_2025Re}%
  \BibitemOpen
  \bibfield  {author} {\bibinfo {author} {\bibfnamefont {Y.}~\bibnamefont
  {Wang}}, \bibinfo {author} {\bibfnamefont {K.}~\bibnamefont {Jiang}},
  \bibinfo {author} {\bibfnamefont {J.}~\bibnamefont {Ying}}, \bibinfo {author}
  {\bibfnamefont {T.}~\bibnamefont {Wu}}, \bibinfo {author} {\bibfnamefont
  {J.}~\bibnamefont {Cheng}}, \bibinfo {author} {\bibfnamefont
  {J.}~\bibnamefont {Hu}},\ and\ \bibinfo {author} {\bibfnamefont
  {X.}~\bibnamefont {Chen}},\ }\bibfield  {title} {\bibinfo {title} {Recent
  progress in nickelate superconductors},\ }\bibfield  {journal} {\bibinfo
  {journal} {National Science Review}\ }\textbf {\bibinfo {volume} {12}},\
  \href {https://doi.org/10.1093/nsr/nwaf373} {10.1093/nsr/nwaf373} (\bibinfo
  {year} {2025})\BibitemShut {NoStop}%
\bibitem [{\citenamefont {Keimer}\ \emph {et~al.}(2015)\citenamefont {Keimer},
  \citenamefont {Kivelson}, \citenamefont {Norman}, \citenamefont {Uchida},\
  and\ \citenamefont {Zaanen}}]{keimer15}%
  \BibitemOpen
  \bibfield  {author} {\bibinfo {author} {\bibfnamefont {B.}~\bibnamefont
  {Keimer}}, \bibinfo {author} {\bibfnamefont {S.~A.}\ \bibnamefont
  {Kivelson}}, \bibinfo {author} {\bibfnamefont {M.~R.}\ \bibnamefont
  {Norman}}, \bibinfo {author} {\bibfnamefont {S.}~\bibnamefont {Uchida}},\
  and\ \bibinfo {author} {\bibfnamefont {J.}~\bibnamefont {Zaanen}},\
  }\bibfield  {title} {\bibinfo {title} {From quantum matter to
  high-temperature superconductivity in copper oxides},\ }\href
  {https://doi.org/10.1038/nature14165} {\bibfield  {journal} {\bibinfo
  {journal} {Nature}\ }\textbf {\bibinfo {volume} {518}},\ \bibinfo {pages}
  {179} (\bibinfo {year} {2015})}\BibitemShut {NoStop}%
\bibitem [{\citenamefont {Lu}\ \emph {et~al.}(2024)\citenamefont {Lu},
  \citenamefont {Pan}, \citenamefont {Yang},\ and\ \citenamefont {Wu}}]{lu24}%
  \BibitemOpen
  \bibfield  {author} {\bibinfo {author} {\bibfnamefont {C.}~\bibnamefont
  {Lu}}, \bibinfo {author} {\bibfnamefont {Z.}~\bibnamefont {Pan}}, \bibinfo
  {author} {\bibfnamefont {F.}~\bibnamefont {Yang}},\ and\ \bibinfo {author}
  {\bibfnamefont {C.}~\bibnamefont {Wu}},\ }\bibfield  {title} {\bibinfo
  {title} {{Interlayer-Coupling-Driven High-Temperature Superconductivity in
  ${\mathrm{La}}_{3}{\mathrm{Ni}}_{2}{\mathrm{O}}_{7}$ under Pressure}},\
  }\href {https://doi.org/10.1103/PhysRevLett.132.146002} {\bibfield  {journal}
  {\bibinfo  {journal} {Phys. Rev. Lett.}\ }\textbf {\bibinfo {volume} {132}},\
  \bibinfo {pages} {146002} (\bibinfo {year} {2024})}\BibitemShut {NoStop}%
\bibitem [{\citenamefont {Qu}\ \emph {et~al.}(2024)\citenamefont {Qu},
  \citenamefont {Qu}, \citenamefont {Chen}, \citenamefont {Wu}, \citenamefont
  {Yang}, \citenamefont {Li},\ and\ \citenamefont {Su}}]{qu24}%
  \BibitemOpen
  \bibfield  {author} {\bibinfo {author} {\bibfnamefont {X.-Z.}\ \bibnamefont
  {Qu}}, \bibinfo {author} {\bibfnamefont {D.-W.}\ \bibnamefont {Qu}}, \bibinfo
  {author} {\bibfnamefont {J.}~\bibnamefont {Chen}}, \bibinfo {author}
  {\bibfnamefont {C.}~\bibnamefont {Wu}}, \bibinfo {author} {\bibfnamefont
  {F.}~\bibnamefont {Yang}}, \bibinfo {author} {\bibfnamefont {W.}~\bibnamefont
  {Li}},\ and\ \bibinfo {author} {\bibfnamefont {G.}~\bibnamefont {Su}},\
  }\bibfield  {title} {\bibinfo {title} {{Bilayer
  ${t\text{\ensuremath{-}}J\text{\ensuremath{-}}J}_{\ensuremath{\perp}}$ Model
  and Magnetically Mediated Pairing in the Pressurized Nickelate
  ${\mathrm{La}}_{3}{\mathrm{Ni}}_{2}{\mathrm{O}}_{7}$}},\ }\href
  {https://doi.org/10.1103/PhysRevLett.132.036502} {\bibfield  {journal}
  {\bibinfo  {journal} {Phys. Rev. Lett.}\ }\textbf {\bibinfo {volume} {132}},\
  \bibinfo {pages} {036502} (\bibinfo {year} {2024})}\BibitemShut {NoStop}%
\bibitem [{\citenamefont {Oh}\ \emph {et~al.}(2026)\citenamefont {Oh},
  \citenamefont {Yang},\ and\ \citenamefont {Zhang}}]{oh2025dop}%
  \BibitemOpen
  \bibfield  {author} {\bibinfo {author} {\bibfnamefont {H.}~\bibnamefont
  {Oh}}, \bibinfo {author} {\bibfnamefont {H.}~\bibnamefont {Yang}},\ and\
  \bibinfo {author} {\bibfnamefont {Y.-H.}\ \bibnamefont {Zhang}},\ }\bibfield
  {title} {\bibinfo {title} {Doping a spin-one mott insulator: possible
  application to bilayer nickelate},\ }\href
  {https://doi.org/10.1088/1367-2630/ae3d5b} {\bibfield  {journal} {\bibinfo
  {journal} {New Journal of Physics}\ }\textbf {\bibinfo {volume} {28}},\
  \bibinfo {pages} {021201} (\bibinfo {year} {2026})}\BibitemShut {NoStop}%
\bibitem [{\citenamefont {Pan}\ \emph {et~al.}(2026)\citenamefont {Pan},
  \citenamefont {Lu}, \citenamefont {Yang},\ and\ \citenamefont
  {Wu}}]{pan2026review}%
  \BibitemOpen
  \bibfield  {author} {\bibinfo {author} {\bibfnamefont {Z.}~\bibnamefont
  {Pan}}, \bibinfo {author} {\bibfnamefont {C.}~\bibnamefont {Lu}}, \bibinfo
  {author} {\bibfnamefont {F.}~\bibnamefont {Yang}},\ and\ \bibinfo {author}
  {\bibfnamefont {C.}~\bibnamefont {Wu}},\ }\href
  {https://arxiv.org/abs/2604.20613} {\bibinfo {title} {Superconductivity in
  bilayer la$_3$ni$_2$o$_7$: A review focusing on the strong-coupling hund's
  rule assisted pairing mechanism}} (\bibinfo {year} {2026}),\ \Eprint
  {https://arxiv.org/abs/2604.20613} {arXiv:2604.20613 [cond-mat.supr-con]}
  \BibitemShut {NoStop}%
\bibitem [{\citenamefont {Bohrdt}\ \emph {et~al.}(2022)\citenamefont {Bohrdt},
  \citenamefont {Homeier}, \citenamefont {Bloch}, \citenamefont {Demler},\ and\
  \citenamefont {Grusdt}}]{bohrdt2022strong}%
  \BibitemOpen
  \bibfield  {author} {\bibinfo {author} {\bibfnamefont {A.}~\bibnamefont
  {Bohrdt}}, \bibinfo {author} {\bibfnamefont {L.}~\bibnamefont {Homeier}},
  \bibinfo {author} {\bibfnamefont {I.}~\bibnamefont {Bloch}}, \bibinfo
  {author} {\bibfnamefont {E.}~\bibnamefont {Demler}},\ and\ \bibinfo {author}
  {\bibfnamefont {F.}~\bibnamefont {Grusdt}},\ }\bibfield  {title} {\bibinfo
  {title} {Strong pairing in mixed-dimensional bilayer antiferromagnetic mott
  insulators},\ }\href {https://doi.org/10.1038/s41567-022-01561-8} {\bibfield
  {journal} {\bibinfo  {journal} {Nature Physics}\ }\textbf {\bibinfo {volume}
  {18}},\ \bibinfo {pages} {651} (\bibinfo {year} {2022})}\BibitemShut
  {NoStop}%
\bibitem [{\citenamefont {Bourgund}\ \emph {et~al.}(2025)\citenamefont
  {Bourgund}, \citenamefont {Chalopin}, \citenamefont {Bojovi{\'c}},
  \citenamefont {Schl{\"o}mer}, \citenamefont {Wang}, \citenamefont {Franz},
  \citenamefont {Hirthe}, \citenamefont {Bohrdt}, \citenamefont {Grusdt},
  \citenamefont {Bloch} \emph {et~al.}}]{bourgund2025formation}%
  \BibitemOpen
  \bibfield  {author} {\bibinfo {author} {\bibfnamefont {D.}~\bibnamefont
  {Bourgund}}, \bibinfo {author} {\bibfnamefont {T.}~\bibnamefont {Chalopin}},
  \bibinfo {author} {\bibfnamefont {P.}~\bibnamefont {Bojovi{\'c}}}, \bibinfo
  {author} {\bibfnamefont {H.}~\bibnamefont {Schl{\"o}mer}}, \bibinfo {author}
  {\bibfnamefont {S.}~\bibnamefont {Wang}}, \bibinfo {author} {\bibfnamefont
  {T.}~\bibnamefont {Franz}}, \bibinfo {author} {\bibfnamefont
  {S.}~\bibnamefont {Hirthe}}, \bibinfo {author} {\bibfnamefont
  {A.}~\bibnamefont {Bohrdt}}, \bibinfo {author} {\bibfnamefont
  {F.}~\bibnamefont {Grusdt}}, \bibinfo {author} {\bibfnamefont
  {I.}~\bibnamefont {Bloch}}, \emph {et~al.},\ }\bibfield  {title} {\bibinfo
  {title} {Formation of individual stripes in a mixed-dimensional cold-atom
  fermi--hubbard system},\ }\href {https://doi.org/10.1038/s41586-024-08270-7}
  {\bibfield  {journal} {\bibinfo  {journal} {Nature}\ }\textbf {\bibinfo
  {volume} {637}},\ \bibinfo {pages} {57} (\bibinfo {year} {2025})}\BibitemShut
  {NoStop}%
\bibitem [{\citenamefont {Miao}\ and\ \citenamefont {Chen}(2026)}]{jianjian26}%
  \BibitemOpen
  \bibfield  {author} {\bibinfo {author} {\bibfnamefont {J.}~\bibnamefont
  {Miao}}\ and\ \bibinfo {author} {\bibfnamefont {W.}~\bibnamefont {Chen}},\
  }\bibfield  {title} {\bibinfo {title} {Weak coupling theory of nickel-based
  327 superconductors},\ }\bibfield  {journal} {\bibinfo  {journal} {Acta
  Physica Sinica}\ }\textbf {\bibinfo {volume} {75}},\ \href
  {https://doi.org/10.7498/aps.75.20251361} {10.7498/aps.75.20251361} (\bibinfo
  {year} {2026})\BibitemShut {NoStop}%
\bibitem [{\citenamefont {Foussats}\ and\ \citenamefont
  {Greco}(2004)}]{foussats04}%
  \BibitemOpen
  \bibfield  {author} {\bibinfo {author} {\bibfnamefont {A.}~\bibnamefont
  {Foussats}}\ and\ \bibinfo {author} {\bibfnamefont {A.}~\bibnamefont
  {Greco}},\ }\bibfield  {title} {\bibinfo {title} {{Large-$N$ expansion based
  on the Hubbard operator path integral representation and its application to
  the $t\text{\ensuremath{-}}J$ model. II. The case for finite $J$}},\ }\href
  {https://doi.org/10.1103/PhysRevB.70.205123} {\bibfield  {journal} {\bibinfo
  {journal} {Phys. Rev. B}\ }\textbf {\bibinfo {volume} {70}},\ \bibinfo
  {pages} {205123} (\bibinfo {year} {2004})}\BibitemShut {NoStop}%
\bibitem [{\citenamefont {Bejas}\ \emph {et~al.}(2012)\citenamefont {Bejas},
  \citenamefont {Greco},\ and\ \citenamefont {Yamase}}]{bejas12}%
  \BibitemOpen
  \bibfield  {author} {\bibinfo {author} {\bibfnamefont {M.}~\bibnamefont
  {Bejas}}, \bibinfo {author} {\bibfnamefont {A.}~\bibnamefont {Greco}},\ and\
  \bibinfo {author} {\bibfnamefont {H.}~\bibnamefont {Yamase}},\ }\bibfield
  {title} {\bibinfo {title} {Possible charge instabilities in two-dimensional
  doped {Mott} insulators},\ }\href
  {https://link.aps.org/doi/10.1103/PhysRevB.86.224509} {\bibfield  {journal}
  {\bibinfo  {journal} {Phys. Rev. B}\ }\textbf {\bibinfo {volume} {86}},\
  \bibinfo {pages} {224509} (\bibinfo {year} {2012})}\BibitemShut {NoStop}%
\bibitem [{\citenamefont {Hubbard}(1963)}]{hubbard63}%
  \BibitemOpen
  \bibfield  {author} {\bibinfo {author} {\bibfnamefont {J.}~\bibnamefont
  {Hubbard}},\ }\bibfield  {title} {\bibinfo {title} {Electron correlations in
  narrow energy bands},\ }\href {https://doi.org/10.1098/rspa.1963.0204}
  {\bibfield  {journal} {\bibinfo  {journal} {Proc. R. Soc. Lond. A:
  Mathematical, Physical and Engineering Sciences}\ }\textbf {\bibinfo {volume}
  {276}},\ \bibinfo {pages} {238} (\bibinfo {year} {1963})}\BibitemShut
  {NoStop}%
\bibitem [{\citenamefont {Bejas}\ \emph {et~al.}(2025)\citenamefont {Bejas},
  \citenamefont {Wu}, \citenamefont {Chakraborty}, \citenamefont {Schnyder},\
  and\ \citenamefont {Greco}}]{bejas25}%
  \BibitemOpen
  \bibfield  {author} {\bibinfo {author} {\bibfnamefont {M.}~\bibnamefont
  {Bejas}}, \bibinfo {author} {\bibfnamefont {X.}~\bibnamefont {Wu}}, \bibinfo
  {author} {\bibfnamefont {D.}~\bibnamefont {Chakraborty}}, \bibinfo {author}
  {\bibfnamefont {A.~P.}\ \bibnamefont {Schnyder}},\ and\ \bibinfo {author}
  {\bibfnamefont {A.}~\bibnamefont {Greco}},\ }\bibfield  {title} {\bibinfo
  {title} {{Out-of-plane bond-order phase, superconductivity, and their
  competition in the
  $t$-${J}_{\ensuremath{\parallel}}$-${J}_{\ensuremath{\perp}}$ model: Possible
  implications for bilayer nickelates}},\ }\href
  {https://doi.org/10.1103/PhysRevB.111.144514} {\bibfield  {journal} {\bibinfo
   {journal} {Phys. Rev. B}\ }\textbf {\bibinfo {volume} {111}},\ \bibinfo
  {pages} {144514} (\bibinfo {year} {2025})}\BibitemShut {NoStop}%
\bibitem [{\citenamefont {Capponi}\ \emph {et~al.}(2004)\citenamefont
  {Capponi}, \citenamefont {Wu},\ and\ \citenamefont {Zhang}}]{WuCJ2004}%
  \BibitemOpen
  \bibfield  {author} {\bibinfo {author} {\bibfnamefont {S.}~\bibnamefont
  {Capponi}}, \bibinfo {author} {\bibfnamefont {C.}~\bibnamefont {Wu}},\ and\
  \bibinfo {author} {\bibfnamefont {S.-C.}\ \bibnamefont {Zhang}},\ }\bibfield
  {title} {\bibinfo {title} {Current carrying ground state in a bilayer model
  of strongly correlated systems},\ }\href
  {https://doi.org/10.1103/PhysRevB.70.220505} {\bibfield  {journal} {\bibinfo
  {journal} {Phys. Rev. B}\ }\textbf {\bibinfo {volume} {70}},\ \bibinfo
  {pages} {220505(R)} (\bibinfo {year} {2004})}\BibitemShut {NoStop}%
\bibitem [{SM()}]{SM}%
  \BibitemOpen
  \href@noop {} {}\bibinfo {note} {See Supplemental Material at [URL will be
  inserted by publisher] for the large-$N$ formalism of the bilayer
  $t$-$J_\perp$-$V$ model, the decoupling between the charge and $z$-BOP
  sectors, the gap equations at the commensurate ordering vector and the
  selection of the $A_\perp$ channel, the Peierls phase, loop currents and
  orbital magnetic moment, the competition between superconductivity and loop
  currents, the analysis for different $J_\perp$, and the absence of
  topologically protected surface states, which includes
  Refs.~\cite{yamase21,merino06,hoang02,koch04,bejas08,hsu91,hu94,%
  tanaka95,kashiwaya00,kuboki14,kuboki20}}\BibitemShut {NoStop}%
\bibitem [{\citenamefont {Luo}\ \emph {et~al.}(2023)\citenamefont {Luo},
  \citenamefont {Hu}, \citenamefont {Wang}, \citenamefont {W\'u},\ and\
  \citenamefont {Yao}}]{luo23}%
  \BibitemOpen
  \bibfield  {author} {\bibinfo {author} {\bibfnamefont {Z.}~\bibnamefont
  {Luo}}, \bibinfo {author} {\bibfnamefont {X.}~\bibnamefont {Hu}}, \bibinfo
  {author} {\bibfnamefont {M.}~\bibnamefont {Wang}}, \bibinfo {author}
  {\bibfnamefont {W.}~\bibnamefont {W\'u}},\ and\ \bibinfo {author}
  {\bibfnamefont {D.-X.}\ \bibnamefont {Yao}},\ }\bibfield  {title} {\bibinfo
  {title} {{Bilayer Two-Orbital Model of
  $\mathrm{L}{\mathrm{a}}_{3}\mathrm{N}{\mathrm{i}}_{2}{\mathrm{O}}_{7}$ under
  Pressure}},\ }\href {https://doi.org/10.1103/PhysRevLett.131.126001}
  {\bibfield  {journal} {\bibinfo  {journal} {Phys. Rev. Lett.}\ }\textbf
  {\bibinfo {volume} {131}},\ \bibinfo {pages} {126001} (\bibinfo {year}
  {2023})}\BibitemShut {NoStop}%
\bibitem [{\citenamefont {Bejas}\ \emph {et~al.}(2017)\citenamefont {Bejas},
  \citenamefont {Yamase},\ and\ \citenamefont {Greco}}]{bejas17}%
  \BibitemOpen
  \bibfield  {author} {\bibinfo {author} {\bibfnamefont {M.}~\bibnamefont
  {Bejas}}, \bibinfo {author} {\bibfnamefont {H.}~\bibnamefont {Yamase}},\ and\
  \bibinfo {author} {\bibfnamefont {A.}~\bibnamefont {Greco}},\ }\bibfield
  {title} {\bibinfo {title} {Dual structure in the charge excitation spectrum
  of electron-doped cuprates},\ }\href
  {https://doi.org/10.1103/PhysRevB.96.214513} {\bibfield  {journal} {\bibinfo
  {journal} {Phys. Rev. B}\ }\textbf {\bibinfo {volume} {96}},\ \bibinfo
  {pages} {214513} (\bibinfo {year} {2017})}\BibitemShut {NoStop}%
\bibitem [{\citenamefont {Affleck}\ and\ \citenamefont
  {Marston}(1988)}]{affleck88a}%
  \BibitemOpen
  \bibfield  {author} {\bibinfo {author} {\bibfnamefont {I.}~\bibnamefont
  {Affleck}}\ and\ \bibinfo {author} {\bibfnamefont {J.~B.}\ \bibnamefont
  {Marston}},\ }\bibfield  {title} {\bibinfo {title} {{Large-n limit of the
  Heisenberg-Hubbard model: Implications for high-${T}_{c}$ superconductors}},\
  }\href {https://link.aps.org/doi/10.1103/PhysRevB.37.3774} {\bibfield
  {journal} {\bibinfo  {journal} {Phys. Rev. B}\ }\textbf {\bibinfo {volume}
  {37}},\ \bibinfo {pages} {3774} (\bibinfo {year} {1988})}\BibitemShut
  {NoStop}%
\bibitem [{\citenamefont {Chakravarty}\ \emph {et~al.}(2001)\citenamefont
  {Chakravarty}, \citenamefont {Laughlin}, \citenamefont {Morr},\ and\
  \citenamefont {Nayak}}]{chakravarty01}%
  \BibitemOpen
  \bibfield  {author} {\bibinfo {author} {\bibfnamefont {S.}~\bibnamefont
  {Chakravarty}}, \bibinfo {author} {\bibfnamefont {R.~B.}\ \bibnamefont
  {Laughlin}}, \bibinfo {author} {\bibfnamefont {D.~K.}\ \bibnamefont {Morr}},\
  and\ \bibinfo {author} {\bibfnamefont {C.}~\bibnamefont {Nayak}},\ }\bibfield
   {title} {\bibinfo {title} {Hidden order in the cuprates},\ }\href
  {https://link.aps.org/doi/10.1103/PhysRevB.63.094503} {\bibfield  {journal}
  {\bibinfo  {journal} {Phys. Rev. B}\ }\textbf {\bibinfo {volume} {63}},\
  \bibinfo {pages} {094503} (\bibinfo {year} {2001})}\BibitemShut {NoStop}%
\bibitem [{\citenamefont {Cappelluti}\ and\ \citenamefont
  {Zeyher}(1999)}]{cappelluti99}%
  \BibitemOpen
  \bibfield  {author} {\bibinfo {author} {\bibfnamefont {E.}~\bibnamefont
  {Cappelluti}}\ and\ \bibinfo {author} {\bibfnamefont {R.}~\bibnamefont
  {Zeyher}},\ }\bibfield  {title} {\bibinfo {title} {{Interplay between
  superconductivity and flux phase in the $t\ensuremath{-}J$ model}},\ }\href
  {https://link.aps.org/doi/10.1103/PhysRevB.59.6475} {\bibfield  {journal}
  {\bibinfo  {journal} {Phys. Rev. B}\ }\textbf {\bibinfo {volume} {59}},\
  \bibinfo {pages} {6475} (\bibinfo {year} {1999})}\BibitemShut {NoStop}%
\bibitem [{\citenamefont {Varma}(1997)}]{varma97}%
  \BibitemOpen
  \bibfield  {author} {\bibinfo {author} {\bibfnamefont {C.~M.}\ \bibnamefont
  {Varma}},\ }\bibfield  {title} {\bibinfo {title} {Non-fermi-liquid states and
  pairing instability of a general model of copper oxide metals},\ }\href
  {https://doi.org/10.1103/PhysRevB.55.14554} {\bibfield  {journal} {\bibinfo
  {journal} {Phys. Rev. B}\ }\textbf {\bibinfo {volume} {55}},\ \bibinfo
  {pages} {14554} (\bibinfo {year} {1997})}\BibitemShut {NoStop}%
\bibitem [{\citenamefont {Kaminski}\ \emph {et~al.}(2002)\citenamefont
  {Kaminski}, \citenamefont {Rosenkranz}, \citenamefont {Fretwell},
  \citenamefont {Campuzano}, \citenamefont {Li}, \citenamefont {Raffy},
  \citenamefont {Cullen}, \citenamefont {You}, \citenamefont {Olson},
  \citenamefont {Varma},\ and\ \citenamefont {H{\"o}chst}}]{kaminski02}%
  \BibitemOpen
  \bibfield  {author} {\bibinfo {author} {\bibfnamefont {A.}~\bibnamefont
  {Kaminski}}, \bibinfo {author} {\bibfnamefont {S.}~\bibnamefont
  {Rosenkranz}}, \bibinfo {author} {\bibfnamefont {H.~M.}\ \bibnamefont
  {Fretwell}}, \bibinfo {author} {\bibfnamefont {J.~C.}\ \bibnamefont
  {Campuzano}}, \bibinfo {author} {\bibfnamefont {Z.}~\bibnamefont {Li}},
  \bibinfo {author} {\bibfnamefont {H.}~\bibnamefont {Raffy}}, \bibinfo
  {author} {\bibfnamefont {W.~G.}\ \bibnamefont {Cullen}}, \bibinfo {author}
  {\bibfnamefont {H.}~\bibnamefont {You}}, \bibinfo {author} {\bibfnamefont
  {C.~G.}\ \bibnamefont {Olson}}, \bibinfo {author} {\bibfnamefont {C.~M.}\
  \bibnamefont {Varma}},\ and\ \bibinfo {author} {\bibfnamefont
  {H.}~\bibnamefont {H{\"o}chst}},\ }\bibfield  {title} {\bibinfo {title}
  {{Spontaneous breaking of time-reversal symmetry in the pseudogap state of a
  high-Tc superconductor}},\ }\href {https://doi.org/10.1038/416610a}
  {\bibfield  {journal} {\bibinfo  {journal} {Nature}\ }\textbf {\bibinfo
  {volume} {416}},\ \bibinfo {pages} {610} (\bibinfo {year}
  {2002})}\BibitemShut {NoStop}%
\bibitem [{\citenamefont {Fauqu\'e}\ \emph {et~al.}(2006)\citenamefont
  {Fauqu\'e}, \citenamefont {Sidis}, \citenamefont {Hinkov}, \citenamefont
  {Pailh\`es}, \citenamefont {Lin}, \citenamefont {Chaud},\ and\ \citenamefont
  {Bourges}}]{fauque06}%
  \BibitemOpen
  \bibfield  {author} {\bibinfo {author} {\bibfnamefont {B.}~\bibnamefont
  {Fauqu\'e}}, \bibinfo {author} {\bibfnamefont {Y.}~\bibnamefont {Sidis}},
  \bibinfo {author} {\bibfnamefont {V.}~\bibnamefont {Hinkov}}, \bibinfo
  {author} {\bibfnamefont {S.}~\bibnamefont {Pailh\`es}}, \bibinfo {author}
  {\bibfnamefont {C.~T.}\ \bibnamefont {Lin}}, \bibinfo {author} {\bibfnamefont
  {X.}~\bibnamefont {Chaud}},\ and\ \bibinfo {author} {\bibfnamefont
  {P.}~\bibnamefont {Bourges}},\ }\bibfield  {title} {\bibinfo {title}
  {Magnetic order in the pseudogap phase of high-${T}_{C}$ superconductors},\
  }\href {https://doi.org/10.1103/PhysRevLett.96.197001} {\bibfield  {journal}
  {\bibinfo  {journal} {Phys. Rev. Lett.}\ }\textbf {\bibinfo {volume} {96}},\
  \bibinfo {pages} {197001} (\bibinfo {year} {2006})}\BibitemShut {NoStop}%
\bibitem [{\citenamefont {Paul}\ and\ \citenamefont
  {Seibold}(2026)}]{paul2026}%
  \BibitemOpen
  \bibfield  {author} {\bibinfo {author} {\bibfnamefont {M.}~\bibnamefont
  {Paul}}\ and\ \bibinfo {author} {\bibfnamefont {G.}~\bibnamefont {Seibold}},\
  }\href {https://arxiv.org/abs/2607.08451} {\bibinfo {title} {Influence of
  electronic correlations on disorder-induced loop currents in high-tc
  superconductors}} (\bibinfo {year} {2026}),\ \Eprint
  {https://arxiv.org/abs/2607.08451} {arXiv:2607.08451 [cond-mat.supr-con]}
  \BibitemShut {NoStop}%
\bibitem [{\citenamefont {Bounoua}\ \emph {et~al.}(2026)\citenamefont
  {Bounoua}, \citenamefont {Liège}, \citenamefont {Sidis},\ and\ \citenamefont
  {Bourges}}]{bounoua_2026}%
  \BibitemOpen
  \bibfield  {author} {\bibinfo {author} {\bibfnamefont {D.}~\bibnamefont
  {Bounoua}}, \bibinfo {author} {\bibfnamefont {W.}~\bibnamefont {Liège}},
  \bibinfo {author} {\bibfnamefont {Y.}~\bibnamefont {Sidis}},\ and\ \bibinfo
  {author} {\bibfnamefont {P.}~\bibnamefont {Bourges}},\ }\bibfield  {title}
  {\bibinfo {title} {Orbital current signature using neutron diffraction},\
  }\bibfield  {journal} {\bibinfo  {journal} {International Journal of Modern
  Physics B}\ }\href {https://doi.org/10.1142/s0217979226400230}
  {10.1142/s0217979226400230} (\bibinfo {year} {2026})\BibitemShut {NoStop}%
\bibitem [{\citenamefont {Jiang}\ \emph {et~al.}(2021)\citenamefont {Jiang},
  \citenamefont {Yin}, \citenamefont {Denner}, \citenamefont {Shumiya},
  \citenamefont {Ortiz}, \citenamefont {Xu}, \citenamefont {Guguchia},
  \citenamefont {He}, \citenamefont {Hossain}, \citenamefont {Liu} \emph
  {et~al.}}]{jiang2021unconventional}%
  \BibitemOpen
  \bibfield  {author} {\bibinfo {author} {\bibfnamefont {Y.-X.}\ \bibnamefont
  {Jiang}}, \bibinfo {author} {\bibfnamefont {J.-X.}\ \bibnamefont {Yin}},
  \bibinfo {author} {\bibfnamefont {M.~M.}\ \bibnamefont {Denner}}, \bibinfo
  {author} {\bibfnamefont {N.}~\bibnamefont {Shumiya}}, \bibinfo {author}
  {\bibfnamefont {B.~R.}\ \bibnamefont {Ortiz}}, \bibinfo {author}
  {\bibfnamefont {G.}~\bibnamefont {Xu}}, \bibinfo {author} {\bibfnamefont
  {Z.}~\bibnamefont {Guguchia}}, \bibinfo {author} {\bibfnamefont
  {J.}~\bibnamefont {He}}, \bibinfo {author} {\bibfnamefont {M.~S.}\
  \bibnamefont {Hossain}}, \bibinfo {author} {\bibfnamefont {X.}~\bibnamefont
  {Liu}}, \emph {et~al.},\ }\bibfield  {title} {\bibinfo {title}
  {Unconventional chiral charge order in kagome superconductor kv3sb5},\
  }\href@noop {} {\bibfield  {journal} {\bibinfo  {journal} {Nature materials}\
  }\textbf {\bibinfo {volume} {20}},\ \bibinfo {pages} {1353} (\bibinfo {year}
  {2021})}\BibitemShut {NoStop}%
\bibitem [{\citenamefont {Mielke~III}\ \emph {et~al.}(2022)\citenamefont
  {Mielke~III}, \citenamefont {Das}, \citenamefont {Yin}, \citenamefont {Liu},
  \citenamefont {Gupta}, \citenamefont {Jiang}, \citenamefont {Medarde},
  \citenamefont {Wu}, \citenamefont {Lei}, \citenamefont {Chang} \emph
  {et~al.}}]{mielke2022time}%
  \BibitemOpen
  \bibfield  {author} {\bibinfo {author} {\bibfnamefont {C.}~\bibnamefont
  {Mielke~III}}, \bibinfo {author} {\bibfnamefont {D.}~\bibnamefont {Das}},
  \bibinfo {author} {\bibfnamefont {J.-X.}\ \bibnamefont {Yin}}, \bibinfo
  {author} {\bibfnamefont {H.}~\bibnamefont {Liu}}, \bibinfo {author}
  {\bibfnamefont {R.}~\bibnamefont {Gupta}}, \bibinfo {author} {\bibfnamefont
  {Y.-X.}\ \bibnamefont {Jiang}}, \bibinfo {author} {\bibfnamefont
  {M.}~\bibnamefont {Medarde}}, \bibinfo {author} {\bibfnamefont
  {X.}~\bibnamefont {Wu}}, \bibinfo {author} {\bibfnamefont {H.~C.}\
  \bibnamefont {Lei}}, \bibinfo {author} {\bibfnamefont {J.}~\bibnamefont
  {Chang}}, \emph {et~al.},\ }\bibfield  {title} {\bibinfo {title}
  {Time-reversal symmetry-breaking charge order in a kagome superconductor},\
  }\href@noop {} {\bibfield  {journal} {\bibinfo  {journal} {Nature}\ }\textbf
  {\bibinfo {volume} {602}},\ \bibinfo {pages} {245} (\bibinfo {year}
  {2022})}\BibitemShut {NoStop}%
\bibitem [{\citenamefont {Guo}\ \emph {et~al.}(2022)\citenamefont {Guo},
  \citenamefont {Putzke}, \citenamefont {Konyzheva}, \citenamefont {Huang},
  \citenamefont {Gutierrez-Amigo}, \citenamefont {Errea}, \citenamefont {Chen},
  \citenamefont {Vergniory}, \citenamefont {Felser}, \citenamefont {Fischer},
  \citenamefont {Neupert},\ and\ \citenamefont {Moll}}]{guo22}%
  \BibitemOpen
  \bibfield  {author} {\bibinfo {author} {\bibfnamefont {C.}~\bibnamefont
  {Guo}}, \bibinfo {author} {\bibfnamefont {C.}~\bibnamefont {Putzke}},
  \bibinfo {author} {\bibfnamefont {S.}~\bibnamefont {Konyzheva}}, \bibinfo
  {author} {\bibfnamefont {X.}~\bibnamefont {Huang}}, \bibinfo {author}
  {\bibfnamefont {M.}~\bibnamefont {Gutierrez-Amigo}}, \bibinfo {author}
  {\bibfnamefont {I.}~\bibnamefont {Errea}}, \bibinfo {author} {\bibfnamefont
  {D.}~\bibnamefont {Chen}}, \bibinfo {author} {\bibfnamefont {M.~G.}\
  \bibnamefont {Vergniory}}, \bibinfo {author} {\bibfnamefont {C.}~\bibnamefont
  {Felser}}, \bibinfo {author} {\bibfnamefont {M.~H.}\ \bibnamefont {Fischer}},
  \bibinfo {author} {\bibfnamefont {T.}~\bibnamefont {Neupert}},\ and\ \bibinfo
  {author} {\bibfnamefont {P.~J.~W.}\ \bibnamefont {Moll}},\ }\bibfield
  {title} {\bibinfo {title} {{Switchable chiral transport in charge-ordered
  kagome metal CsV3Sb5}},\ }\href {https://doi.org/10.1038/s41586-022-05127-9}
  {\bibfield  {journal} {\bibinfo  {journal} {Nature}\ }\textbf {\bibinfo
  {volume} {611}},\ \bibinfo {pages} {461} (\bibinfo {year}
  {2022})}\BibitemShut {NoStop}%
\bibitem [{\citenamefont {Wang}\ \emph
  {et~al.}(2026{\natexlab{a}})\citenamefont {Wang}, \citenamefont {Krix},
  \citenamefont {Tkachenko}, \citenamefont {Tkachenko}, \citenamefont {Chen},
  \citenamefont {Farrer}, \citenamefont {Ritchie}, \citenamefont {Sushkov},
  \citenamefont {Hamilton},\ and\ \citenamefont {Klochan}}]{wangka}%
  \BibitemOpen
  \bibfield  {author} {\bibinfo {author} {\bibfnamefont {D.~Q.}\ \bibnamefont
  {Wang}}, \bibinfo {author} {\bibfnamefont {Z.}~\bibnamefont {Krix}}, \bibinfo
  {author} {\bibfnamefont {O.~A.}\ \bibnamefont {Tkachenko}}, \bibinfo {author}
  {\bibfnamefont {V.~A.}\ \bibnamefont {Tkachenko}}, \bibinfo {author}
  {\bibfnamefont {C.}~\bibnamefont {Chen}}, \bibinfo {author} {\bibfnamefont
  {I.}~\bibnamefont {Farrer}}, \bibinfo {author} {\bibfnamefont {D.~A.}\
  \bibnamefont {Ritchie}}, \bibinfo {author} {\bibfnamefont {O.~P.}\
  \bibnamefont {Sushkov}}, \bibinfo {author} {\bibfnamefont {A.~R.}\
  \bibnamefont {Hamilton}},\ and\ \bibinfo {author} {\bibfnamefont
  {O.}~\bibnamefont {Klochan}},\ }\bibfield  {title} {\bibinfo {title}
  {Correlated insulator in the kagome flat band of a two-dimensional
  electrostatic crystal},\ }\href {https://doi.org/10.1038/s41567-026-03291-7}
  {\bibfield  {journal} {\bibinfo  {journal} {Nature Physics}\ }\textbf
  {\bibinfo {volume} {22}},\ \bibinfo {pages} {1079} (\bibinfo {year}
  {2026}{\natexlab{a}})}\BibitemShut {NoStop}%
\bibitem [{\citenamefont {Fu}\ \emph {et~al.}(2025)\citenamefont {Fu},
  \citenamefont {Zhan}, \citenamefont {D{\"u}rrnagel}, \citenamefont {Hohmann},
  \citenamefont {Thomale}, \citenamefont {Hu}, \citenamefont {Wang},
  \citenamefont {Zhou},\ and\ \citenamefont {Wu}}]{fu2025exotic}%
  \BibitemOpen
  \bibfield  {author} {\bibinfo {author} {\bibfnamefont {R.}~\bibnamefont
  {Fu}}, \bibinfo {author} {\bibfnamefont {J.}~\bibnamefont {Zhan}}, \bibinfo
  {author} {\bibfnamefont {M.}~\bibnamefont {D{\"u}rrnagel}}, \bibinfo {author}
  {\bibfnamefont {H.}~\bibnamefont {Hohmann}}, \bibinfo {author} {\bibfnamefont
  {R.}~\bibnamefont {Thomale}}, \bibinfo {author} {\bibfnamefont
  {J.}~\bibnamefont {Hu}}, \bibinfo {author} {\bibfnamefont {Z.}~\bibnamefont
  {Wang}}, \bibinfo {author} {\bibfnamefont {S.}~\bibnamefont {Zhou}},\ and\
  \bibinfo {author} {\bibfnamefont {X.}~\bibnamefont {Wu}},\ }\bibfield
  {title} {\bibinfo {title} {Exotic charge-density waves and superconductivity
  on the kagome lattice},\ }\href@noop {} {\bibfield  {journal} {\bibinfo
  {journal} {National Science Review}\ }\textbf {\bibinfo {volume} {12}},\
  \bibinfo {pages} {nwaf414} (\bibinfo {year} {2025})}\BibitemShut {NoStop}%
\bibitem [{\citenamefont {Tazai}\ \emph {et~al.}(2026)\citenamefont {Tazai},
  \citenamefont {Yamakawa},\ and\ \citenamefont {Kontani}}]{Tazai_2026}%
  \BibitemOpen
  \bibfield  {author} {\bibinfo {author} {\bibfnamefont {R.}~\bibnamefont
  {Tazai}}, \bibinfo {author} {\bibfnamefont {Y.}~\bibnamefont {Yamakawa}},\
  and\ \bibinfo {author} {\bibfnamefont {H.}~\bibnamefont {Kontani}},\
  }\bibfield  {title} {\bibinfo {title} {Nematic and chiral superconductivity
  emerging within the loop-current phase in kagome metals},\ }\bibfield
  {journal} {\bibinfo  {journal} {Nature Communications}\ }\href
  {https://doi.org/10.1038/s41467-026-75259-3} {10.1038/s41467-026-75259-3}
  (\bibinfo {year} {2026})\BibitemShut {NoStop}%
\bibitem [{\citenamefont {Zhan}\ \emph {et~al.}(2026)\citenamefont {Zhan},
  \citenamefont {Hohmann}, \citenamefont {D\"urrnagel}, \citenamefont {Fu},
  \citenamefont {Zhou}, \citenamefont {Wang}, \citenamefont {Thomale},
  \citenamefont {Wu},\ and\ \citenamefont {Hu}}]{jun26}%
  \BibitemOpen
  \bibfield  {author} {\bibinfo {author} {\bibfnamefont {J.}~\bibnamefont
  {Zhan}}, \bibinfo {author} {\bibfnamefont {H.}~\bibnamefont {Hohmann}},
  \bibinfo {author} {\bibfnamefont {M.}~\bibnamefont {D\"urrnagel}}, \bibinfo
  {author} {\bibfnamefont {R.}~\bibnamefont {Fu}}, \bibinfo {author}
  {\bibfnamefont {S.}~\bibnamefont {Zhou}}, \bibinfo {author} {\bibfnamefont
  {Z.}~\bibnamefont {Wang}}, \bibinfo {author} {\bibfnamefont {R.}~\bibnamefont
  {Thomale}}, \bibinfo {author} {\bibfnamefont {X.}~\bibnamefont {Wu}},\ and\
  \bibinfo {author} {\bibfnamefont {J.}~\bibnamefont {Hu}},\ }\bibfield
  {title} {\bibinfo {title} {Loop current order on the kagome lattice},\ }\href
  {https://doi.org/10.1103/5vyy-rj6v} {\bibfield  {journal} {\bibinfo
  {journal} {Phys. Rev. Lett.}\ }\textbf {\bibinfo {volume} {136}},\ \bibinfo
  {pages} {126001} (\bibinfo {year} {2026})}\BibitemShut {NoStop}%
\bibitem [{\citenamefont {Jeong}\ \emph {et~al.}(2017)\citenamefont {Jeong},
  \citenamefont {Sidis}, \citenamefont {Louat}, \citenamefont {Brouet},\ and\
  \citenamefont {Bourges}}]{jeong17}%
  \BibitemOpen
  \bibfield  {author} {\bibinfo {author} {\bibfnamefont {J.}~\bibnamefont
  {Jeong}}, \bibinfo {author} {\bibfnamefont {Y.}~\bibnamefont {Sidis}},
  \bibinfo {author} {\bibfnamefont {A.}~\bibnamefont {Louat}}, \bibinfo
  {author} {\bibfnamefont {V.}~\bibnamefont {Brouet}},\ and\ \bibinfo {author}
  {\bibfnamefont {P.}~\bibnamefont {Bourges}},\ }\bibfield  {title} {\bibinfo
  {title} {{Time-reversal symmetry breaking hidden order in Sr2(Ir,Rh)O4}},\
  }\href {https://doi.org/10.1038/ncomms15119} {\bibfield  {journal} {\bibinfo
  {journal} {Nature Communications}\ }\textbf {\bibinfo {volume} {8}},\
  \bibinfo {pages} {15119} (\bibinfo {year} {2017})}\BibitemShut {NoStop}%
\bibitem [{\citenamefont {Fittipaldi}\ \emph {et~al.}(2021)\citenamefont
  {Fittipaldi}, \citenamefont {Hartmann}, \citenamefont {Mercaldo},
  \citenamefont {Komori}, \citenamefont {Bj{\o}rlig}, \citenamefont {Kyung},
  \citenamefont {Yasui}, \citenamefont {Miyoshi}, \citenamefont {Olde~Olthof},
  \citenamefont {Palomares~Garcia}, \citenamefont {Granata}, \citenamefont
  {Keren}, \citenamefont {Higemoto}, \citenamefont {Suter}, \citenamefont
  {Prokscha}, \citenamefont {Romano}, \citenamefont {Noce}, \citenamefont
  {Kim}, \citenamefont {Maeno}, \citenamefont {Scheer}, \citenamefont
  {Kalisky}, \citenamefont {Robinson}, \citenamefont {Cuoco}, \citenamefont
  {Salman}, \citenamefont {Vecchione},\ and\ \citenamefont
  {Di~Bernardo}}]{fittipaldi21}%
  \BibitemOpen
  \bibfield  {author} {\bibinfo {author} {\bibfnamefont {R.}~\bibnamefont
  {Fittipaldi}}, \bibinfo {author} {\bibfnamefont {R.}~\bibnamefont
  {Hartmann}}, \bibinfo {author} {\bibfnamefont {M.~T.}\ \bibnamefont
  {Mercaldo}}, \bibinfo {author} {\bibfnamefont {S.}~\bibnamefont {Komori}},
  \bibinfo {author} {\bibfnamefont {A.}~\bibnamefont {Bj{\o}rlig}}, \bibinfo
  {author} {\bibfnamefont {W.}~\bibnamefont {Kyung}}, \bibinfo {author}
  {\bibfnamefont {Y.}~\bibnamefont {Yasui}}, \bibinfo {author} {\bibfnamefont
  {T.}~\bibnamefont {Miyoshi}}, \bibinfo {author} {\bibfnamefont {L.~A.~B.}\
  \bibnamefont {Olde~Olthof}}, \bibinfo {author} {\bibfnamefont {C.~M.}\
  \bibnamefont {Palomares~Garcia}}, \bibinfo {author} {\bibfnamefont
  {V.}~\bibnamefont {Granata}}, \bibinfo {author} {\bibfnamefont
  {I.}~\bibnamefont {Keren}}, \bibinfo {author} {\bibfnamefont
  {W.}~\bibnamefont {Higemoto}}, \bibinfo {author} {\bibfnamefont
  {A.}~\bibnamefont {Suter}}, \bibinfo {author} {\bibfnamefont
  {T.}~\bibnamefont {Prokscha}}, \bibinfo {author} {\bibfnamefont
  {A.}~\bibnamefont {Romano}}, \bibinfo {author} {\bibfnamefont
  {C.}~\bibnamefont {Noce}}, \bibinfo {author} {\bibfnamefont {C.}~\bibnamefont
  {Kim}}, \bibinfo {author} {\bibfnamefont {Y.}~\bibnamefont {Maeno}}, \bibinfo
  {author} {\bibfnamefont {E.}~\bibnamefont {Scheer}}, \bibinfo {author}
  {\bibfnamefont {B.}~\bibnamefont {Kalisky}}, \bibinfo {author} {\bibfnamefont
  {J.~W.~A.}\ \bibnamefont {Robinson}}, \bibinfo {author} {\bibfnamefont
  {M.}~\bibnamefont {Cuoco}}, \bibinfo {author} {\bibfnamefont
  {Z.}~\bibnamefont {Salman}}, \bibinfo {author} {\bibfnamefont
  {A.}~\bibnamefont {Vecchione}},\ and\ \bibinfo {author} {\bibfnamefont
  {A.}~\bibnamefont {Di~Bernardo}},\ }\bibfield  {title} {\bibinfo {title}
  {{Unveiling unconventional magnetism at the surface of Sr2RuO4}},\ }\href
  {https://doi.org/10.1038/s41467-021-26020-5} {\bibfield  {journal} {\bibinfo
  {journal} {Nature Communications}\ }\textbf {\bibinfo {volume} {12}},\
  \bibinfo {pages} {5792} (\bibinfo {year} {2021})}\BibitemShut {NoStop}%
\bibitem [{\citenamefont {Ji}\ \emph {et~al.}(2026)\citenamefont {Ji},
  \citenamefont {Xie}, \citenamefont {Chen}, \citenamefont {Zhou},
  \citenamefont {Pan}, \citenamefont {Wang}, \citenamefont {Huang},
  \citenamefont {Ge}, \citenamefont {Liu}, \citenamefont {Zhang}, \citenamefont
  {Wang}, \citenamefont {Xue}, \citenamefont {Chen},\ and\ \citenamefont
  {Wang}}]{ji2026}%
  \BibitemOpen
  \bibfield  {author} {\bibinfo {author} {\bibfnamefont {H.}~\bibnamefont
  {Ji}}, \bibinfo {author} {\bibfnamefont {Z.}~\bibnamefont {Xie}}, \bibinfo
  {author} {\bibfnamefont {Y.}~\bibnamefont {Chen}}, \bibinfo {author}
  {\bibfnamefont {G.}~\bibnamefont {Zhou}}, \bibinfo {author} {\bibfnamefont
  {L.}~\bibnamefont {Pan}}, \bibinfo {author} {\bibfnamefont {H.}~\bibnamefont
  {Wang}}, \bibinfo {author} {\bibfnamefont {H.}~\bibnamefont {Huang}},
  \bibinfo {author} {\bibfnamefont {J.}~\bibnamefont {Ge}}, \bibinfo {author}
  {\bibfnamefont {Y.}~\bibnamefont {Liu}}, \bibinfo {author} {\bibfnamefont
  {G.-M.}\ \bibnamefont {Zhang}}, \bibinfo {author} {\bibfnamefont
  {Z.}~\bibnamefont {Wang}}, \bibinfo {author} {\bibfnamefont {Q.-K.}\
  \bibnamefont {Xue}}, \bibinfo {author} {\bibfnamefont {Z.}~\bibnamefont
  {Chen}},\ and\ \bibinfo {author} {\bibfnamefont {J.}~\bibnamefont {Wang}},\
  }\href {https://arxiv.org/abs/2508.16412} {\bibinfo {title} {{Time-reversal
  symmetry breaking superconductivity with electronic glass in nickelate (La,
  Pr, Sm)3Ni2O7 films}}} (\bibinfo {year} {2026}),\ \Eprint
  {https://arxiv.org/abs/2508.16412} {arXiv:2508.16412 [cond-mat.supr-con]}
  \BibitemShut {NoStop}%
\bibitem [{\citenamefont {Morisseau}\ \emph {et~al.}(2026)\citenamefont
  {Morisseau}, \citenamefont {Rhodes}, \citenamefont {Wahl},\ and\
  \citenamefont {Marques}}]{morisseau26}%
  \BibitemOpen
  \bibfield  {author} {\bibinfo {author} {\bibfnamefont {V.}~\bibnamefont
  {Morisseau}}, \bibinfo {author} {\bibfnamefont {L.~C.}\ \bibnamefont
  {Rhodes}}, \bibinfo {author} {\bibfnamefont {P.}~\bibnamefont {Wahl}},\ and\
  \bibinfo {author} {\bibfnamefont {C.~A.}\ \bibnamefont {Marques}},\ }\href
  {https://arxiv.org/abs/2607.20030} {\bibinfo {title} {How to measure loop
  currents in scanning tunneling microscopy}} (\bibinfo {year} {2026}),\
  \Eprint {https://arxiv.org/abs/2607.20030} {arXiv:2607.20030
  [cond-mat.str-el]} \BibitemShut {NoStop}%
\bibitem [{\citenamefont {Zeyher}\ and\ \citenamefont
  {Greco}(2003)}]{zeyher03}%
  \BibitemOpen
  \bibfield  {author} {\bibinfo {author} {\bibfnamefont {R.}~\bibnamefont
  {Zeyher}}\ and\ \bibinfo {author} {\bibfnamefont {A.}~\bibnamefont {Greco}},\
  }\bibfield  {title} {\bibinfo {title} {{Competition between superconductivity
  and the pseudogap phase in the t–J model}},\ }\href
  {https://doi.org/https://doi.org/10.1002/pssb.200301675} {\bibfield
  {journal} {\bibinfo  {journal} {physica status solidi (b)}\ }\textbf
  {\bibinfo {volume} {236}},\ \bibinfo {pages} {343} (\bibinfo {year}
  {2003})}\BibitemShut {NoStop}%
\bibitem [{\citenamefont {Greco}\ and\ \citenamefont {Zeyher}(2004)}]{greco04}%
  \BibitemOpen
  \bibfield  {author} {\bibinfo {author} {\bibfnamefont {A.}~\bibnamefont
  {Greco}}\ and\ \bibinfo {author} {\bibfnamefont {R.}~\bibnamefont {Zeyher}},\
  }\bibfield  {title} {\bibinfo {title} {$c$-axis tunneling spectra in
  high-${T}_{c}$ superconductors in the presence of a $d$-charge-density
  wave},\ }\href {https://doi.org/10.1103/PhysRevB.70.024518} {\bibfield
  {journal} {\bibinfo  {journal} {Phys. Rev. B}\ }\textbf {\bibinfo {volume}
  {70}},\ \bibinfo {pages} {024518} (\bibinfo {year} {2004})}\BibitemShut
  {NoStop}%
\bibitem [{\citenamefont {Hao}\ \emph {et~al.}(2025)\citenamefont {Hao},
  \citenamefont {Wang}, \citenamefont {Sun}, \citenamefont {Yang},
  \citenamefont {Mao}, \citenamefont {Yan}, \citenamefont {Sun}, \citenamefont
  {Zhang}, \citenamefont {Han}, \citenamefont {Gu}, \citenamefont {Zhou},
  \citenamefont {Ji},\ and\ \citenamefont {Nie}}]{hao2025}%
  \BibitemOpen
  \bibfield  {author} {\bibinfo {author} {\bibfnamefont {B.}~\bibnamefont
  {Hao}}, \bibinfo {author} {\bibfnamefont {M.}~\bibnamefont {Wang}}, \bibinfo
  {author} {\bibfnamefont {W.}~\bibnamefont {Sun}}, \bibinfo {author}
  {\bibfnamefont {Y.}~\bibnamefont {Yang}}, \bibinfo {author} {\bibfnamefont
  {Z.}~\bibnamefont {Mao}}, \bibinfo {author} {\bibfnamefont {S.}~\bibnamefont
  {Yan}}, \bibinfo {author} {\bibfnamefont {H.}~\bibnamefont {Sun}}, \bibinfo
  {author} {\bibfnamefont {H.}~\bibnamefont {Zhang}}, \bibinfo {author}
  {\bibfnamefont {L.}~\bibnamefont {Han}}, \bibinfo {author} {\bibfnamefont
  {Z.}~\bibnamefont {Gu}}, \bibinfo {author} {\bibfnamefont {J.}~\bibnamefont
  {Zhou}}, \bibinfo {author} {\bibfnamefont {D.}~\bibnamefont {Ji}},\ and\
  \bibinfo {author} {\bibfnamefont {Y.}~\bibnamefont {Nie}},\ }\bibfield
  {title} {\bibinfo {title} {{Superconductivity in Sr-doped La3Ni2O7 thin
  films}},\ }\href {https://doi.org/10.1038/s41563-025-02327-2} {\bibfield
  {journal} {\bibinfo  {journal} {Nature Materials}\ }\textbf {\bibinfo
  {volume} {24}},\ \bibinfo {pages} {1756} (\bibinfo {year}
  {2025})}\BibitemShut {NoStop}%
\bibitem [{\citenamefont {Wang}\ \emph
  {et~al.}(2026{\natexlab{b}})\citenamefont {Wang}, \citenamefont {Hao},
  \citenamefont {Sun}, \citenamefont {Yan}, \citenamefont {Sun}, \citenamefont
  {Zhang}, \citenamefont {Gu},\ and\ \citenamefont {Nie}}]{Wang_2026}%
  \BibitemOpen
  \bibfield  {author} {\bibinfo {author} {\bibfnamefont {M.}~\bibnamefont
  {Wang}}, \bibinfo {author} {\bibfnamefont {B.}~\bibnamefont {Hao}}, \bibinfo
  {author} {\bibfnamefont {W.}~\bibnamefont {Sun}}, \bibinfo {author}
  {\bibfnamefont {S.}~\bibnamefont {Yan}}, \bibinfo {author} {\bibfnamefont
  {S.}~\bibnamefont {Sun}}, \bibinfo {author} {\bibfnamefont {H.}~\bibnamefont
  {Zhang}}, \bibinfo {author} {\bibfnamefont {Z.}~\bibnamefont {Gu}},\ and\
  \bibinfo {author} {\bibfnamefont {Y.}~\bibnamefont {Nie}},\ }\bibfield
  {title} {\bibinfo {title} {{Superconducting Dome in
  ${\mathrm{La}}_{3\ensuremath{-}x}{\mathrm{Sr}}_{x}{\mathrm{Ni}}_{2}{\mathrm{O}}_{7\ensuremath{-}\ensuremath{\delta}}$
  Thin Films}},\ }\href {https://doi.org/10.1103/qrkk-l2ng} {\bibfield
  {journal} {\bibinfo  {journal} {Phys. Rev. Lett.}\ }\textbf {\bibinfo
  {volume} {136}},\ \bibinfo {pages} {066002} (\bibinfo {year}
  {2026}{\natexlab{b}})}\BibitemShut {NoStop}%
\bibitem [{\citenamefont {Oh}\ \emph {et~al.}(2025)\citenamefont {Oh},
  \citenamefont {Zhou},\ and\ \citenamefont {Zhang}}]{oh25}%
  \BibitemOpen
  \bibfield  {author} {\bibinfo {author} {\bibfnamefont {H.}~\bibnamefont
  {Oh}}, \bibinfo {author} {\bibfnamefont {B.}~\bibnamefont {Zhou}},\ and\
  \bibinfo {author} {\bibfnamefont {Y.-H.}\ \bibnamefont {Zhang}},\ }\bibfield
  {title} {\bibinfo {title} {{Type-II $t\text{\ensuremath{-}}J$ model in charge
  transfer regime in bilayer
  ${\mathrm{La}}_{3}{\mathrm{Ni}}_{2}{\mathrm{O}}_{7}$ and trilayer
  ${\mathrm{La}}_{4}{\mathrm{Ni}}_{3}{\mathrm{O}}_{10}$}},\ }\href
  {https://doi.org/10.1103/PhysRevB.111.L020504} {\bibfield  {journal}
  {\bibinfo  {journal} {Phys. Rev. B}\ }\textbf {\bibinfo {volume} {111}},\
  \bibinfo {pages} {L020504} (\bibinfo {year} {2025})}\BibitemShut {NoStop}%
\bibitem [{\citenamefont {Shen}\ \emph {et~al.}(2026)\citenamefont {Shen},
  \citenamefont {Zhou}, \citenamefont {Miao}, \citenamefont {Li}, \citenamefont
  {Ou}, \citenamefont {Chen}, \citenamefont {Wang}, \citenamefont {Luan},
  \citenamefont {Sun}, \citenamefont {Feng}, \citenamefont {Yong},
  \citenamefont {Li}, \citenamefont {Xu}, \citenamefont {Lv}, \citenamefont
  {Nie}, \citenamefont {Wang}, \citenamefont {Huang}, \citenamefont {Sun},
  \citenamefont {Xue}, \citenamefont {He},\ and\ \citenamefont
  {Chen}}]{Shen_2026}%
  \BibitemOpen
  \bibfield  {author} {\bibinfo {author} {\bibfnamefont {J.}~\bibnamefont
  {Shen}}, \bibinfo {author} {\bibfnamefont {G.}~\bibnamefont {Zhou}}, \bibinfo
  {author} {\bibfnamefont {Y.}~\bibnamefont {Miao}}, \bibinfo {author}
  {\bibfnamefont {P.}~\bibnamefont {Li}}, \bibinfo {author} {\bibfnamefont
  {Z.}~\bibnamefont {Ou}}, \bibinfo {author} {\bibfnamefont {Y.}~\bibnamefont
  {Chen}}, \bibinfo {author} {\bibfnamefont {Z.}~\bibnamefont {Wang}}, \bibinfo
  {author} {\bibfnamefont {R.}~\bibnamefont {Luan}}, \bibinfo {author}
  {\bibfnamefont {H.}~\bibnamefont {Sun}}, \bibinfo {author} {\bibfnamefont
  {Z.}~\bibnamefont {Feng}}, \bibinfo {author} {\bibfnamefont {X.}~\bibnamefont
  {Yong}}, \bibinfo {author} {\bibfnamefont {Y.}~\bibnamefont {Li}}, \bibinfo
  {author} {\bibfnamefont {L.}~\bibnamefont {Xu}}, \bibinfo {author}
  {\bibfnamefont {W.}~\bibnamefont {Lv}}, \bibinfo {author} {\bibfnamefont
  {Z.}~\bibnamefont {Nie}}, \bibinfo {author} {\bibfnamefont {H.}~\bibnamefont
  {Wang}}, \bibinfo {author} {\bibfnamefont {H.}~\bibnamefont {Huang}},
  \bibinfo {author} {\bibfnamefont {Y.-J.}\ \bibnamefont {Sun}}, \bibinfo
  {author} {\bibfnamefont {Q.-K.}\ \bibnamefont {Xue}}, \bibinfo {author}
  {\bibfnamefont {J.}~\bibnamefont {He}},\ and\ \bibinfo {author}
  {\bibfnamefont {Z.}~\bibnamefont {Chen}},\ }\bibfield  {title} {\bibinfo
  {title} {{Nodeless superconducting gap and electron-boson coupling in
  (La,Pr,Sm)$_3$Ni$_2$O$_7$ films}},\ }\href
  {https://doi.org/10.1126/science.adw8329} {\bibfield  {journal} {\bibinfo
  {journal} {Science}\ }\textbf {\bibinfo {volume} {392}},\ \bibinfo {pages}
  {1396} (\bibinfo {year} {2026})}\BibitemShut {NoStop}%
\bibitem [{\citenamefont {Craco}\ and\ \citenamefont {Leoni}(2024)}]{craco24}%
  \BibitemOpen
  \bibfield  {author} {\bibinfo {author} {\bibfnamefont {L.}~\bibnamefont
  {Craco}}\ and\ \bibinfo {author} {\bibfnamefont {S.}~\bibnamefont {Leoni}},\
  }\bibfield  {title} {\bibinfo {title} {{Strange metal and
  coherence-incoherence crossover in pressurized
  ${\mathrm{La}}_{3}{\mathrm{Ni}}_{2}{\mathrm{O}}_{7}$}},\ }\href
  {https://doi.org/10.1103/PhysRevB.109.165116} {\bibfield  {journal} {\bibinfo
   {journal} {Phys. Rev. B}\ }\textbf {\bibinfo {volume} {109}},\ \bibinfo
  {pages} {165116} (\bibinfo {year} {2024})}\BibitemShut {NoStop}%
\bibitem [{\citenamefont {Zhang}\ \emph {et~al.}(2024)\citenamefont {Zhang},
  \citenamefont {Su}, \citenamefont {Huang}, \citenamefont {Shan},
  \citenamefont {Sun}, \citenamefont {Huo}, \citenamefont {Ye}, \citenamefont
  {Zhang}, \citenamefont {Yang}, \citenamefont {Xu}, \citenamefont {Su},
  \citenamefont {Li}, \citenamefont {Smidman}, \citenamefont {Wang},
  \citenamefont {Jiao},\ and\ \citenamefont {Yuan}}]{zhang23}%
  \BibitemOpen
  \bibfield  {author} {\bibinfo {author} {\bibfnamefont {Y.}~\bibnamefont
  {Zhang}}, \bibinfo {author} {\bibfnamefont {D.}~\bibnamefont {Su}}, \bibinfo
  {author} {\bibfnamefont {Y.}~\bibnamefont {Huang}}, \bibinfo {author}
  {\bibfnamefont {Z.}~\bibnamefont {Shan}}, \bibinfo {author} {\bibfnamefont
  {H.}~\bibnamefont {Sun}}, \bibinfo {author} {\bibfnamefont {M.}~\bibnamefont
  {Huo}}, \bibinfo {author} {\bibfnamefont {K.}~\bibnamefont {Ye}}, \bibinfo
  {author} {\bibfnamefont {J.}~\bibnamefont {Zhang}}, \bibinfo {author}
  {\bibfnamefont {Z.}~\bibnamefont {Yang}}, \bibinfo {author} {\bibfnamefont
  {Y.}~\bibnamefont {Xu}}, \bibinfo {author} {\bibfnamefont {Y.}~\bibnamefont
  {Su}}, \bibinfo {author} {\bibfnamefont {R.}~\bibnamefont {Li}}, \bibinfo
  {author} {\bibfnamefont {M.}~\bibnamefont {Smidman}}, \bibinfo {author}
  {\bibfnamefont {M.}~\bibnamefont {Wang}}, \bibinfo {author} {\bibfnamefont
  {L.}~\bibnamefont {Jiao}},\ and\ \bibinfo {author} {\bibfnamefont
  {H.}~\bibnamefont {Yuan}},\ }\bibfield  {title} {\bibinfo {title}
  {{High-temperature superconductivity with zero resistance and strange-metal
  behaviour in La3Ni2O7-$\delta$}},\ }\href
  {https://doi.org/10.1038/s41567-024-02515-y} {\bibfield  {journal} {\bibinfo
  {journal} {Nature Physics}\ }\textbf {\bibinfo {volume} {20}},\ \bibinfo
  {pages} {1269} (\bibinfo {year} {2024})}\BibitemShut {NoStop}%
\bibitem [{\citenamefont {Feng}\ \emph {et~al.}(2024)\citenamefont {Feng},
  \citenamefont {Han}, \citenamefont {Song}, \citenamefont {Long},
  \citenamefont {Hou}, \citenamefont {Zhang}, \citenamefont {Mu},\ and\
  \citenamefont {Shan}}]{fengNd}%
  \BibitemOpen
  \bibfield  {author} {\bibinfo {author} {\bibfnamefont {J.-J.}\ \bibnamefont
  {Feng}}, \bibinfo {author} {\bibfnamefont {T.}~\bibnamefont {Han}}, \bibinfo
  {author} {\bibfnamefont {J.-P.}\ \bibnamefont {Song}}, \bibinfo {author}
  {\bibfnamefont {M.-S.}\ \bibnamefont {Long}}, \bibinfo {author}
  {\bibfnamefont {X.-Y.}\ \bibnamefont {Hou}}, \bibinfo {author} {\bibfnamefont
  {C.-J.}\ \bibnamefont {Zhang}}, \bibinfo {author} {\bibfnamefont {Q.-G.}\
  \bibnamefont {Mu}},\ and\ \bibinfo {author} {\bibfnamefont {L.}~\bibnamefont
  {Shan}},\ }\bibfield  {title} {\bibinfo {title} {{Unaltered density wave
  transition and pressure-induced signature of superconductivity in Nd-doped
  ${\mathrm{La}}_{3}{\mathrm{Ni}}_{2}{\mathrm{O}}_{7}$}},\ }\href
  {https://doi.org/10.1103/PhysRevB.110.L100507} {\bibfield  {journal}
  {\bibinfo  {journal} {Phys. Rev. B}\ }\textbf {\bibinfo {volume} {110}},\
  \bibinfo {pages} {L100507} (\bibinfo {year} {2024})}\BibitemShut {NoStop}%
\bibitem [{\citenamefont {Zhong}\ \emph
  {et~al.}(2026{\natexlab{a}})\citenamefont {Zhong}, \citenamefont {Chen},
  \citenamefont {Qiu}, \citenamefont {Li}, \citenamefont {Huang}, \citenamefont
  {Ma}, \citenamefont {Huo}, \citenamefont {Dong}, \citenamefont {Deng},
  \citenamefont {He}, \citenamefont {Han}, \citenamefont {Sun},\ and\
  \citenamefont {Wang}}]{zhongNd}%
  \BibitemOpen
  \bibfield  {author} {\bibinfo {author} {\bibfnamefont {Q.}~\bibnamefont
  {Zhong}}, \bibinfo {author} {\bibfnamefont {J.}~\bibnamefont {Chen}},
  \bibinfo {author} {\bibfnamefont {Z.}~\bibnamefont {Qiu}}, \bibinfo {author}
  {\bibfnamefont {J.}~\bibnamefont {Li}}, \bibinfo {author} {\bibfnamefont
  {X.}~\bibnamefont {Huang}}, \bibinfo {author} {\bibfnamefont
  {P.}~\bibnamefont {Ma}}, \bibinfo {author} {\bibfnamefont {M.}~\bibnamefont
  {Huo}}, \bibinfo {author} {\bibfnamefont {H.}~\bibnamefont {Dong}}, \bibinfo
  {author} {\bibfnamefont {S.}~\bibnamefont {Deng}}, \bibinfo {author}
  {\bibfnamefont {L.}~\bibnamefont {He}}, \bibinfo {author} {\bibfnamefont
  {Y.}~\bibnamefont {Han}}, \bibinfo {author} {\bibfnamefont {H.}~\bibnamefont
  {Sun}},\ and\ \bibinfo {author} {\bibfnamefont {M.}~\bibnamefont {Wang}},\
  }\bibfield  {title} {\bibinfo {title} {{Evolution of superconductivity
  evidence in pressurized
  ${\mathrm{La}}_{3\ensuremath{-}x}{\mathrm{Sm}}_{x}{\mathrm{Ni}}_{2}{\mathrm{O}}_{7}$}},\
  }\href {https://doi.org/10.1103/pzv7-1nlr} {\bibfield  {journal} {\bibinfo
  {journal} {Phys. Rev. B}\ }\textbf {\bibinfo {volume} {113}},\ \bibinfo
  {pages} {174512} (\bibinfo {year} {2026}{\natexlab{a}})}\BibitemShut
  {NoStop}%
\bibitem [{\citenamefont {Qiu}\ \emph {et~al.}(2025)\citenamefont {Qiu},
  \citenamefont {Chen}, \citenamefont {Semenok}, \citenamefont {Zhong},
  \citenamefont {Zhou}, \citenamefont {Li}, \citenamefont {Ma}, \citenamefont
  {Huang}, \citenamefont {Huo}, \citenamefont {Xie}, \citenamefont {Chen},
  \citenamefont {kwang Mao}, \citenamefont {Struzhkin}, \citenamefont {Sun},\
  and\ \citenamefont {Wang}}]{qiu2025}%
  \BibitemOpen
  \bibfield  {author} {\bibinfo {author} {\bibfnamefont {Z.}~\bibnamefont
  {Qiu}}, \bibinfo {author} {\bibfnamefont {J.}~\bibnamefont {Chen}}, \bibinfo
  {author} {\bibfnamefont {D.~V.}\ \bibnamefont {Semenok}}, \bibinfo {author}
  {\bibfnamefont {Q.}~\bibnamefont {Zhong}}, \bibinfo {author} {\bibfnamefont
  {D.}~\bibnamefont {Zhou}}, \bibinfo {author} {\bibfnamefont {J.}~\bibnamefont
  {Li}}, \bibinfo {author} {\bibfnamefont {P.}~\bibnamefont {Ma}}, \bibinfo
  {author} {\bibfnamefont {X.}~\bibnamefont {Huang}}, \bibinfo {author}
  {\bibfnamefont {M.}~\bibnamefont {Huo}}, \bibinfo {author} {\bibfnamefont
  {T.}~\bibnamefont {Xie}}, \bibinfo {author} {\bibfnamefont {X.}~\bibnamefont
  {Chen}}, \bibinfo {author} {\bibfnamefont {H.}~\bibnamefont {kwang Mao}},
  \bibinfo {author} {\bibfnamefont {V.}~\bibnamefont {Struzhkin}}, \bibinfo
  {author} {\bibfnamefont {H.}~\bibnamefont {Sun}},\ and\ \bibinfo {author}
  {\bibfnamefont {M.}~\bibnamefont {Wang}},\ }\href
  {https://arxiv.org/abs/2510.12359} {\bibinfo {title} {{Interlayer coupling
  enhanced superconductivity near 100 K in La$_{3-x}$Nd$_x$Ni$_2$O$_7$}}}
  (\bibinfo {year} {2025}),\ \Eprint {https://arxiv.org/abs/2510.12359}
  {arXiv:2510.12359 [cond-mat.supr-con]} \BibitemShut {NoStop}%
\bibitem [{\citenamefont {Zhong}\ \emph
  {et~al.}(2026{\natexlab{b}})\citenamefont {Zhong}, \citenamefont {Chen},
  \citenamefont {Qiu}, \citenamefont {Li}, \citenamefont {Huang}, \citenamefont
  {Ma}, \citenamefont {Huo}, \citenamefont {Dong}, \citenamefont {Deng},
  \citenamefont {He}, \citenamefont {Han}, \citenamefont {Sun},\ and\
  \citenamefont {Wang}}]{zhong2025e}%
  \BibitemOpen
  \bibfield  {author} {\bibinfo {author} {\bibfnamefont {Q.}~\bibnamefont
  {Zhong}}, \bibinfo {author} {\bibfnamefont {J.}~\bibnamefont {Chen}},
  \bibinfo {author} {\bibfnamefont {Z.}~\bibnamefont {Qiu}}, \bibinfo {author}
  {\bibfnamefont {J.}~\bibnamefont {Li}}, \bibinfo {author} {\bibfnamefont
  {X.}~\bibnamefont {Huang}}, \bibinfo {author} {\bibfnamefont
  {P.}~\bibnamefont {Ma}}, \bibinfo {author} {\bibfnamefont {M.}~\bibnamefont
  {Huo}}, \bibinfo {author} {\bibfnamefont {H.}~\bibnamefont {Dong}}, \bibinfo
  {author} {\bibfnamefont {S.}~\bibnamefont {Deng}}, \bibinfo {author}
  {\bibfnamefont {L.}~\bibnamefont {He}}, \bibinfo {author} {\bibfnamefont
  {Y.}~\bibnamefont {Han}}, \bibinfo {author} {\bibfnamefont {H.}~\bibnamefont
  {Sun}},\ and\ \bibinfo {author} {\bibfnamefont {M.}~\bibnamefont {Wang}},\
  }\bibfield  {title} {\bibinfo {title} {Evolution of superconductivity
  evidence in pressurized
  ${\mathrm{la}}_{3\ensuremath{-}x}{\mathrm{sm}}_{x}{\mathrm{ni}}_{2}{\mathrm{o}}_{7}$},\
  }\href {https://doi.org/10.1103/pzv7-1nlr} {\bibfield  {journal} {\bibinfo
  {journal} {Phys. Rev. B}\ }\textbf {\bibinfo {volume} {113}},\ \bibinfo
  {pages} {174512} (\bibinfo {year} {2026}{\natexlab{b}})}\BibitemShut
  {NoStop}%
\bibitem [{\citenamefont {Gao}\ \emph {et~al.}(2026)\citenamefont {Gao},
  \citenamefont {Zhou}, \citenamefont {Guo}, \citenamefont {Xu}, \citenamefont
  {Chen}, \citenamefont {Han}, \citenamefont {Xu}, \citenamefont {Wu},\ and\
  \citenamefont {Qian}}]{gao2026}%
  \BibitemOpen
  \bibfield  {author} {\bibinfo {author} {\bibfnamefont {Y.}~\bibnamefont
  {Gao}}, \bibinfo {author} {\bibfnamefont {W.}~\bibnamefont {Zhou}}, \bibinfo
  {author} {\bibfnamefont {W.~H.}\ \bibnamefont {Guo}}, \bibinfo {author}
  {\bibfnamefont {C.}~\bibnamefont {Xu}}, \bibinfo {author} {\bibfnamefont
  {H.~F.}\ \bibnamefont {Chen}}, \bibinfo {author} {\bibfnamefont {Z.~D.}\
  \bibnamefont {Han}}, \bibinfo {author} {\bibfnamefont {X.}~\bibnamefont
  {Xu}}, \bibinfo {author} {\bibfnamefont {Y.}~\bibnamefont {Wu}},\ and\
  \bibinfo {author} {\bibfnamefont {B.}~\bibnamefont {Qian}},\ }\bibfield
  {title} {\bibinfo {title} {{Enhancement of metallicity by Na doping in
  ${\mathrm{La}}_{3}{\mathrm{Ni}}_{2}{\mathrm{O}}_{7+\ensuremath{\delta}}$}},\
  }\href {https://doi.org/10.1103/fcwr-3jrg} {\bibfield  {journal} {\bibinfo
  {journal} {Phys. Rev. B}\ }\textbf {\bibinfo {volume} {114}},\ \bibinfo
  {pages} {L020504} (\bibinfo {year} {2026})}\BibitemShut {NoStop}%
\bibitem [{\citenamefont {Mo}\ and\ \citenamefont {Wú}(2026)}]{shi26}%
  \BibitemOpen
  \bibfield  {author} {\bibinfo {author} {\bibfnamefont {S.-C.}\ \bibnamefont
  {Mo}}\ and\ \bibinfo {author} {\bibfnamefont {W.}~\bibnamefont {Wú}},\
  }\href {https://arxiv.org/abs/2605.30297} {\bibinfo {title} {Electron doping
  of $\mathrm{La_3Ni_2O_7}$ thin films: Candidate metal dopants and their
  potential impact on superconductivity}} (\bibinfo {year} {2026}),\ \Eprint
  {https://arxiv.org/abs/2605.30297} {arXiv:2605.30297 [cond-mat.supr-con]}
  \BibitemShut {NoStop}%
\bibitem [{\citenamefont {Fan}\ \emph {et~al.}(2026)\citenamefont {Fan},
  \citenamefont {Ma}, \citenamefont {Chesi}, \citenamefont {Wu},\ and\
  \citenamefont {Ma}}]{fan2026}%
  \BibitemOpen
  \bibfield  {author} {\bibinfo {author} {\bibfnamefont {Z.}~\bibnamefont
  {Fan}}, \bibinfo {author} {\bibfnamefont {R.}~\bibnamefont {Ma}}, \bibinfo
  {author} {\bibfnamefont {S.}~\bibnamefont {Chesi}}, \bibinfo {author}
  {\bibfnamefont {C.}~\bibnamefont {Wu}},\ and\ \bibinfo {author}
  {\bibfnamefont {T.}~\bibnamefont {Ma}},\ }\bibfield  {title} {\bibinfo
  {title} {Superconductivity with macroscopic time-reversal symmetry breaking
  in the presence of loop-current fluctuations},\ }\href
  {https://doi.org/10.1088/0256-307X/43/8/080702} {\bibfield  {journal}
  {\bibinfo  {journal} {Chinese Physics Letters}\ }\textbf {\bibinfo {volume}
  {43}},\ \bibinfo {pages} {080702} (\bibinfo {year} {2026})}\BibitemShut
  {NoStop}%
\bibitem [{\citenamefont {Yamase}\ \emph {et~al.}(2021)\citenamefont {Yamase},
  \citenamefont {Bejas},\ and\ \citenamefont {Greco}}]{yamase21}%
  \BibitemOpen
  \bibfield  {author} {\bibinfo {author} {\bibfnamefont {H.}~\bibnamefont
  {Yamase}}, \bibinfo {author} {\bibfnamefont {M.}~\bibnamefont {Bejas}},\ and\
  \bibinfo {author} {\bibfnamefont {A.}~\bibnamefont {Greco}},\ }\bibfield
  {title} {\bibinfo {title} {{Electron self-energy from quantum charge
  fluctuations in the layered $t\ensuremath{-}J$ model with long-range Coulomb
  interaction}},\ }\href {https://doi.org/10.1103/PhysRevB.104.045141}
  {\bibfield  {journal} {\bibinfo  {journal} {Phys. Rev. B}\ }\textbf {\bibinfo
  {volume} {104}},\ \bibinfo {pages} {045141} (\bibinfo {year}
  {2021})}\BibitemShut {NoStop}%
\bibitem [{\citenamefont {Merino}\ \emph {et~al.}(2006)\citenamefont {Merino},
  \citenamefont {Greco}, \citenamefont {Drichko},\ and\ \citenamefont
  {Dressel}}]{merino06}%
  \BibitemOpen
  \bibfield  {author} {\bibinfo {author} {\bibfnamefont {J.}~\bibnamefont
  {Merino}}, \bibinfo {author} {\bibfnamefont {A.}~\bibnamefont {Greco}},
  \bibinfo {author} {\bibfnamefont {N.}~\bibnamefont {Drichko}},\ and\ \bibinfo
  {author} {\bibfnamefont {M.}~\bibnamefont {Dressel}},\ }\bibfield  {title}
  {\bibinfo {title} {Non-fermi liquid behavior in nearly charge ordered layered
  metals},\ }\href {https://doi.org/10.1103/PhysRevLett.96.216402} {\bibfield
  {journal} {\bibinfo  {journal} {Phys. Rev. Lett.}\ }\textbf {\bibinfo
  {volume} {96}},\ \bibinfo {pages} {216402} (\bibinfo {year}
  {2006})}\BibitemShut {NoStop}%
\bibitem [{\citenamefont {Hoang}\ and\ \citenamefont
  {Thalmeier}(2002)}]{hoang02}%
  \BibitemOpen
  \bibfield  {author} {\bibinfo {author} {\bibfnamefont {A.~T.}\ \bibnamefont
  {Hoang}}\ and\ \bibinfo {author} {\bibfnamefont {P.}~\bibnamefont
  {Thalmeier}},\ }\bibfield  {title} {\bibinfo {title} {Coherent potential
  approximation for charge ordering in the extended hubbard model},\ }\href
  {https://doi.org/10.1088/0953-8984/14/26/304} {\bibfield  {journal} {\bibinfo
   {journal} {Journal of Physics: Condensed Matter}\ }\textbf {\bibinfo
  {volume} {14}},\ \bibinfo {pages} {6639–6646} (\bibinfo {year}
  {2002})}\BibitemShut {NoStop}%
\bibitem [{\citenamefont {Koch}\ and\ \citenamefont {Zeyher}(2004)}]{koch04}%
  \BibitemOpen
  \bibfield  {author} {\bibinfo {author} {\bibfnamefont {E.}~\bibnamefont
  {Koch}}\ and\ \bibinfo {author} {\bibfnamefont {R.}~\bibnamefont {Zeyher}},\
  }\bibfield  {title} {\bibinfo {title} {Renormalization of the electron-phonon
  coupling in the one-band hubbard model},\ }\href
  {https://doi.org/10.1103/PhysRevB.70.094510} {\bibfield  {journal} {\bibinfo
  {journal} {Phys. Rev. B}\ }\textbf {\bibinfo {volume} {70}},\ \bibinfo
  {pages} {094510} (\bibinfo {year} {2004})}\BibitemShut {NoStop}%
\bibitem [{\citenamefont {Bejas}\ \emph {et~al.}(2008)\citenamefont {Bejas},
  \citenamefont {Greco}, \citenamefont {Muramatsu},\ and\ \citenamefont
  {Foussats}}]{bejas08}%
  \BibitemOpen
  \bibfield  {author} {\bibinfo {author} {\bibfnamefont {M.}~\bibnamefont
  {Bejas}}, \bibinfo {author} {\bibfnamefont {A.}~\bibnamefont {Greco}},
  \bibinfo {author} {\bibfnamefont {A.}~\bibnamefont {Muramatsu}},\ and\
  \bibinfo {author} {\bibfnamefont {A.}~\bibnamefont {Foussats}},\ }\bibfield
  {title} {\bibinfo {title} {{One-particle spectral properties of the
  $t\text{\ensuremath{-}}J\text{\ensuremath{-}}V$ model on the triangular
  lattice near charge order}},\ }\href
  {https://doi.org/10.1103/PhysRevB.77.075131} {\bibfield  {journal} {\bibinfo
  {journal} {Phys. Rev. B}\ }\textbf {\bibinfo {volume} {77}},\ \bibinfo
  {pages} {075131} (\bibinfo {year} {2008})}\BibitemShut {NoStop}%
\bibitem [{\citenamefont {Hsu}\ \emph {et~al.}(1991)\citenamefont {Hsu},
  \citenamefont {Marston},\ and\ \citenamefont {Affleck}}]{hsu91}%
  \BibitemOpen
  \bibfield  {author} {\bibinfo {author} {\bibfnamefont {T.~C.}\ \bibnamefont
  {Hsu}}, \bibinfo {author} {\bibfnamefont {J.~B.}\ \bibnamefont {Marston}},\
  and\ \bibinfo {author} {\bibfnamefont {I.}~\bibnamefont {Affleck}},\
  }\bibfield  {title} {\bibinfo {title} {{Two observable features of the
  staggered-flux phase at nonzero doping}},\ }\href
  {https://doi.org/10.1103/PhysRevB.43.2866} {\bibfield  {journal} {\bibinfo
  {journal} {Phys. Rev. B}\ }\textbf {\bibinfo {volume} {43}},\ \bibinfo
  {pages} {2866} (\bibinfo {year} {1991})}\BibitemShut {NoStop}%
\bibitem [{\citenamefont {Hu}(1994)}]{hu94}%
  \BibitemOpen
  \bibfield  {author} {\bibinfo {author} {\bibfnamefont {C.-R.}\ \bibnamefont
  {Hu}},\ }\bibfield  {title} {\bibinfo {title} {Midgap surface states as a
  novel signature for $d_{x_a^2-x_b^2}$-wave superconductivity},\ }\href
  {https://doi.org/10.1103/PhysRevLett.72.1526} {\bibfield  {journal} {\bibinfo
   {journal} {Phys. Rev. Lett.}\ }\textbf {\bibinfo {volume} {72}},\ \bibinfo
  {pages} {1526} (\bibinfo {year} {1994})}\BibitemShut {NoStop}%
\bibitem [{\citenamefont {Tanaka}\ and\ \citenamefont
  {Kashiwaya}(1995)}]{tanaka95}%
  \BibitemOpen
  \bibfield  {author} {\bibinfo {author} {\bibfnamefont {Y.}~\bibnamefont
  {Tanaka}}\ and\ \bibinfo {author} {\bibfnamefont {S.}~\bibnamefont
  {Kashiwaya}},\ }\bibfield  {title} {\bibinfo {title} {Theory of tunneling
  spectroscopy of $d$-wave superconductors},\ }\href
  {https://doi.org/10.1103/PhysRevLett.74.3451} {\bibfield  {journal} {\bibinfo
   {journal} {Phys. Rev. Lett.}\ }\textbf {\bibinfo {volume} {74}},\ \bibinfo
  {pages} {3451} (\bibinfo {year} {1995})}\BibitemShut {NoStop}%
\bibitem [{\citenamefont {Kashiwaya}\ and\ \citenamefont
  {Tanaka}(2000)}]{kashiwaya00}%
  \BibitemOpen
  \bibfield  {author} {\bibinfo {author} {\bibfnamefont {S.}~\bibnamefont
  {Kashiwaya}}\ and\ \bibinfo {author} {\bibfnamefont {Y.}~\bibnamefont
  {Tanaka}},\ }\bibfield  {title} {\bibinfo {title} {Tunnelling effects on
  surface bound states in unconventional superconductors},\ }\href
  {https://doi.org/10.1088/0034-4885/63/10/202} {\bibfield  {journal} {\bibinfo
   {journal} {Rep. Prog. Phys.}\ }\textbf {\bibinfo {volume} {63}},\ \bibinfo
  {pages} {1641} (\bibinfo {year} {2000})}\BibitemShut {NoStop}%
\bibitem [{\citenamefont {Kuboki}(2014)}]{kuboki14}%
  \BibitemOpen
  \bibfield  {author} {\bibinfo {author} {\bibfnamefont {K.}~\bibnamefont
  {Kuboki}},\ }\bibfield  {title} {\bibinfo {title} {{Flux Phase as Possible
  Time-Reversal Symmetry Breaking Surface States of High-Tc Cuprate
  Superconductors}},\ }\href {https://doi.org/10.7566/JPSJ.83.054703}
  {\bibfield  {journal} {\bibinfo  {journal} {Journal of the Physical Society
  of Japan}\ }\textbf {\bibinfo {volume} {83}},\ \bibinfo {pages} {054703}
  (\bibinfo {year} {2014})}\BibitemShut {NoStop}%
\bibitem [{\citenamefont {Kuboki}(2020)}]{kuboki20}%
  \BibitemOpen
  \bibfield  {author} {\bibinfo {author} {\bibfnamefont {K.}~\bibnamefont
  {Kuboki}},\ }\bibfield  {title} {\bibinfo {title} {{Spontaneous Magnetic
  Field and Local Density of States near a Time-Reversal Symmetry Broken
  Surface State of YBCO}},\ }\href {https://doi.org/10.7566/JPSJ.89.073703}
  {\bibfield  {journal} {\bibinfo  {journal} {Journal of the Physical Society
  of Japan}\ }\textbf {\bibinfo {volume} {89}},\ \bibinfo {pages} {073703}
  (\bibinfo {year} {2020})}\BibitemShut {NoStop}%
\end{thebibliography}%

\clearpage 

\onecolumngrid 


\begin{center}
  \textbf{\large Supplemental Material for\\
``Competing Interlayer Loop Currents and Superconductivity \\ in the Bilayer $t$-$J_\perp$-$V$ Model''}\\[.2cm]

\hspace{-.2cm}Luciano Zinni,$^{1}$
Fabricio G\'omez,$^{1}$
Jun Zhan,$^{2}$
Mat\'{\i}as Bejas,$^{3}$
Xianxin Wu,$^{4}$
Andreas P. Schnyder,$^{5}$
and Andr\'es Greco$^{3}$\\[.2cm]

{\small \itshape
$^1$Facultad de Ciencias Exactas, Ingenier\'{\i}a y Agrimensura. Avenida Pellegrini 250, 2000 Rosario, Argentina\\[.1cm]
$^2$Department of Physics, Nagoya University, Nagoya 464-8602, Japan\\
$^3$Facultad de Ciencias Exactas, Ingenier\'{\i}a y Agrimensura and Instituto de F\'{\i}sica Rosario (UNR-CONICET),\\Avenida Pellegrini 250, 2000 Rosario, Argentina\\
$^{4}$Institute of Theoretical Physics, Chinese Academy of Sciences, Beijing, China\\
$^{5}$Max Planck Institute for Solid State Research, Heisenbergstrasse 1, 70569 Stuttgart, Germany
}
\end{center}


%
%

\vspace{0.6cm}

\setcounter{equation}{0}
\setcounter{figure}{0}
\setcounter{table}{0}
\setcounter{page}{1}
\setcounter{section}{0}

\renewcommand{\theequation}{S\arabic{equation}}
\renewcommand{\thefigure}{S\arabic{figure}}
\renewcommand{\thetable}{S\arabic{table}}
\renewcommand{\thesection}{S\arabic{section}}
\renewcommand{\thepage}{S\arabic{page}} 

\setcounter{secnumdepth}{3}
\renewcommand{\thesection}{S\arabic{section}}


\title{Supplemental Material for\\
``Competing Interlayer Loop Currents and Superconductivity \\ in the Bilayer $t$-$J_\perp$-$V$ Model''}

\section{Large-$N$ formalism for the bilayer $t$-$\Jperp$-$V$ model}
\label{sec:formalism}


The strongly correlated bilayer $t$-$J$-$\Jperp$ model, in which the $d_{x^2-y^2}$ orbitals are the only active degrees of freedom was originaly presentend in Refs.~\cite{lu24,qu24}. 
Here we focus on the relevant out-of-plane spin exchange interaction $J_\perp$, i.e., we did not consider the in-plane exchange interaction $J$. The Hamiltonian reads
\begin{align}
H_{t\text{-}J_\perp} =& \sum_{i,j,\sigma,\alpha} t_{ij}\,
\tilde{c}^{\dagger}_{i\sigma,\alpha}\tilde{c}_{j\sigma,\alpha}+ t_{\perp}\!\sum_{i,\sigma,\alpha}\!
\tilde{c}^{\dagger}_{i\sigma,\alpha}\tilde{c}_{i\sigma,\bar{\alpha}}
+\frac{\Jperp}{2}\!\sum_{i,\alpha}\!
\Big(\vec{S}_{i,\alpha}\cdot\vec{S}_{i,\bar{\alpha}}
-\tfrac{1}{4}n_{i,\alpha}n_{i,\bar{\alpha}}\Big)
-\mu\sum_{i,\alpha} n_{i,\alpha}\,.
\label{eq:Hfull}
\end{align}
The index $\alpha = 1,2$ labels the two planes and $\bar{\alpha}$ denotes the plane opposite to $\alpha$, while $i$ and $j$ run over the sites of the square lattice of each plane.
The hopping $t_{ij}$ takes a value $t$ between first- and $t'$ between second-nearest neighbors on each plane, $\Jperp$ is the exchange interaction between sites at different planes, the operators $\tilde{c}^{\dagger}_{i\sigma,\alpha}$ ($\tilde{c}_{i\sigma,\alpha}$) create (annihilate) electrons in the restricted Fock space with no double occupancy, $n_{i,\alpha}$, $\vec{S}_{i,\alpha}$, and $\mu$ are the density, spin operators, and the chemical potential, respectively.

To the Hamiltonian (\ref{eq:Hfull}) we will add the nearest-neighbor repulsion
inside each plane, and the repulsion between sites in different planes for the
reasons detailed in Sec.~\ref{sec:6x6}, which read
\begin{equation}
  H_V =
  V_{\parallel}\!\!\sum_{\langle i,j\rangle,\alpha}\! n_{i,\alpha}n_{j,\alpha}
+ V_{\perp}\sum_{i} n_{i,1}n_{i,2}\,,
\label{eq:HV}
\end{equation}
\noindent where $\langle i,j \rangle$ indicates a nearest-neighbor pair of sites. Thus, $H=H_{t\text{-}J_\perp}+H_V$ defines the Hamiltonian of the $t$-$J_\perp$-$V$ model. 

Two obstacles stand in the way of a direct treatment of Eq.~(\ref{eq:Hfull}):
the projected operators $\tilde{c}$ do not obey canonical anticommutation relations, and there is no small parameter to expand in.
We remove both at once by using the large-$N$ scheme built on the path-integral representation of the Hubbard operators~\cite{foussats04,bejas12,yamase21}, in the bilayer version
formulated in Ref.~\cite{bejas25}.
The projected operators are identified with Hubbard operators~\cite{hubbard63}, $\tilde{c}^{\dagger}_{i\sigma,\alpha}=\hat{X}^{\sigma0}_{i,\alpha}$ and $\tilde{c}_{i\sigma,\alpha}=\hat{X}^{0\sigma}_{i,\alpha}$, which are fermionlike; the spin and density operators are bilinear in the bosonlike $\hat{X}^{\sigma\sigma'}_{i,\alpha}$, and the remaining bosonlike operator $\hat{X}^{00}_{i,\alpha}$ counts the doped holes.
The constraint of no double occupancy is thereby carried by the algebra of the $\hat{X}$'s instead of being imposed on top of a wider Hilbert space, and no auxiliary particles, gauge fixing, or holon condensation are needed.
Inside the path integral the fermionlike and bosonlike operators become Grassmann and ordinary complex variables, respectively, and from now on we drop the hat symbol.

The large-$N$ expansion for the $t$-$J$-$\Jperp$ model was extensively discussed in Ref.~\cite{bejas25} and here, for clarity, we focus on the particular case for the $t$-$\Jperp$-$V$ model. 
The expansion parameter is obtained by letting the spin index run over $p = 1,\dots,N$ rather than over $\sigma = \uparrow,\downarrow$, with the couplings rescaled as $t/N$, $t'/N$, $t_{\perp}/N$, $\Jperp/N$, $V_{\parallel}/N$, and $V_{\perp}/N$ so that all terms of Eq.~(\ref{eq:Hfull}) survive at the same order when $N\to\infty$.
One of the advantages of this method is that it treats all possible
charge excitations on an equal footing.

The exchange term is bilinear in the bosonlike variables $X^{pp'}_{i,\alpha}$, which obey
\begin{equation}
X^{pp'}_{i,\alpha}=\frac{X^{p0}_{i,\alpha}X^{0p'}_{i,\alpha}}{X^{00}_{i,\alpha}}\, ,
\label{eq:Xrel}
\end{equation}
so that it can be written entirely in terms of the fermionlike variables. It is
convenient to rescale the latter as
\begin{equation}
f^{\dagger}_{ip,\alpha}=\frac{1}{\sqrt{N\dop/2}}\,X^{p0}_{i,\alpha},
\qquad
f_{ip,\alpha}=\frac{1}{\sqrt{N\dop/2}}\,X^{0p}_{i,\alpha},
\label{eq:ffields}
\end{equation}
with $\dop$ being the hole doping measured from half filling.
The bosonlike variable $X^{00}_{i,\alpha}$ and the Lagrange multiplier $\lambda_{i,\alpha}$ that enforces the constraint $X^{00}_{i,\alpha}+\sum_{p}X^{pp}_{i,\alpha} = N/2$ separately in each plane are split into a static value and a fluctuation,
\begin{equation}
X^{00}_{i,\alpha} = N\frac{\dop}{2}\bigl(1+\delta R_{i,\alpha}\bigr),
\qquad
\lambda_{i,\alpha}=\lambda_{0}+\delta\lambda_{i,\alpha},
\label{eq:RandLambda}
\end{equation}
so that $\delta R_{i,\alpha}$ denotes the fluctuation of the hole density at site $i$ in plane $\alpha$, and $\lambda_{0}$ can be reabsorbed into $\mu$.

It is worth stressing at this point that the densities entering the
Coulomb terms are not independent variables: the same constraint gives $n_{i,\alpha} = \sum_{p}X^{pp}_{i,\alpha}=N/2-X^{00}_{i,\alpha}$, so that $V_{\parallel}$ and $V_{\perp}$ can only act on the fields $\delta R_{i,\alpha}$.
This is the microscopic reason why the Coulomb repulsions will be confined to the on-site charge sector of the theory.

Once Eqs.~(\ref{eq:Xrel}) and (\ref{eq:ffields}) are inserted, the exchange term becomes quartic in the fermion variables.
This is linearized by Hubbard-Stratonovich transformations, which
introduce the bond fields
\begin{equation}
\Delta'_{i}=\frac{\Jperp}{2N}\sum_{p}
\frac{f^{\dagger}_{ip,1}f_{ip,2}}
{\sqrt{(1+\delta R_{i,1})(1+\delta R_{i,2})}}\,.
\label{eq:HSfields}
\end{equation}
%
Expanding the fields around their common static value gives
\begin{equation}
\Delta'_{i}=\chi'\bigl(1+r_{\perp,i}+\ii A_{\perp,i}\bigr),
\label{eq:param}
\end{equation}
where $\chi'$ is the static amplitude, taken real, and the fluctuations have been decomposed into their real part $r_{\perp,i}$ and their imaginary part $A_{\perp,i}$.
The last ingredient is the expansion of the residual factors
$(1+\delta R_{i,\alpha})^{-1/2}$ in powers of $\delta R_{i,\alpha}$: every extra power adds one fermion-boson interaction and one power of $1/N$, so the series is truncated at the order demanded by the quantity under study.
What remains is an effective theory of fermions, bosons, and the vertices linking them.\\

{\bf Fermionic propagator and mean-field equations}\\

To leading order the fermions of the spinor $(f_{\kvec p,1},f_{\kvec p,2})$ propagate according to the $2\times2$ Green's function, which is $O(1)$,
\begin{equation}
G^{(0)}_{\alpha\beta}(\kvec,\ii\nu_n)=
\begin{pmatrix}
\ii\nu_n-\varepsilon^{\parallel}_{\kvec} & -\varepsilon_{\perp}\\[2pt]
-\varepsilon_{\perp} & \ii\nu_n-\varepsilon^{\parallel}_{\kvec}
\end{pmatrix}^{-1},
\label{eq:G0}
\end{equation}
whose diagonal and off-diagonal entries are governed by
\begin{equation}
\varepsilon^{\parallel}_{\kvec}
=-2t\frac{\dop}{2}\bigl(\cos k_x+\cos k_y\bigr)
-4t'\frac{\dop}{2}\cos k_x\cos k_y - \mu,
\qquad
\varepsilon_{\perp}=t_{\perp}\frac{\dop}{2}-\chi',
\label{eq:dispersion}
\end{equation}
with $\nu_n$ a fermionic Matsubara frequency, and $\kvec = (k_x, k_y)$ an in-plane momentum.
Note how the bare hoppings are renormalized by correlations.
Diagonalizing Eq.~(\ref{eq:G0}) yields the bonding and antibonding bands
\begin{equation}
\varepsilon^{\pm}_{\kvec} = \varepsilon^{\parallel}_{\kvec} \pm \varepsilon_{\perp}\,,
\label{eq:bands}
\end{equation}
which are the only fermionic input needed later on.
At each doping $\dop$ the quantities $\mu$, and $\chi'$ follow from the coupled saddle-point equations
\begin{equation}
\chi'=\frac{\Jperp}{2N_s}\sum_{\kvec,\ii\nu_n}G^{(0)}_{12}(\kvec,\ii\nu_n),
\qquad
1-\dop=\frac{2}{N_s}\sum_{\kvec,\ii\nu_n}G^{(0)}_{11}(\kvec,\ii\nu_n),
\label{eq:selfcons}
\end{equation}
where $N_s$ is the number of sites of each plane and $G^{(0)}_{11}$,
$G^{(0)}_{12}$ are the corresponding elements of Eq.~(\ref{eq:G0}). \\

{\bf The $6\times6$ bare bosonic propagator}\\

Gathering all the fluctuating fields introduced in Eqs.~(\ref{eq:RandLambda}) and (\ref{eq:param}), the bosonic sector of the bilayer is spanned by six components,
\begin{equation}
\delta X^{a} = \bigl(\delta R_1,\;\delta\lambda_1,\;
\delta R_2,\;\delta\lambda_2,\;
r_{\perp},\;A_{\perp}\bigr),
\label{eq:boson14}
\end{equation}
the site index being omitted for clarity.
The first four entries are the on-site charge fluctuations of each plane together with the multipliers that keep the no-double-occupancy constraint,
and the last two are the vertical bond fluctuations produced by
$\Jperp$.
Applying the Feynman rules of the effective theory~\cite{foussats04,bejas12,bejas25} one obtains a $6\times6$ bare bosonic propagator $D^{(0)}_{ab}(\qvec,\ii\omega_n)$ of order $1/N$, which in the basis of Eq.~(\ref{eq:boson14}) is real, symmetric, and block structured,
\begin{equation}
\bigl[D^{(0)}_{ab}(\qvec,\ii\omega_{n})\bigr]^{-1}=
N\begin{pmatrix}
D^{(0)}_{A} & D^{(0)}_{B} & 0\\
D^{(0)}_{B} & D^{(0)}_{A} & 0\\
0 & 0 & D^{(0)}_{D}
\end{pmatrix},
\label{eq:D014}
\end{equation}
$\omega_n$ being a bosonic Matsubara frequency.
The first two block rows and columns refer to $(\delta R_{\alpha},\delta\lambda_{\alpha})$ of each plane, and the last one to the vertical bonds.
The blocks read
\begin{equation}
\setlength{\arraycolsep}{3pt}
\begin{gathered}
D^{(0)}_{A}=
\begin{pmatrix}
\dfrac{\dop^{2}}{2}\dfrac{V(\qvec)}{2} & \dfrac{\dop}{2}\\[10pt]
\dfrac{\dop}{2} & 0
\end{pmatrix},
\qquad
D^{(0)}_{B}=
\begin{pmatrix}
\dfrac{\dop^{2}}{2}\Bigl[\dfrac{V'(\qvec)}{2}-J'(\qvec)\Bigr] & 0\\[10pt]
0 & 0
\end{pmatrix},
\qquad
D^{(0)}_{D}=
\begin{pmatrix}
\dfrac{4\chi'^{2}}{\Jperp} & 0\\[8pt]
0 & \dfrac{4\chi'^{2}}{\Jperp}
\end{pmatrix},
\end{gathered}
\label{eq:D0blocks}
\end{equation}
with $V(\qvec) = 2V_{\parallel}\bigl(\cos q_x+\cos q_y\bigr)$,
$V'(\qvec) = V_{\perp}$, $J'(\qvec)=\frac{\Jperp}{4}$, and
$\qvec = (q_x,q_y)$ is an in-plane momentum.

The interactions enter Eq.~(\ref{eq:D0blocks}) only through the
$\delta R_{\alpha}$-$\delta R_{\beta}$ entries, and always in the combination
$V'(\qvec)/2-J'(\qvec)$ between the planes. The exchange therefore enters the
charge channel with the opposite sign to the Coulomb repulsion, i.e., as an
effective attraction which originates in the correlations and not in the Coulomb
terms; it is this attraction that drives phase separation in the $t$-$J$ model at
low doping~\cite{bejas12}, while $V_{\parallel}$ and $V_{\perp}$ act against it.
The bond block $D^{(0)}_{D}$ is instead diagonal and free of any Coulomb
contribution: it is fixed by the ratio between the square of the bond amplitude
and the exchange, $4\chi'^{2}/\Jperp$.\\

{\bf Vertices}\\

Fermions and bosons interact through vertices with three and four legs.
The three-legs vertex $\Lambda_{\alpha\beta,a}$ transports a fermion from plane $\alpha$ to plane $\beta$ with one boson $\delta X^{a}$ attached; its nonzero components are
\begin{align}
\Lambda_{\alpha\alpha,a}(\kvec,\qvec)=-\Bigl[&
\frac{\ii\nu_n+\ii\nu'_n}{2}+\mu,\;1\;\Bigr]
\label{eq:threeleg1}
\end{align}
when the boson lives in the same plane as the fermion,
$a = \delta R_{\alpha},\,\delta\lambda_{\alpha}$, and
\begin{equation}
\Lambda_{1 2,a}(\kvec,\qvec)=
-\Bigl(\frac{\chi'}{2},\,\frac{\chi'}{2},\,-\chi',\,-\ii\chi'\Bigr),
\qquad
\Lambda_{2 1,a}(\kvec,\qvec)=
-\Bigl(\frac{\chi'}{2},\,\frac{\chi'}{2},\,-\chi',\,+\ii\chi'\Bigr),
\label{eq:threeleg2}
\end{equation}
when the fermion changes plane, $a = \delta R_1,\,\delta R_2,\,r_{\perp},\,A_{\perp}$.
Note that $\Lambda_{1 2,a}$ involves the charge fields $\delta R_{\alpha}$ together with the vertical bond fields $r_{\perp}$ and $A_{\perp}$.
The two sectors are thus coupled through the fermion loop, a point that becomes relevant in Sec.~\ref{sec:6x6}, where we examine to what extent the interlayer bond-order parameter ($\zBOP$) can be treated separately from the charge excitations.
The component associated with $A_{\perp}$ is the only one that
reverses its sign when the planes are interchanged.

The four-legs vertex $\Lambda_{\alpha\beta,ab}$ describes a fermion propagating from plane $\alpha$ to plane $\beta$ while interacting with two bosons $\delta X^{a}$ and $\delta X^{b}$.
Its nonzero components are
\begin{equation}
\Lambda_{\alpha\alpha,ab}=
\begin{pmatrix}
F_{\qvec\qvec'} & 1/2\\[2pt]
1/2 & 0
\end{pmatrix},
\qquad
F_{\qvec\qvec'}=\frac{\ii\nu_n+\ii\nu'_n}{2}+\mu
\label{eq:fourleg1}
\end{equation}
for $a,b = \delta R_{\alpha},\,\delta\lambda_{\alpha}$, and
\begin{equation}
\Lambda_{1 2,ab}=-\frac{\chi'}{4}
\begin{pmatrix}
-3/2 & -1/2 & 1 & \ii\\
-1/2 & -3/2 & 1 & \ii\\
1 & 1 & 0 & 0\\
\ii & \ii & 0 & 0
\end{pmatrix}
\label{eq:fourleg3}
\end{equation}
for $a,b = \delta R_1,\,\delta R_2,\,r_{\perp},\,A_{\perp}$; the entries of $\Lambda_{2 1,ab}$ that involve $A_{\perp}$ (last row and last column) carry the opposite sign, again by the argument given above.
All vertices are $O(1)$.\\

{\bf Bosonic self-energy and Dyson equation}\\

At the order we work, two diagrams contribute to the bosonic self-energy $\Pi_{ab}(\qvec,\ii\omega_n)$: a particle-hole bubble closed with two three-legs vertices, and a tadpole closed with a single four-legs vertex,
\begin{equation}
\Pi_{ab}(\qvec,\ii\omega_{n})=
\frac{N}{N_s}\sum_{k}\sum_{\alpha\beta\gamma\rho}
\Lambda_{\alpha\beta,a}\,G^{(0)}_{\beta\gamma}(k)\,
\Lambda_{\gamma\rho,b}\,G^{(0)}_{\rho\alpha}(k-q)
+\frac{N}{N_s}\sum_{k}\sum_{\alpha\beta}
\Lambda_{\alpha\beta,ab}\,G^{(0)}_{\beta\alpha}(k),
\label{eq:Pi}
\end{equation}
where we abbreviate $k\equiv(\kvec,\ii\nu_n)$, $q\equiv(\qvec,\ii\omega_n)$, and $\sum_{k}\equiv T\sum_{\nu_n}\sum_{\kvec}$.
The explicit $N$ counts the spin components circulating around the closed fermion line, so that $\Pi_{ab}$ is of order $O(N)$, the same order as $[D^{(0)}_{ab}]^{-1}$.
Carrying out the Matsubara sums, every component of $\Pi_{ab}$ reduces to the elementary particle-hole bubbles
\begin{equation}
g^{\alpha\beta}(\qvec,\ii\omega_{n})=
\frac{n_{F}(\varepsilon^{\alpha}_{\kvec-\qvec})-n_{F}(\varepsilon^{\beta}_{\kvec})}
{\ii\omega_{n}+\varepsilon^{\alpha}_{\kvec-\qvec}-\varepsilon^{\beta}_{\kvec}},
\qquad \alpha,\beta=\pm,
\label{eq:gab}
\end{equation}
weighted by the vertex factors of the corresponding channel, with $n_{F}$ the Fermi function and $\varepsilon^{\pm}_{\kvec}$ the bands of Eq.~(\ref{eq:bands}).
Intraband combinations ($\alpha=\beta$) and interband ones
($\alpha\neq\beta$) enter with different weights, which is what will later distinguish the $\rperp$ from the $\Aperp$ channel.
The $\zBOP$ block of $\Pi_{ab}$ is written out explicitly in Ref.~\cite{bejas25}; the remaining components follow from Eqs.~(\ref{eq:threeleg1})-(\ref{eq:fourleg3}) by direct
evaluation and are not reproduced here.
The dressed propagator, which goes beyond the mean-field level, then follows from the Dyson equation
\begin{equation}
D^{-1}_{ab}(\qvec,\ii\omega_{n})=
\bigl[D^{(0)}_{ab}(\qvec,\ii\omega_{n})\bigr]^{-1}
-\Pi_{ab}(\qvec,\ii\omega_{n}).
\label{eq:Dyson}
\end{equation}
The power of $1/N$ carried by any computed quantity is obtained by counting the propagators and vertices that build it \cite{foussats04}.

\section{Decoupling between the charge and $\zBOP$ sectors}
\label{sec:6x6}

The bosonic propagator of Eq.~(\ref{eq:D014}) contains two sectors: a $4\times4$
block spanned by $a,b=1$-$4$, which describes the on-site charge excitations of the
two planes together with the constraint, and a $2\times2$ block spanned by
$a,b=5$-$6$, which describes the vertical bond excitations, i.e., the $\zBOP$. At the
bare level the two are block diagonal, and, as discussed in
Sec.~\ref{sec:formalism}, the Coulomb repulsions can only act on the fields
$\delta R_{i,\alpha}$ and are therefore confined to the charge block: $\Jperp$
alone fixes the stiffness $4\chi'^{2}/\Jperp$ of the $\zBOP$ block. The two sectors
communicate only through the off-diagonal components of the self-energy,
$\Pi_{\delta R_{\alpha}a}$ and $\Pi_{\delta\lambda_{\alpha}a}$ with
$a=\rperp,\Aperp$, generated by the interlayer vertex of
Eq.~(\ref{eq:threeleg2}) with an amplitude set by $\chi'$. In this section we first
fix $V_{\parallel}$ and $V_{\perp}$ so that the homogeneous state is stable, and
then show that this residual coupling is weak enough to be immaterial, so that the
$\zBOP$ is insensitive to the charge sector and, in particular, to the Coulomb
repulsions.

Instabilities of the homogeneous state follow from the condition
$\det D^{-1}_{ab}(\qvec,\omega=0)=0$. The values of $T$, $\dop$ and $\qvec$ at
which it is first satisfied define the corresponding critical line, and the
eigenvector of the vanishing eigenvalue identifies the sector in which the
instability takes place.\\

{\bf Charge instabilities and the choice of $V_{\parallel}$ and $V_{\perp}$}\\

\begin{figure}[ht]
\centering
\includegraphics[]{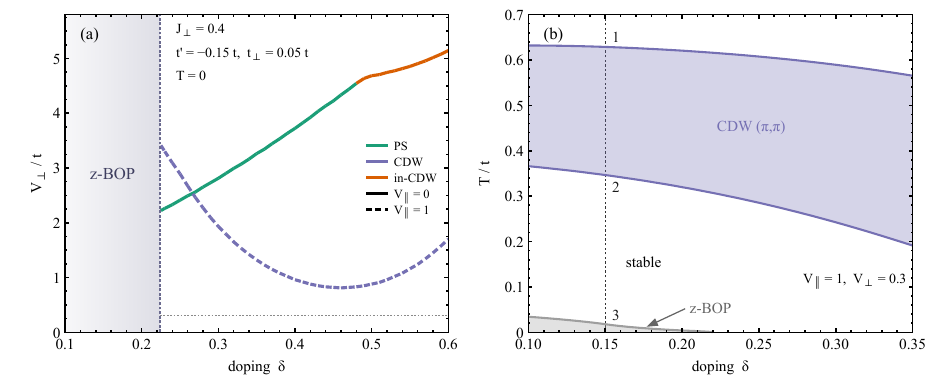}
\caption{(a) Critical interlayer Coulomb repulsion $V_{\perp}$ for the different
charge instabilities as a function of doping $\dop$, at $T=0$. Colors label the
instability---phase separation (PS), commensurate CDW (CDW), and incommensurate CDW
(in-CDW)---and the line style distinguishes the two values of $V_{\parallel}$
given in the legend; the homogeneous state is stable below all of them. The CDW
boundary has a minimum $V_{\perp}\simeq0.8t$ around $\dop\simeq0.45$, so that the
value of $V_{\perp}=0.3 t$ used in the main text (dotted horizontal line) lies below every instability
line at any doping. The shaded strip indicates the doping range where the $\zBOP$
develops, $\dop\lesssim0.22$, which is independent of $V_{\perp}$. (b) Stability
map of the homogeneous solution in the $T$-$\dop$ plane. The shaded area is the
re-entrant window in which the charge sector is unstable towards a CDW at
$\qvec=(\pi,\pi)$, bounded by the lines labeled $1$ and $2$. Outside
this window the homogeneous correlated metal is stable. The line labeled $3$ bounds
the region where the $\zBOP$ develops, which extends from there down to $T=0$.}
\label{fig:charge}
\end{figure}

The charge sector does become unstable, but only outside the region of interest.
Figure~\ref{fig:charge}(a) shows, at $T=0$, the critical interlayer repulsion at
which each charge instability sets in as a function of doping. The homogeneous
state is stable below all these lines, and the value of $V_{\perp}$ used in the
main text lies below them at every doping, so that no vestige of charge order is
left. The doping range where the $\zBOP$ develops appears as a vertical strip,
$\dop\lesssim0.22$, whose extent is independent of $V_{\perp}$.

Figure~\ref{fig:charge}(b) completes the picture in the $T$-$\dop$ plane. The charge-density-wave (CDW)
instability is re-entrant in temperature: the charge sector is unstable only inside
a window bounded by the two lines labeled $1$ and $2$, and the homogeneous state is
recovered both on cooling below $2$ and on heating above $1$. Such a re-entrant
temperature dependence is well known, and has been obtained in the Hubbard model
within the weak-coupling RPA~\cite{merino06}, the coherent-potential
approximation~\cite{hoang02}, and the Kotliar-Ruckenstein slave-boson
approach~\cite{koch04}, as well as in the $t$-$J$-$V$ model~\cite{bejas08}. In our
case the lower boundary lies above $T\simeq0.2t$, one order of magnitude above the
scale at which the $\zBOP$ and superconductivity develop, so that both phases
discussed in the main text emerge from a stable homogeneous correlated metal.

The eigenvectors identify the sector of each instability unambiguously. On both
boundaries of the CDW window, labeled $1$ and $2$ in Fig.~\ref{fig:charge}(b), the
eigenvector of the vanishing eigenvalue takes the form $(1,a,-1,-a,0,0)$ in the
basis of Eq.~(\ref{eq:boson14}), with $a$ between $0.11$ and $0.19$: the weight
lies entirely in the charge sector, with the two planes modulated out of phase, and
the vertical bond components vanish. That the multipliers $\delta\lambda_{\alpha}$
enter with a finite amplitude is expected, since they are coupled to
$\delta R_{\alpha}$ already at the bare level [Eq.~(\ref{eq:D0blocks})]. On the
boundary of the $\zBOP$ region, labeled $3$, the eigenvector is instead
$(0,0,0,0,0,1)$: the instability lives in the $\zBOP$ block and, within it, purely
in the imaginary component $\Aperp$.\\

{\bf The $\zBOP$ critical line}\\

Having fixed the Coulomb parameters, we now compare the $\zBOP$ instability line
obtained with and without the charge sector. Within the $2\times2$ block the
instability is signaled by the divergence of the corresponding susceptibility.
Projecting the $2\times2$ sector of $D_{ab}$
on the eigenvectors $(1,0)$ and $(0,1)$ gives the two
channels~\cite{bejas25},
\begin{equation}
\chi^{\rperp}(\qvec,\ii\omega_{n})=
\left[\frac{4\chi'^{2}}{\Jperp}
-\Pi_{\rperp\rperp}(\qvec,\ii\omega_{n})\right]^{-1},
\qquad
\chi^{\Aperp}(\qvec,\ii\omega_{n})=
\left[\frac{4\chi'^{2}}{\Jperp}
-\Pi_{\Aperp\Aperp}(\qvec,\ii\omega_{n})\right]^{-1},
\label{eq:chi_zbop}
\end{equation}
with
\begin{equation}
\Pi_{\rperp\rperp}=-\chi'^{2}\sum_{\kvec}
\bigl(g^{--}+g^{++}\bigr),
\qquad
\Pi_{\Aperp\Aperp}=-\chi'^{2}\sum_{\kvec}
\bigl(g^{-+}+g^{+-}\bigr),
\label{eq:Pi_zbop}
\end{equation}
written in terms of Eq.~(\ref{eq:gab}).
The two channels differ only in which combinations enter, intraband for $\rperp$
and interband for $\Aperp$, as anticipated in Sec.~\ref{sec:formalism}; the
critical temperature is the one at which the corresponding bracket in Eq. (\ref{eq:chi_zbop}) vanishes.

\begin{figure}[ht]
\centering
\includegraphics[]{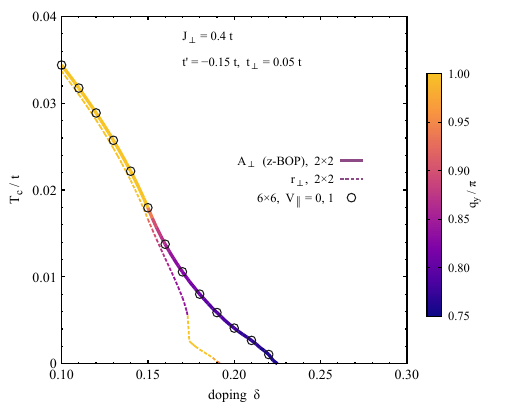}
\caption{Critical temperature of the $\zBOP$ instability as a function of doping
$\dop$. The solid (dashed) line follows from the divergence of the susceptibility
$\chi^{\Aperp}$ ($\chi^{\rperp}$) of Eq.~(\ref{eq:chi_zbop}); the color gives the
component $q_y$ of the ordering vector $\qc=(\pi,q_y)$ (right scale). Open circles
are obtained from $\det D^{-1}_{ab}=0$ for the full $6\times6$ propagator, for both
$V_{\parallel}=0$ and $V_{\parallel}=1.0t$. The three determinations coincide,
showing that neither the charge sector nor the Coulomb repulsion affects the
$\zBOP$. The ordering vector is incommensurate at the endpoint of the line at
$T=0$, $\dop\simeq0.22$, and moves towards $(\pi,\pi)$ with increasing
temperature.}
\label{fig:Tc_zBOP}
\end{figure}

Figure~\ref{fig:Tc_zBOP} shows the resulting critical temperature as a function of
doping, together with the one obtained from the condition
$\det D^{-1}_{ab}=0$ applied to the full $6\times6$ propagator, for
$V_{\parallel}=0$ and for $V_{\parallel}=1.0t$. The $6\times6$ calculation gives
exactly the same critical line as the $2\times2$ one, for both values of
$V_{\parallel}$: the three determinations fall on top of each other over the whole
doping range. Switching on the in-plane Coulomb repulsion, and letting the $\zBOP$
mix with the charge sector, leaves the critical line unchanged.

The same figure shows two further features. The color of the lines gives
the ordering vector along the critical line. It is incommensurate,
$\qc=(\pi,q_y)$ with $q_y\simeq0.76\pi$, at the QCP, and
moves towards $(\pi,\pi)$ with increasing temperature, becoming commensurate
already at $T=0.02t$, the temperature at which the gap equations of
Sec.~\ref{sec:gapeqs} are solved. In addition, the $\rperp$ and $\Aperp$ channels
are nearly degenerate at high temperature and separate as the critical line is
followed towards $T=0$, the $\Aperp$ one having the higher critical temperature at
any given doping and reaching $T=0$ at a larger doping, consistently with the
selection of the purely imaginary solution discussed in Sec.~\ref{sec:gapeqs}.\\

{\bf Dual structure}\\

Two consequences follow. First, as in the usual $t$-$J$ model in two
dimensions~\cite{bejas17}, the theory possesses a dual structure that separates the
on-site charge excitations from the bond sector, here the vertical bonds generated
by $\Jperp$; the $6\times6$ problem may be replaced by its $2\times2$ $\zBOP$ block
without any loss. Second, since $V_{\parallel}$ and $V_{\perp}$ enter only the
charge block, the Coulomb interaction is irrelevant for the formation and the
stability of the $\zBOP$ and the LCs.

\section{Gap equations at the commensurate ordering vector and self-selection of
the $\Aperp$ channel}
\label{sec:gapeqs}
\label{sec:selection}

At the commensurate $\qc=(\pi,\pi)$ the condensed $\zBOP$ is described within the
four-component spinor\cite{bejas25}
\begin{equation}
\psi^{\dagger}_{\kvec}
=\bigl(f^{\dagger}_{\kvec\sigma,1},\,
f^{\dagger}_{\kvec\sigma,2},\,
f^{\dagger}_{\kvec+\qc\,\sigma,1},\,
f^{\dagger}_{\kvec+\qc\,\sigma,2}\bigr),
\label{eq:spinor4}
\end{equation}
where the subscripts $1$ and $2$ label the layers. For $\Jperp/t=0.4$ the leading
$\zBOP$ instability is already commensurate at $T/t=0.02$, the temperature at
which the gap equations below are solved, so that the $(\pi,\pi)$ construction is
exact there. Since $\qc=(\pi,\pi)$ doubles the unit cell in the $xy$ plane, the
construction of Eq.~(\ref{eq:spinor4}) already pairs each $\kvec$ with its image
$\kvec+\qc$, and the momentum sums in the gap equations must be restricted to the
reduced Brillouin zone (RBZ), defined as the rhombus $|k_x|+|k_y|\leq\pi$
centered at the origin, in order to avoid double counting. In
Ref.~\cite{bejas25} the corresponding sums were carried out over the full
Brillouin zone; we have checked that both prescriptions give the same gap value,
and we adopt the RBZ as the formally correct summation domain.

Allowing the coupling to be a
general complex number, $\phi=\phi_{\rperp}+\ii\,\phi_{\Aperp}$, where $\phi_{\rperp}$ ($\phi_{\Aperp}$) is the order parameter associated with the interplane bond order $\langle f^{\dagger}_{\kvec\sigma,1}f_{\kvec+\qc\,\sigma,2}\rangle$ and
describes a modulation of the interlayer hopping magnitude (phase) with wave vector
$\qc$, the inverse Green's
function for each spin projection is 
\begin{equation}
G^{-1}(\kvec,\ii\nu_{n})=
\begin{pmatrix}
\ii\nu_{n}-\varepsilon^{\parallel}_{\kvec} & -\varepsilon_{\perp} & 0 & -\phi\\[2pt]
-\varepsilon_{\perp} & \ii\nu_{n}-\varepsilon^{\parallel}_{\kvec} & -\phi^{*} & 0\\[2pt]
0 & -\phi & \ii\nu_{n}-\varepsilon^{\parallel}_{\kvec+\qc} & -\varepsilon_{\perp}\\[2pt]
-\phi^{*} & 0 & -\varepsilon_{\perp} & \ii\nu_{n}-\varepsilon^{\parallel}_{\kvec+\qc}
\end{pmatrix}.
\label{eq:complexG}
\end{equation}

\noindent Since $\phi$ is complex, the gap equation splits into two coupled equations for $\phi_{\rperp}$ and $\phi_{\Aperp}$, so that neither channel is imposed from the outset. Following Ref.~\cite{bejas25}, the complex gap is obtained self-consistently from
\begin{equation}
\phi=-\frac{\Jperp}{2}\sum_{\kvec\in\mathrm{RBZ},\,\ii\nu_{n}}G_{14}(\kvec,\ii\nu_{n}).
\label{eq:complex}
\end{equation}

\begin{figure}[!htbp]
\centering
\includegraphics[]{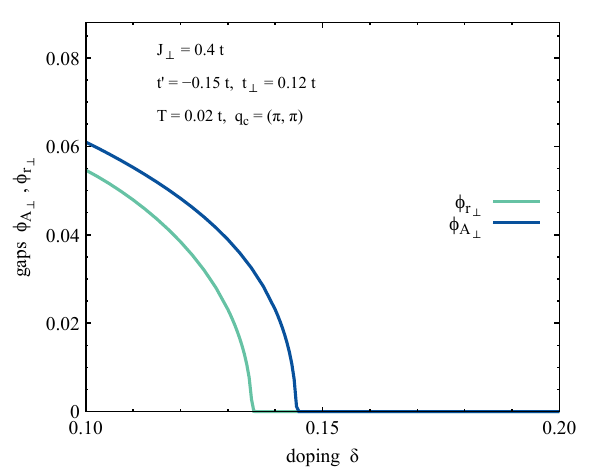}
\caption{$\zBOP$ gaps $\phi_{\rperp}$ and $\phi_{\Aperp}$ as a function of doping
$\dop$ at $T/t=0.02$ for the enhanced interlayer hopping $t_\perp/t=0.12$,
computed separately by restricting $\phi$ in Eq.~(\ref{eq:complex}) to be real or
purely imaginary (all other parameters as in the main text: $\Jperp/t=0.4$,
$t'/t=-0.15$). The larger $t_\perp$ lifts the near-degeneracy seen at
$t_\perp/t=0.05$: the two gaps separate clearly, with $\phi_{\Aperp}$ remaining
above $\phi_{\rperp}$ throughout.}
\label{fig:tperp}
\end{figure}

Equation~(\ref{eq:complex}) can also be solved separately in each channel, by
setting one of the two components of $\phi$ to zero. That choice is preserved by
the gap equation: with $\phi$ real the right-hand side is real, and with $\phi$
purely imaginary it is purely imaginary, so that the equation closes on the
surviving component and each gap is obtained independently, with no competition
between them. These are the gaps compared in Fig.~2(a) of the main text and in
Fig.~\ref{fig:tperp}.

Solving Eq.~(\ref{eq:complex}) in its
unconstrained form at $T/t=0.02$, the
self-consistent solution converges to a purely imaginary gap: $\phi_{\rperp}$
flows to zero and only the $\Aperp$ component $\phi_{\Aperp}$ survives. This selection is set by the competition between the two channels, which are nearly
degenerate but not exactly so: as shown in Fig.~\ref{fig:Tc_zBOP}, the $\Aperp$
instability has the slightly higher critical temperature, and correspondingly a
slightly larger gap [Fig.~2(a) of the main text]. Since both order at the same wave vector
$\qc=(\pi,\pi)$, the two gaps cannot grow independently: the opening of one
reconstructs the Fermi surface and removes the spectral weight that would drive
the other. The marginally leading $\Aperp$ channel therefore sets in first and,
once its gap opens, preempts the $\rperp$ channel, relegating $\phi_{\rperp}$ to
zero. The purely imaginary solution thus reflects this competition rather than an
exact decoupling of the sectors. 

The near-degeneracy of the two channels is itself controlled by the interlayer
hopping $t_\perp$: it is close at the physical value $t_\perp/t=0.05$ used in the
main text and is progressively lifted as $t_\perp$ increases.
Figure~\ref{fig:tperp} shows the $\rperp$ and $\Aperp$ gaps, computed separately,
at the larger value $t_\perp/t=0.12$: the two curves separate clearly and
$\phi_{\Aperp}$ stays above $\phi_{\rperp}$ over the whole doping range. This
exposes the marginal advantage of the $\Aperp$ sector that, at the physical value
$t_\perp/t=0.05$, underlies the selection of the purely imaginary solution.

\section{Peierls phase, loop currents, and orbital magnetic moment}
\label{sec:currents}
\label{sec:moment}

As shown in Sec.~\ref{sec:gapeqs}, the self-consistent solution of
Eq.~(\ref{eq:complex}) is purely imaginary, $\phi=\ii\,\phi_{\Aperp}$. In the
Green's function of Eq.~(\ref{eq:complexG}) the interlayer coupling is independent of $\kvec$ with amplitude $\varepsilon_{\perp}$ and connects $\kvec$ with
$\kvec+\qc$ with amplitude $\ii\phi_{\Aperp}$. In real space these two
contributions combine into a site-dependent interlayer hopping,
\begin{equation}
t^{\mathrm{eff}}_{\perp}(i)=\varepsilon_{\perp}+\ii\,\phi_{\Aperp}\,\epsilon_{i}
=\bigl|t^{\mathrm{eff}}_{\perp}\bigr|\,e^{\ii\theta_{i}},
\qquad
\epsilon_{i}=(-1)^{i_x+i_y},
\label{eq:Peierls}
\end{equation}
with
\begin{equation}
\bigl|t^{\mathrm{eff}}_{\perp}\bigr|
=\sqrt{\varepsilon_{\perp}^{2}+\phi_{\Aperp}^{2}},
\qquad
\theta_{i}=\epsilon_{i}\,\theta_{0},
\qquad
\theta_{0}=\arctan\!\left(\frac{\phi_{\Aperp}}{\varepsilon_{\perp}}\right),
\label{eq:theta0}
\end{equation}
i.e., a sublattice-staggered Peierls phase on the vertical bonds.
Here, $i_x$ and $i_y$ are the lattice indices of site $i$.
The in-plane
hoppings $t$ and $t'$ remain real, and $|t^{\mathrm{eff}}_{\perp}|$ is site
independent, so the $\Aperp$ condensate does not modulate the bond charge: it is a
pure current wave.

\begin{figure}[!htbp]
\centering
\begin{tikzpicture}[
  scale=1.3,
  >={Stealth[length=2.5mm,width=2mm]},
  site/.style={circle,fill=black,inner sep=1.3pt},
  bondL/.style={thick,-{Stealth[length=2.5mm,width=2mm]}}
]
  \begin{scope}[shift={(0,0)}]
    \node[site] at (-1.25,-0.8) {}; \node[below left]  at (-1.25,-0.8) {$(i,1)$};
    \node[site] at ( 1.25,-0.8) {}; \node[below right] at ( 1.25,-0.8) {$(j,1)$};
    \node[site] at (-1.25, 0.8) {}; \node[above left]  at (-1.25, 0.8) {$(i,2)$};
    \node[site] at ( 1.25, 0.8) {}; \node[above right] at ( 1.25, 0.8) {$(j,2)$};
    \node[anchor=east] at (-1.9,-0.8) {Layer 1};
    \node[anchor=east] at (-1.9, 0.8) {Layer 2};
    \draw[bondL] (-1.25,-0.8) -- (-1.25, 0.8);
    \node[left]  at (-1.4,0) {$\theta_{i} = +\theta_{0}$};
    \draw[bondL] (-1.25, 0.8) -- ( 1.25, 0.8);
    \node[above] at (0, 0.85) {$0$};
    \draw[bondL] ( 1.25, 0.8) -- ( 1.25,-0.8);
    \node[right] at ( 1.4,0) {$\theta_{j} = -\theta_{0}$};
    \draw[bondL] ( 1.25,-0.8) -- (-1.25,-0.8);
    \node[below] at (0,-0.85) {$0$};
    \node at (0,0) {$\Phi_{\text{plaq}} = 2\theta_{0}$};
  \end{scope}
  \begin{scope}[shift={(-4.5,-0.8)}]
    \draw[->,thick] (0,0) -- (0.6,0) node[right]{$x$};
    \draw[->,thick] (0,0) -- (0,0.6) node[above]{$z$};
  \end{scope}
\end{tikzpicture}
\caption{Vertical $xz$-plaquette spanning two nearest-neighbor in-plane sites $i$
and $j$ in layers $1$ and $2$. The interlayer bonds at sites $i$ and $j$ carry
opposite Peierls phases $\theta_{i}=+\theta_{0}$ and $\theta_{j}=-\theta_{0}$, and
both contribute $+\theta_{0}$ to the directed loop integral, while the in-plane
bonds are phase-free. The reversal of the interlayer phase at $j$ relative to $i$
is what prevents the two contributions from canceling and yields the plaquette
flux $\Phi_{\mathrm{plaq}}=2\theta_{0}$ of Eq.~(\ref{eq:flux}).}
\label{fig:plaquette}
\end{figure}
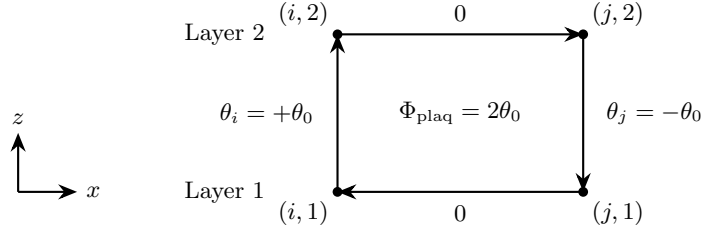

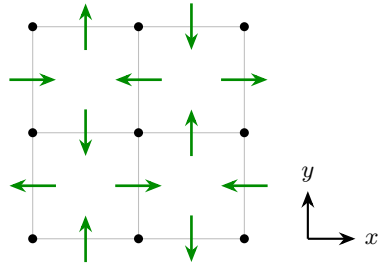
\begin{figure}[!htbp]
\centering
\begin{tikzpicture}[
  scale=1.4,
  >={Stealth[length=2.2mm,width=1.7mm]},
  site/.style={circle,fill=black,inner sep=1.2pt},
  ghostbond/.style={black!25,thin},
  flux/.style={->,thick,green!55!black,line width=1.0pt}
]
  \foreach \i in {0,1,2}{
    \foreach \j in {0,1}{
      \draw[ghostbond] (\i,\j) -- (\i,\j+1);
    }
  }
  \foreach \j in {0,1,2}{
    \foreach \i in {0,1}{
      \draw[ghostbond] (\i,\j) -- (\i+1,\j);
    }
  }
  \foreach \i in {0,1,2}{
    \foreach \j in {0,1,2}{
      \node[site] at (\i,\j) {};
    }
  }
  \draw[flux] (0.5,-0.22) -- (0.5, 0.22);   
  \draw[flux] (1.5, 0.22) -- (1.5,-0.22);   
  \draw[flux] (0.5, 1.22) -- (0.5, 0.78);   
  \draw[flux] (1.5, 0.78) -- (1.5, 1.22);   
  \draw[flux] (0.5, 1.78) -- (0.5, 2.22);   
  \draw[flux] (1.5, 2.22) -- (1.5, 1.78);   
  \draw[flux] ( 0.22, 0.5) -- (-0.22, 0.5); 
  \draw[flux] ( 0.78, 0.5) -- ( 1.22, 0.5); 
  \draw[flux] ( 2.22, 0.5) -- ( 1.78, 0.5); 
  \draw[flux] (-0.22, 1.5) -- ( 0.22, 1.5); 
  \draw[flux] ( 1.22, 1.5) -- ( 0.78, 1.5); 
  \draw[flux] ( 1.78, 1.5) -- ( 2.22, 1.5); 
  \begin{scope}[shift={(2.6,0)}]
    \draw[->,thick] (0,0) -- (0.45,0) node[right]{$x$};
    \draw[->,thick] (0,0) -- (0,0.45) node[above]{$y$};
  \end{scope}
\end{tikzpicture}
\caption{Top-down view of the in-plane magnetic field $\mathbf{B}$ generated by the $\Aperp$ $\zBOP$ at $\qc=(\pi,\pi)$. Each green arrow gives the direction of $\mathbf{B}$ at the center of the vertical plaquette whose projection onto the $xy$ plane is the corresponding bond: $\mathbf{B}$ is perpendicular to the bond, along $\pm\hat{y}$ for horizontal bonds ($xz$-plaquettes) and along $\pm\hat{x}$ for vertical bonds ($yz$-plaquettes), with the sign forming a staggered pattern that doubles the unit cell. The spatially averaged $\mathbf{B}$ vanishes.}
\label{fig:Btopdown}
\end{figure}

The Peierls phase produces a magnetic flux through the vertical plaquettes.
Consider the $xz$-plaquette of Fig.~\ref{fig:plaquette}, spanned by two
nearest-neighbor in-plane sites $i$ and $j$ with $\epsilon_{j}=-\epsilon_{i}$, and
therefore $\theta_{j}=-\theta_{i}$. Traversing the loop
$(i,1)\to(i,2)\to(j,2)\to(j,1)\to(i,1)$ gives the plaquette flux
\begin{equation}
\Phi_{\mathrm{plaq}}=\theta_{i}+0+(-\theta_{j})+0=2\,\theta_{0}\, .
\label{eq:flux}
\end{equation}
The flux alternates in sign from plaquette to plaquette and its spatial average
vanishes. Were $\phi_{\Aperp}$ uniform, the two interlayer contributions would
cancel and the flux would vanish; it is the staggering imposed by $\qc=(\pi,\pi)$
that makes it finite. The same analysis holds for the $yz$-plaquettes, with the
plaquette normal along $\pm\hat{x}$ instead of $\pm\hat{y}$
(Fig.~\ref{fig:Btopdown}).\\

{\bf In-plane current from $G_{13}$ and loop closure}\\

The $4\times4$ construction of Eq.~(\ref{eq:complexG}) automatically provides an in-plane current, through the off-diagonal element $G_{13}$, associated with the
intra-plane bond order at the ordering vector $\qc$,
$\langle f^{\dagger}_{\kvec\sigma,1}f_{\kvec+\qc\,\sigma,1}\rangle$. Had the
condensate been real, $G_{13}$ would be real by symmetry; the imaginary coupling
activates its imaginary part,
\begin{equation}
G_{13}(\kvec,\ii\nu_{n})\propto
\frac{\ii\,\varepsilon_{\perp}\,\phi_{\Aperp}\,
\bigl(\varepsilon^{\parallel}_{\kvec+\qc}-\varepsilon^{\parallel}_{\kvec}\bigr)}
{\det G^{-1}(\kvec,\ii\nu_{n})},
\label{eq:G13SM}
\end{equation}
which is finite whenever $\phi_{\Aperp}\neq0$, provided the two layers are
coherently coupled ($\varepsilon_{\perp}\neq0$) and the two folded bands are
inequivalent
($\varepsilon^{\parallel}_{\kvec+\qc}\neq\varepsilon^{\parallel}_{\kvec}$).
It follows that the in-plane current accompanying the $\zBOP$
is~\cite{hsu91}
\begin{equation}
\bigl\langle J^{\,x(y)}_{13}\bigr\rangle
=\!\!\sum_{\kvec\in\mathrm{RBZ},\,\ii\nu_{n}}\!\!
v_{x(y)}\!\Bigl(\kvec+\frac{\qc}{2}\Bigr)\,G_{13}(\kvec,\ii\nu_{n}),
\label{eq:J13}
\end{equation}
with $\mathbf{v}(\kvec)=\nabla_{\kvec}\,\varepsilon^{\parallel}_{\kvec}$ obtained
from Eq.~(\ref{eq:dispersion}),
\begin{equation}
v_{x}(\kvec)=t\dop\sin k_{x}+2\,t'\dop\,\sin k_{x}\cos k_{y},
\qquad
v_{y}(\kvec)=t\dop\sin k_{y}+2\,t'\dop\,\cos k_{x}\sin k_{y}.
\label{eq:vel}
\end{equation}
The interlayer amplitude $\varepsilon_{\perp}$ does not depend on $\kvec$ and does
not contribute to $\mathbf{v}$. For a particle-hole pair with momentum
transfer $\qc$ the group velocity of the in-plane dispersion is evaluated at the
midpoint $\kvec+\qc/2$ of the bond. In the $\Aperp$ channel
$\mathrm{Re}\,\langle G_{13}\rangle=0$, so the current is carried entirely by the
imaginary parts, and the quantity we report is the magnitude
\begin{equation}
\bigl|\langle J_{13}\rangle\bigr|
=\sqrt{\bigl[\mathrm{Im}\,\langle J^{\,x}_{13}\rangle\bigr]^{2}
+\bigl[\mathrm{Im}\,\langle J^{\,y}_{13}\rangle\bigr]^{2}}\;,
\label{eq:Jmod}
\end{equation}
with $\langle J^{x}_{13}\rangle=\langle J^{y}_{13}\rangle$, as required by the
$C_{4}$ symmetry of the $\qc=(\pi,\pi)$ state.

Being modulated with $\qc$, this current averages to zero over the plane, so there
is no net transport current. It is the in-plane current that closes the interlayer
loops sketched in Fig.~1 of the main text.\\

{\bf Orbital magnetic moment and local magnetic field}\\

The currents above are given in units of $t$. We now restore $e$ and $\hbar$ in
order to estimate the orbital magnetic moment carried by a single loop and the
magnetic field it generates.

Since the midpoint velocity $t\dop\cos k_{x}$ is twice the nearest-neighbor
amplitude $t\dop/2$ of Eq.~(\ref{eq:dispersion}), the current of an individual
in-plane bond is half the momentum sum of Eq.~(\ref{eq:J13}),
\begin{equation}
\langle j^{\parallel}\rangle=\frac{2e}{\hbar}\,
\mathrm{Im}\bigl\langle J^{x}_{13}\bigr\rangle ,
\label{eq:jpar}
\end{equation}
which already includes the sum over the two spin components. The four in-plane
bonds meeting at a site feed the vertical one, so
$\langle j^{z}\rangle=4\langle j^{\parallel}\rangle$.

The moment cannot be read off directly from the bond currents, because the four
bonds of a vertical plaquette do not carry the same current: the vertical ones
carry $\langle j^{z}\rangle$ and the in-plane ones $\langle j^{z}\rangle/4$. We
therefore view the pattern of Fig.~1 of the main text as a set of elementary loops,
one for each vertical plaquette. Each loop carries a single current
$j_{\mathrm{loop}}$ around its four segments, and the current of a bond is the sum
of the loops sharing it. An in-plane bond belongs to one plaquette only, and thus
carries $j_{\mathrm{loop}}$. A vertical bond is shared by the four plaquettes
attached to it, which circulate through it in the same direction
because the four
neighboring sites belong to the opposite sublattice; it thus carries
$4j_{\mathrm{loop}}$. The two readings give
\begin{equation}
j_{\mathrm{loop}}=\langle j^{\parallel}\rangle
=\frac{\langle j^{z}\rangle}{4}\, .
\label{eq:jloop}
\end{equation}

Each vertical plaquette is then a rectangular loop of sides $a$ and $d$, the
in-plane and intrabilayer lattice constants, carrying the current
$j_{\mathrm{loop}}$. Its orbital moment, and the field obtained by adding the
Biot-Savart contributions of the four straight segments at the geometric center~\cite{hsu91},
are
\begin{equation}
m=j_{\mathrm{loop}}\,a\,d,
\qquad
B=\frac{2\mu_{0}\,j_{\mathrm{loop}}}{\pi}\,\frac{\sqrt{a^{2}+d^{2}}}{a\,d}\, ,
\label{eq:mandB}
\end{equation}
with $\mathbf{B}$ along $\pm\hat{y}$ for the $xz$ plaquettes and along $\pm\hat{x}$
for the $yz$ ones, alternating in sign with the sublattice index, so that its
spatial average vanishes. 

We note that in the large-$N$ formalism the hopping parameters and the interaction
strengths are rescaled by $1/N$ (Sec.~\ref{sec:formalism}). When $N$ is set to its
physical value $N=2$, the natural unit of energy is $t/2$, for which we take $t/2\simeq0.5$~eV~\cite{luo23} for the
nearest-neighbor hopping of the $d_{x^{2}-y^{2}}$ orbital. For the in-plane and intrabilayer lattice constants we take $a=3.83$~\AA\ and $d=4.3$~\AA~\cite{sun23}. At $\dop=0.13$, where
$\phi_{\Aperp}=0.0414\,t\simeq41$~meV and
$|\langle J_{13}\rangle|=5.6\times10^{-4}t$,
\begin{equation}
\langle j^{z}\rangle=7.8\times10^{-7}\,\mathrm{A},
\quad
j_{\mathrm{loop}}=1.9\times10^{-7}\,\mathrm{A},
\quad
m=3.2\times10^{-26}\,\mathrm{J/T}=3.5\times10^{-3}\,\mu_{B},
\quad
B=5.4\;\mathrm{G}.
\label{eq:numbers}
\end{equation}
The moment follows $|\langle J_{13}\rangle|$ and therefore decreases with doping,
from $5.9\times10^{-3}\mu_{B}$ ($B=9.2$~G) at $\dop=0.10$ down to zero at the QCP
[Fig.~2(c) of the main text]. In the normal state ($\phi_{\Aperp}=0$) all these
quantities vanish identically, so each of them is a direct signature of the order
parameter.

\section{Competition between superconductivity and interlayer loop currents}
\label{sec:competition}

Having discussed the $\zBOP$ separately, we now treat it together with
superconductivity. We introduce the eight-spinor field~\cite{bejas25}
\begin{equation}
\psi^{\dagger}_{\kvec}=\bigl(
f^{\dagger}_{\kvec\uparrow,1},\;f^{\dagger}_{\kvec\uparrow,2},\;
f_{-\kvec\downarrow,1},\;f_{-\kvec\downarrow,2},
f^{\dagger}_{\kvec+\qc\uparrow,1},\;f^{\dagger}_{\kvec+\qc\uparrow,2},\;
f_{-\kvec-\qc\downarrow,1},\;f_{-\kvec-\qc\downarrow,2}\bigr),
\label{eq:spinorcomp}
\end{equation}
where $\qc$ is the ordering momentum of the $\zBOP$. The inverse of the
$8\times8$ Green function can be written as
\begin{equation}
G^{-1}(\kvec,\ii\nu_{n})=
\begin{pmatrix}
A & B\\
B & C
\end{pmatrix},
\label{eq:G8x8}
\end{equation}
with $A$, $B$, and $C$ the following $4\times4$ matrices
\begin{equation}
A=
\begin{pmatrix}
\ii\nu_{n}-\varepsilon^{\parallel}_{\kvec} & -\varepsilon_{\perp} & 0 & -\Dperp\\[2pt]
-\varepsilon_{\perp} & \ii\nu_{n}-\varepsilon^{\parallel}_{\kvec} & -\Dperp & 0\\[2pt]
0 & -\Dperp & \ii\nu_{n}+\varepsilon^{\parallel}_{\kvec} & \varepsilon_{\perp}\\[2pt]
-\Dperp & 0 & \varepsilon_{\perp} & \ii\nu_{n}+\varepsilon^{\parallel}_{\kvec}
\end{pmatrix},
\label{eq:GA}
\end{equation}
\begin{equation}
B=
\begin{pmatrix}
0 & -\phi & 0 & 0\\[2pt]
-\phi^{*} & 0 & 0 & 0\\[2pt]
0 & 0 & 0 & \phi^{*}\\[2pt]
0 & 0 & \phi & 0
\end{pmatrix},
\label{eq:GB}
\end{equation}
and $C$ has the form of $A$ with $\kvec+\qc$ instead of $\kvec$. Here
$\varepsilon^{\parallel}_{\kvec}$ and $\varepsilon_{\perp}$ are given by
Eq.~(\ref{eq:dispersion}), and $\Dperp$ and $\phi$ are the out-of-plane
superconducting and $\zBOP$ gap, respectively. Only the out-of-plane
superconducting gap enters Eq.~(\ref{eq:GA}), since the in-plane $d$-wave
channel is absent for $\Jpar=0$. As in Sec.~\ref{sec:selection} the $\zBOP$ gap
is purely imaginary, $\phi=\ii\,\phi_{\Aperp}$ with $\phi_{\Aperp}$ real, so
that $\phi^{*}=-\phi$ and all the nonzero entries of $B$ reduce to
$\pm\ii\,\phi_{\Aperp}$. The gap equations for $\Dperp$ and $\phi$ 
are~\cite{bejas25}
\begin{equation}
\Dperp=-\frac{\Jperp}{4}\sum_{\kvec,\,\ii\nu_{n}}
G_{14}(\kvec,\ii\nu_{n}),
\label{eq:Dperpcomp}
\end{equation}
and
\begin{equation}
\phi=-\frac{\Jperp}{2}\sum_{\kvec,\,\ii\nu_{n}}
G_{16}(\kvec,\ii\nu_{n}),
\label{eq:phicomp}
\end{equation}
and must be solved self consistently.

Equation~(\ref{eq:G8x8}) holds for a commensurate momentum; an incommensurate one
requires including the additional symmetry-related momenta. The ordering vector
used here, $\qc=(\pi,0.76\pi)$, stays close to $(\pi,\pi)$, so the $8\times8$
construction remains a good approximation and we keep it, as in
Ref.~\cite{bejas25}. The momentum sums in Eqs.~(\ref{eq:Dperpcomp}) and
(\ref{eq:phicomp}) then run over the full Brillouin zone, and not over the RBZ as
in Sec.~\ref{sec:gapeqs}.

\section{Analysis for different $\Jperp$}
\label{sec:Jperp}

Throughout the paper we have used $\Jperp=0.4t$. In this section we show that
$\Jperp$ sets the temperature and doping scales but does not change the physics.

\begin{figure}[ht]
\centering
\includegraphics[]{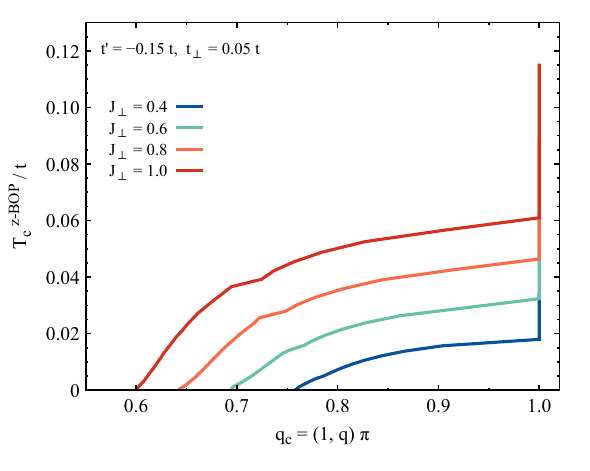}
\caption{$\zBOP$ instability line in the $(\qc,T_{c}^{\zBOP})$ plane, for the
values of $\Jperp$ indicated in the legend. Each curve is parametrized by the
doping, and ends on the horizontal axis at the quantum critical point. The
vertical segment at $q=1$ is the commensurate plateau $\qc=(\pi,\pi)$, obtained
at low doping; on approaching the quantum critical point the ordering vector
incommensurates and the curve leaves $q=1$.}
\label{fig:qcT}
\end{figure}

Figure~\ref{fig:qcT} shows the $\zBOP$ instability line in the
$(\qc,T_{c}^{\zBOP})$ plane for several values of $\Jperp$; each curve is
parametrized by the doping. At $T=0$, the ordering vector is incommensurate. It moves towards $(\pi,\pi)$ as the
temperature increases, and above a value of $T_{c}^{\zBOP}$ that grows with
$\Jperp$ the instability becomes commensurate: that is the vertical segment at
$q=1$. The incommensurability
takes place along the $(\pi, 0)$-$(\pi,\pi)$ direction,
with $q_{x}=\pi$ in every case, so that $\qc=(1,q)\pi$.

Increasing $\Jperp$ raises the critical temperature and pushes the whole structure
to larger doping. The maximum $T_{c}^{\zBOP}$ grows from $0.034t$ at
$\Jperp=0.4t$ to $0.115t$ at $\Jperp=1.0t$, the commensurate plateau extends from
$\dop\lesssim0.15$ to $\dop\lesssim0.26$, and the quantum critical point moves from
$\dop^{*}=0.22$ to $\dop^{*}=0.39$. The ordering vector at $T=0$ moves away from
$(\pi,\pi)$ as $\Jperp$ grows, from $q=0.76$ at $\Jperp=0.4t$, the value used in
Sec.~\ref{sec:competition}, down to $q=0.60$ at $\Jperp=1.0t$.

\begin{figure}[ht]
\centering
\includegraphics[]{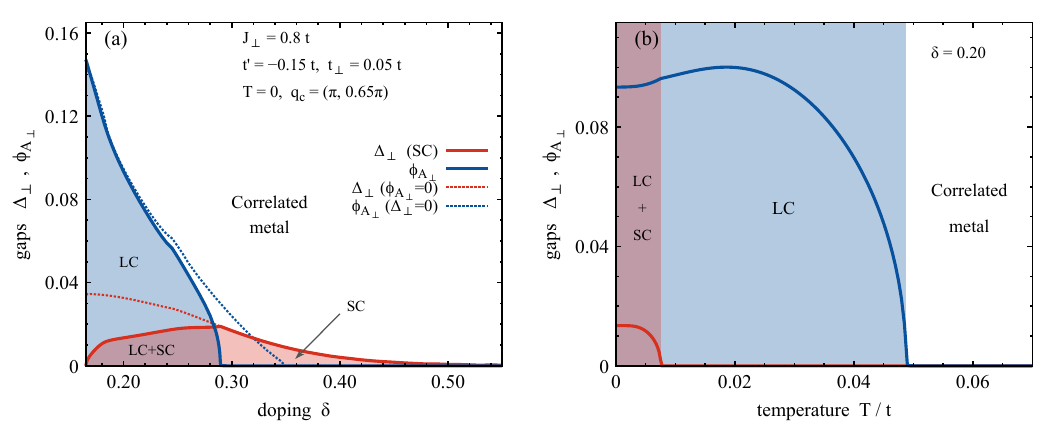}
\caption{Same as Fig.~3 of the main text, for $\Jperp=0.8t$.}
\label{fig:Jp08}
\end{figure}

Figure~\ref{fig:Jp08} repeats for $\Jperp=0.8t$ the analysis of Fig.~3 of the main
text. The superconducting gap again shows a dome, with its optimal doping close to
the $\zBOP$ quantum critical point, and the phase diagram again contains a region
of pure loop currents, a coexistence region, a region of pure superconductivity,
and a correlated metal at large doping. What changes are the scales: both gaps and
both critical temperatures increase, and the $\zBOP$ now extends up to
$\dop^{*}=0.35$ without competition instead of $0.22$. The interplay between the
loop-current phase and superconductivity is therefore a robust feature of the
model, and not a consequence of the particular value of $\Jperp$.

\section{Absence of topologically protected surface states}
\label{sec:edge}

We ask here whether the coexisting LC and superconducting state supports protected
surface states, which would produce a zero-bias peak in tunneling.

Since $q_{x}=\pi$ exactly, the modulation of the interlayer bond field at the site of
in-plane coordinates $(x,y)$ factorizes as
\begin{equation}
\cos(q_{x}x+q_{y}y+\theta)=(-1)^{x}\cos(q_{y}y+\theta),
\label{eq:factorize}
\end{equation}
so that the $(-1)^{x}$ doubles the unit cell along $x$ while the second factor is a
row-dependent amplitude. We therefore keep $k_{x}$ as a good quantum number, open the
system along $y$ over $N_{y}$ rows, and diagonalize the resulting $8N_{y}\times8N_{y}$
Hamiltonian built from Eqs.~(\ref{eq:G8x8})--(\ref{eq:GB}) at the self-consistent
gaps. The phase $\theta$ fixes the offset of the current modulation relative to the
first row, that is, the termination of the crystal, and is scanned below.

At fixed $k_{\parallel}$ the problem perpendicular to a specular surface is a chain
with the Bogoliubov--de Gennes structure
$\mathcal{H}=\varepsilon\,\tau_{z}+\Delta\,\tau_{x}$, where $\tau_{i}$ are Pauli
matrices in Nambu space, $\varepsilon$ is the normal-state dispersion along the surface
normal and $\Delta$ the pair potential, both at the given $k_{\parallel}$; $\Delta$ is
kept general at this stage. For a quasiparticle energy $E$ inside the gap,
$\varepsilon=\pm\ii\,\Omega$ with $\Omega=\sqrt{|\Delta|^{2}-E^{2}}$, so that two
solutions $z^{\,y}\chi_{\pm}$ with $|z|<1$ decay into the bulk, with
\begin{equation}
\chi_{\pm}=\bigl(\Delta_{\pm},\;E\mp\ii\,\Omega_{\pm}\bigr)^{T},
\qquad
\Omega_{\pm}=\sqrt{|\Delta_{\pm}|^{2}-E^{2}},
\label{eq:chipm}
\end{equation}
where $\Delta_{+}$ and $\Delta_{-}$ denote $\Delta$ evaluated at the incident and at
the specularly reflected momentum, which differ because the surface maps
$k_{\perp}\to-k_{\perp}$. A hard wall requires $c_{+}\chi_{+}+c_{-}\chi_{-}=0$ and hence
$\det[\chi_{+}|\chi_{-}]=0$, which for $|\Delta_{+}|=|\Delta_{-}|\equiv\Delta_{0}$ and a
phase difference $\zeta=\arg\Delta_{+}-\arg\Delta_{-}$ gives the bound-state
energy
\begin{equation}
E_{b}(k_{\parallel})=\pm\,\Delta_{0}\,
\cos\!\left[\frac{\zeta(k_{\parallel})}{2}\right].
\label{eq:Eb}
\end{equation}
For $\zeta=\pi$ this is the zero-energy condition
$\Delta(\mathbf{k}_{\rm inc})\Delta(\mathbf{k}_{\rm ref})<0$ of
Refs.~\cite{hu94,tanaka95,kashiwaya00}.

In the present model $\Jperp$ acts on the vertical bond only, and the in-plane
$d$-wave channel is absent since we did not consider the in-plane exchange interaction $J$, so $\Dperp$ enters Eq.~(\ref{eq:GA}) without a form factor.
Then $\Delta_{+}=\Delta_{-}=\Dperp$, $\zeta\equiv0$ and $E_{b}=\pm\Dperp$: the
bound state lies at the gap edge for any surface orientation and any doping. We keep
the order parameters at their bulk values and do not re-solve them near the surface,
where in a $d$-wave superconductor they are known to reconstruct~\cite{kuboki14};
this does not affect the conclusion, since $\zeta$ is set by the relative phase of the
two gaps and not by their magnitude, so any spatial variation of the on-site $\Dperp$
still gives $\zeta\equiv0$. The
$s^{\pm}$ character does not change this, since the sign change is between the
bonding and antibonding bands and not along the Fermi surface of either, and specular
reflection does not mix them. The LCs cannot generate zero modes either: for a single
site with two layers and no dispersion, Eqs.~(\ref{eq:GA})--(\ref{eq:GB}) give
\begin{equation}
E=\pm\sqrt{\varepsilon_{\perp}^{2}+\phi_{\Aperp}^{2}+\Dperp^{2}} ,
\label{eq:quadrature}
\end{equation}
so the three channels add in quadrature and $\phi_{\Aperp}$ can only increase $|E|$.
The situation differs from that of a $d$-wave superconductor near a (110) surface,
where a locally induced flux phase splits the zero-energy Andreev bound
state~\cite{kuboki20}: there the currents shift a state that the pairing itself has
produced, whereas here there is no such state to shift.

A second and independent argument constrains the chiral channel. By
Eq.~(\ref{eq:factorize}) a translation by one site along $x$ reverses the current
modulation, and so does time reversal $\mathcal{T}$, so that
\begin{equation}
\Theta=\mathcal{T}\circ T_{(1,0)}
\label{eq:theta}
\end{equation}
is an exact symmetry for any $\qc=(\pi,q_{y})$, commensurate or not. Being
antiunitary, $\Theta$ maps the Berry curvature onto minus itself and forces the Chern
number $C$ to vanish throughout the coexistence region; it also forbids an $s+\ii s'$
admixture, and the self-consistent solution returns an imaginary part of $\Dperp$
equal to zero. A vanishing $C$ excludes chiral branches but not flat bands protected by other invariants, such as the zero-energy band of a $d$-wave
(110) surface, where the bulk is nodal; those are excluded by Eq.~(\ref{eq:Eb}).

\begin{figure}[!htbp]
\centering
\includegraphics[]{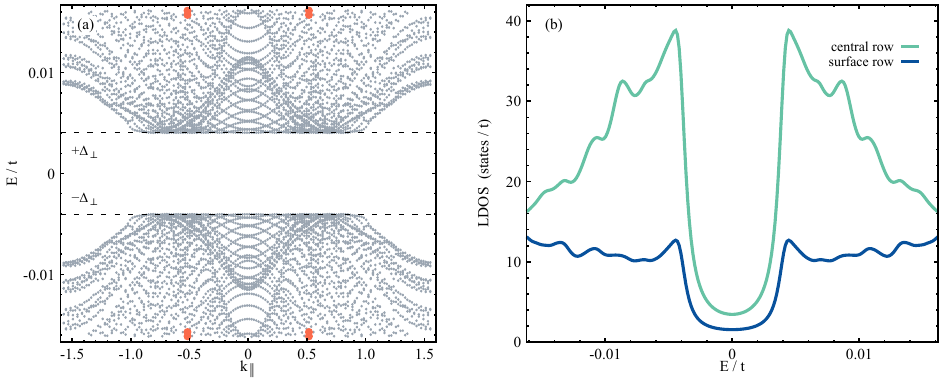}
\caption{(a) Quasiparticle spectrum of a ribbon of $N_{y}=80$ rows at $\dop=0.15$,
with the self-consistent gaps of Fig.~3 of the main text
($\Dperp=4.046\times10^{-3}$, $\phi_{\Aperp}=2.854\times10^{-2}$,
$\qc=(\pi,0.76\pi)$, $T=0$). Dashed lines mark $\pm\Dperp$. Gray dots are extended
states; orange dots are the levels with more than $50\%$ of their weight on the
outermost rows, which appear only at $|E|\simeq3.9\,\Dperp$ and are surface
resonances. (b) Local density of states on the outermost and on the central row. The surface row carries less spectral weight,
having fewer neighbors, but shows the same gap and no zero-bias structure.}
\label{fig:edge}
\end{figure}

Figure~\ref{fig:edge} shows the ribbon spectrum and the local density of states at
$\dop=0.15$, obtained with no adjustable parameters. We find
\begin{equation}
\frac{\min|E|_{\rm ribbon}}{\Delta_{\rm bulk}}=1.0 ,
\label{eq:ratio}
\end{equation}
with $\Delta_{\rm bulk}=\Dperp$, as expected for an on-site gap of uniform magnitude
over the Fermi surface. None of the $6784$ levels with $|E|<3.5\,\Dperp$ carries more
than $50\%$ of its weight on the outermost rows, and the surface density of states
shows the same gap as the bulk one.

Since $\theta\to\theta+\pi$ reverses all the currents and leaves the spectrum
unchanged, the inequivalent range is $\theta\in[0,\pi)$. Scanning it at
$\theta=0,\,0.5,\,1.0,\,1.5,\,2.0,\,2.5$ and $3.0$ gives
$\min|E|/\Delta_{\rm bulk}=1.0$ in every case, and the same holds over the whole
coexistence window $\dop\in[0.11,0.2125]$, with $C=0$ at every doping. Repeating the
mean field, the gap equations and the ribbon at the commensurate $\qc=(\pi,\pi)$
changes the state quantitatively but neither the surface spectrum nor the conclusions
of this section.

\end{document}